\documentclass[twocolumn]{aastex62}

\usepackage{amsmath}
\usepackage{lipsum}

\def\aap{A\&A}

\def\apjs{ApJS}
\def\apjl{ApJL}

\newcommand{\expf}[1]{{{\rm e}^{#1}}}

\newcommand{\vdata}{\mathbf{d}}
\newcommand{\vmap}{\mathbf{m}}
\newcommand{\vamp}{\mathbf{a}}
\newcommand{\mP}{\mathbf{P}}
\newcommand{\mF}{\mathbf{F}}
\newcommand{\mZ}{\mathbf{Z}}
\newcommand{\ninv}{\mathbf{N}^{-1}}
\newcommand{\T}{^\mathrm{T}}

\usepackage{multirow}
\received{xxx}
\revised{yyy}
\accepted{zzz}
\submitjournal{ApJ}

\begin{document}

%%%%%%%%%%%%%%%%%%%%%%%%%%%%%%%%%%%%%%%%%%%%%%%%%%%%%%%%%%%%%%%%%%%%%%%%%%%%%%%%

\title{Constraining the Anomalous Microwave Emission Mechanism in the S140 Star Forming Region \\ with Spectroscopic Observations Between 4 and 8~GHz at the Green Bank Telescope}

%%%%%%%%%%%%%%%%%%%%%%%%%%%%%%%%%%%%%%%%%%%%%%%%%%%%%%%%%%%%%%%%%%%%%%%%%%%%%%%%

\correspondingauthor{Maximilian H. Abitbol}
\email{mha2125@columbia.edu}

%%%%%%%%%%%%%%%%%%%%%%%%%%%%%%%%%%%%%%%%%%%%%%%%%%%%%%%%%%%%%%%%%%%%%%%%%%%%%%%%

\author{Maximilian H. Abitbol}
\author{Bradley R. Johnson}
\author{Glenn Jones}
\affiliation{Department of Physics, Columbia University, New York, NY 10027, USA}

\author{Clive Dickinson}
\author{Stuart Harper}
\affiliation{Jodrell Bank Centre for Astrophysics, Alan Turing Building, School of Physics and Astronomy, The University of Manchester, Oxford Road, Manchester, M13 9PL, U.K.}

%%%%%%%%%%%%%%%%%%%%%%%%%%%%%%%%%%%%%%%%%%%%%%%%%%%%%%%%%%%%%%%%%%%%%%%%%%%%%%%%

\begin{abstract}
Anomalous microwave emission (AME) is a category of Galactic signals that cannot be explained by synchrotron radiation, thermal dust emission, or optically thin free-free radiation.
Spinning dust is one variety of AME that could be partially polarized and therefore relevant for ongoing and future cosmic microwave background polarization studies.
The {\it Planck} satellite mission identified candidate AME regions in approximately $1^\circ$ patches that were found to have spectra generally consistent with spinning dust grain models.
The spectra for one of these regions, G107.2+5.2, was also consistent with optically thick free-free emission because of a lack of measurements between 2 and 20~GHz.
Follow-up observations were needed.
Therefore, we used the C-band receiver (4 to 8~GHz) and the VEGAS spectrometer at the Green Bank Telescope to constrain the AME mechanism.
For the study described in this paper, we produced three band averaged maps at 4.575, 5.625, and 6.125~GHz and used aperture photometry to measure the spectral flux density in the region relative to the background. 
We found if the spinning dust description is correct, then the spinning dust signal peaks at $30.9 \pm 1.4$~GHz, and it explains the excess emission. 
The morphology and spectrum together suggest the spinning dust grains are concentrated near S140, which is a star forming region inside our chosen photometry aperture.
If the AME is sourced by optically thick free-free radiation, then the region would have to contain HII with an emission measure of $5.27^{+2.5}_{-1.5}\times 10^8$~$\rm{cm^{-6}\,pc}$ and a physical extent of $1.01^{+0.21}_{-0.20} \times 10^{-2}$\,pc.
This result suggests the HII would have to be ultra or hyper compact to remain an AME candidate.
\end{abstract}

%%%%%%%%%%%%%%%%%%%%%%%%%%%%%%%%%%%%%%%%%%%%%%%%%%%%%%%%%%%%%%%%%%%%%%%%%%%%%%%%

\keywords{radio continuum: ISM - radiation mechanisms: general - ISM: dust - ISM: HII regions - ISM: clouds - cosmology: cosmic background radiation }

%%%%%%%%%%%%%%%%%%%%%%%%%%%%%%%%%%%%%%%%%%%%%%%%%%%%%%%%%%%%%%%%%%%%%%%%%%%%%%%%

\section{Introduction}
\label{sec:introduction}

%%%%%%%%%%%%%%%%%%%%%%%%%%%%%%%%%%%%%%%%%%%%%%%%%%%%%%%%%%%%%%%%%%%%%%%%%%%%%%%%

% What is AME?

Diffuse Galactic signals obscure our view of the cosmic microwave background (CMB).
Ongoing and future CMB polarization studies will likely be limited by these Galactic foreground signals~\citep{Errard2016}.
Component separation analysis methods currently being used for CMB polarization studies commonly consider only Galactic dust emission and synchrotron radiation~\citep[see][for example]{Wehus2016}.
There may be additional signals to consider as well.

Diffuse Galactic microwave signals that are not synchrotron radiation, optically thin free-free emission, or thermal dust emission are commonly referred to as anomalous microwave emission (AME)~\citep{Dickinson2018}.
AME was first detected near the north celestial pole~\citep{Leitch1997}.
Since then, evidence for AME has been reported in many other regions as well~\citep[see][and references therein]{Harper2015}.
The reported AME signals have been detected between approximately 10 and 60~GHz, and active AME research is focused on understanding the emission mechanisms~\citep{hensley2017, draine2016}.
The emission mechanism models that are currently being considered include (i) flat-spectrum synchrotron radiation~\citep{Bennett2003, Kogut1996}, (ii) optically thick free-free emission from, for example, ultra compact HII (UCHII) regions~\citep{Dickinson2013, Kurtz2002}, (iii) thermal magnetic dust emission~\citep{Draine1999}, and (iv) emission from rapidly rotating dust grains that have an electric dipole moment~\citep{Draine1998a, Draine1998b}.

%%%%%%%%%%%%%%%%%%%%%%%%%%%%%%%%%%%%%%%%%%%%%%%%%%%%%%%%%%%%%%%%%%%%%%%%%%%%%%%%

\begin{figure}
\includegraphics[width=\columnwidth]{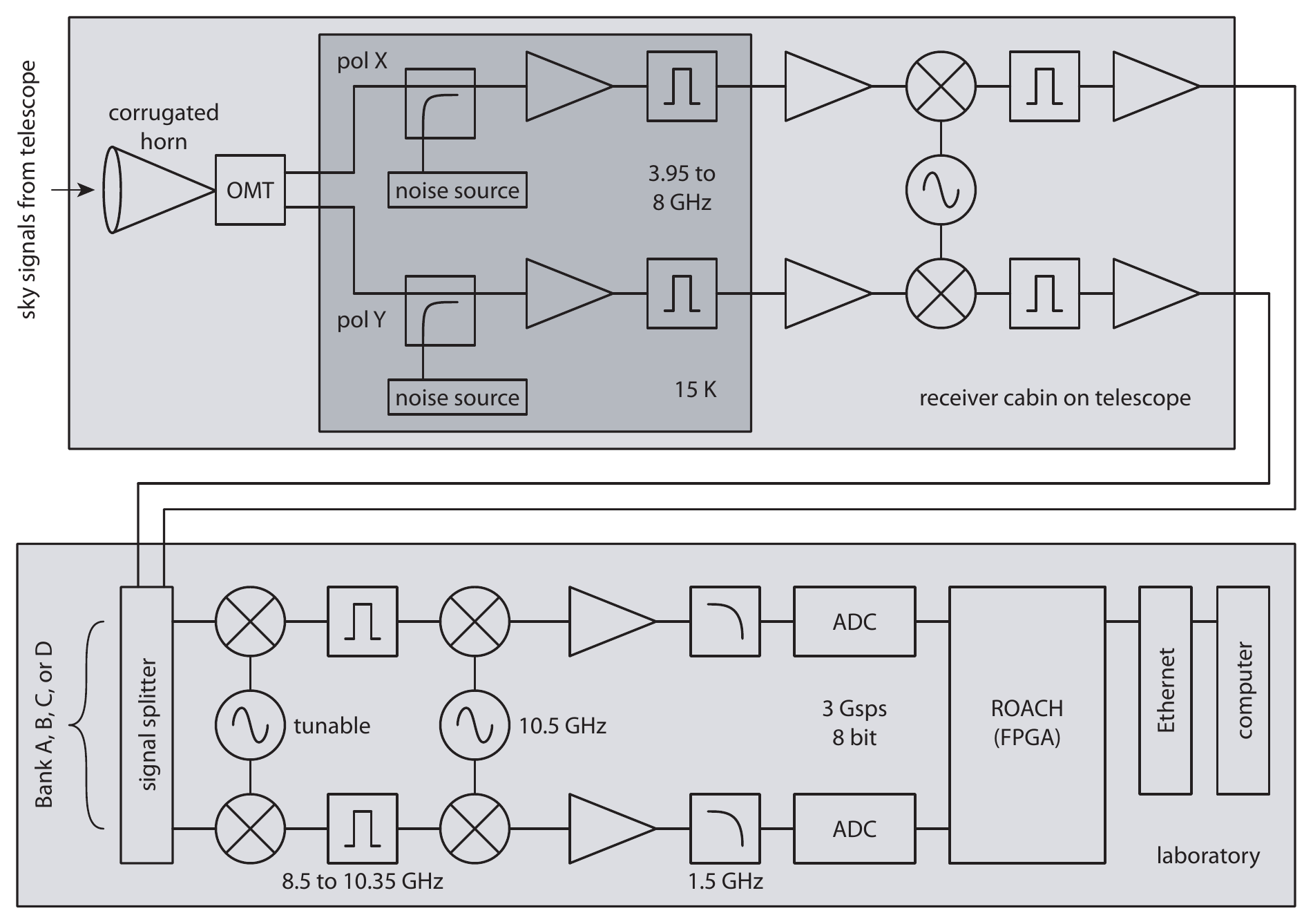}
\caption{
A schematic of the GBT instrument we used for this study.
The C-band receiver elements in the receiver cabin on the telescope are shown in the box on the top.
The digital spectrometer elements in the laboratory are shown in the box on the bottom.
For clarity, just one spectrometer bank is shown.
More details are given in Section~\ref{sec:spectrometer}.
}
\label{fig:receiver}
\end{figure}

%%%%%%%%%%%%%%%%%%%%%%%%%%%%%%%%%%%%%%%%%%%%%%%%%%%%%%%%%%%%%%%%%%%%%%%%%%%%%%%%

Spinning dust grains are expected to produce linearly polarized signals~\cite[see][for example]{Lazarian2000}, and the theoretical emission spectrum for spinning dust grains can extend up to frequencies above 80~GHz, where the CMB polarization anisotropy is commonly being observed.
Therefore, spinning-dust emission could be a third important polarized Galactic foreground signal that should be considered for CMB polarization studies~\citep{Hervias-Caimapo2016, Remazeilles2016, Armitage-Caplan2012}.
Observational evidence to date suggests the AME signal can be partially polarized at the level of a few percent or less~\citep{Genova-Santos2017, Planck2016, Dickinson2011}.
However, this detected level of polarization is still appreciable because the CMB polarization anisotropy signals are polarized at a level of $\sim10^{-6}$ or less~\citep[see][and references therein]{Staggs2018}.
More investigation is required to see if polarized AME would bias future CMB polarization anisotropy measurements.

% What did we do for this project?

Active spinning-dust research focuses on searching for and characterizing regions with spinning dust signal.
Discovering spinning-dust regions is challenging because they need to be detected both spectroscopically and morphologically~\citep{battistelli2015, paladini2015}.
Using multi-wavelength analyses members of the Planck Collaboration have identified several regions that could contain spinning-dust signal~\citep{PlanckXX, PlanckXV}.
However, there are limited observations between 2 and 20~GHz~\citep{Genova-Santos2015}, so there is some remaining uncertainty in the AME emission mechanism in these regions.
As a result, these Planck-discovered regions are excellent targets for follow-up spinning-dust studies.
One target is near the star-forming region S140~\citep{Sharpless1959}, and it is centered on $(l, b) = (107.2^\circ, 5.20^\circ$), which we will refer to in this paper as G107.2+5.20.
Previous analysis of this region showed that both spinning dust and UCHII models fit the data well~\citep{perrott2013, PlanckXV}.
In an effort to further constrain the emission mechanism in this region and possibly expand the catalog of known spinning-dust regions, we made spectropolarimetric measurements of the region using the the 100-m Green Bank Telescope (GBT) in West Virginia~\citep{Jewell2004}.
Specifically, we used the C-band receiver (4 to 8~GHz) and the Versatile GBT Astronomical Spectrometer (VEGAS), which is a digital back-end~\citep{Prestage2015}.
During our 18 hours of observing (10 hours mapping and 8 hours calibrating) we measured all four Stokes parameters of a nearly circular region centered on G107.2+5.20.

% How is the paper organized?

In this paper, we first describe the instrument and the observations in Section~\ref{sec:observations}.
The analysis methods are described in Section~\ref{sec:datareduction}.
%
%This paper focuses on using our intensity measurements (the Stokes $I$ parameter) in conjunction with other measurements for spectroscopic determination of the emission mechanism of the signal.
%
Our measurements of the spatial morphology of the intensity of the region (the Stokes $I$ parameter) and the derived spectroscopic results are presented in Section~\ref{sec:results}.
Our polarization results (the Stokes $Q$, $U$, and $V$ maps) will be published in a future paper. \vfill

%%%%%%%%%%%%%%%%%%%%%%%%%%%%%%%%%%%%%%%%%%%%%%%%%%%%%%%%%%%%%%%%%%%%%%%%%%%%%%%%

\begin{table}
\centering
\caption{
Definition of the four spectral banks.
Each bank is divided into 16,384 channels that are 91.552~kHz wide yielding the raw bandwidth, $\Delta \nu_r$.
The subscript $c$ denotes center frequency.
We ultimately used 6,400 channels in each bank (see Section~\ref{sec:rfi}), so the selected bandwidth for map making is $\Delta \nu_s$.
The estimated beam full-width at half-maximum (FWHM) for each bank is listed as well.
}
\label{table:spectral_banks}
\begin{tabular}{ccccc}
\hline 
Bank & $\nu_c$   & $\Delta \nu_r$  & $\Delta \nu_s$ & FWHM$_c$     \\ [-1mm]
     & [\,GHz\,] & [\,GHz\,]      & [\,GHz\,]       & [\,arcmin\,] \\ [1mm] \hline
A    & 4.575     & 3.975 - 5.225  & 4.407 - 4.993   & 2.75         \\
B    & 5.625     & 5.025 - 6.275  & 5.457 - 6.043   & 2.25         \\
C    & 6.125     & 5.525 - 6.775  & 5.957 - 6.543   & 2.05         \\ 
D    & 7.175     & 6.575 - 7.825  & 7.007 - 7.593   & 1.75         \\ [1mm] \hline
\end{tabular}
\end{table}

%%%%%%%%%%%%%%%%%%%%%%%%%%%%%%%%%%%%%%%%%%%%%%%%%%%%%%%%%%%%%%%%%%%%%%%%%%%%%%%%

%%%%%%%%%%%%%%%%%%%%%%%%%%%%%%%%%%%%%%%%%%%%%%%%%%%%%%%%%%%%%%%%%%%%%%%%%%%%%%%%

\begin{table*}
\centering
\caption{
Data sets used in this study.
We used a circular aperture with a radius of 45$^{\prime}$ to determine the spectral flux density~(SFD).
}
\label{tab:datasets}
\begin{tabular}{ccccc}
\hline
Experiment        & Frequency & Beam FWHM & Aperture SFD    & Reference           \\ [-0.5mm]
                  & [GHz]     & [arcmin]  & [Jy]            &                     \\ [1mm]
\hline
%Haslam            & 0.408     & 51.0      & - & \cite{haslamupdate} \\
CGPS              & 0.408     & 2.8       & $17.0 \pm 3 $ & \cite{cgps2017}     \\
Reich             & 1.42      & 36.0      & $18.9 \pm 2 $ & \cite{reich2001}    \\
GBT (Bank A)      & 4.575     & 2.75      & $18.1 \pm 2 $ & This work           \\
GBT (Bank B)      & 5.625     & 2.24      & $17.5 \pm 2 $ & `` ''               \\
GBT (Band C)      & 6.125     & 2.05      & $17.7 \pm 2 $ & `` ''               \\
Planck            & 28.4      & 32.3      & $30.3 \pm 1 $ & \cite{planck2015I}  \\
Planck            & 44.1      & 27.1      & $26.8 \pm 1 $ & `` ''               \\
Planck            & 70.4      & 13.3      & $26.1 \pm 1 $ & `` ''               \\
Planck            & 100       & 9.7       & - & `` ''               \\
Planck            & 143       & 7.3       & $88.7 \pm 5 $ & `` ''               \\
Planck            & 217       & 5.0       & - & `` ''               \\
Planck            & 353       & 4.8       & $1,550 \pm 70 $ & `` ''               \\
Planck            & 545       & 4.7       & $5,190 \pm 200 $ & `` ''               \\
Planck            & 857       & 4.3       & $18,100 \pm 700 $ & `` ''               \\
DIRBE             & 1249      & 39.5      & $44,000 \pm 1,000 $ & \cite{dirbe1998}    \\
DIRBE             & 2141      & 40.4      & $74,600 \pm 2,000 $ & `` ''               \\
DIRBE             & 2997      & 41.0      & $41,900 \pm 800 $ & `` ''               \\
IRIS ($100~\mu$m) & 3000      & 4.3       & - & \cite{iris2005}     \\
IRIS ($60~\mu$m)  & 5000      & 4.0       & - & `` ''               \\
IRIS ($25~\mu$m)  & 12000     & 3.8       & - & `` ''               \\
IRIS ($12~\mu$m)  & 25000     & 3.8       & - & `` ''               \\ [1mm]
\hline
\end{tabular}
\end{table*}

%%%%%%%%%%%%%%%%%%%%%%%%%%%%%%%%%%%%%%%%%%%%%%%%%%%%%%%%%%%%%%%%%%%%%%%%%%%%%%%%

\section{Observations}
\label{sec:observations}

%%%%%%%%%%%%%%%%%%%%%%%%%%%%%%%%%%%%%%%%%%%%%%%%%%%%%%%%%%%%%%%%%%%%%%%%%%%%%%%%

\subsection{Receiver and Spectrometer}
\label{sec:spectrometer}

%%%%%%%%%%%%%%%%%%%%%%%%%%%%%%%%%%%%%%%%%%%%%%%%%%%%%%%%%%%%%%%%%%%%%%%%%%%%%%%%

GBT is a fully steerable off-axis Gregorian reflecting antenna designed for observations below approximately 115~GHz. 
The prime focus of the parabolic primary mirror is directed into a receiver cabin using an elliptical secondary mirror.
The C-band receiver we used for our observations is mounted in this receiver cabin.
The unblocked aperture diameter is 100~m, so the beam size for our observations was between 1.8 and 2.8~arcmin, depending on frequency.
The VEGAS back-end electronics used to measure the spectra are housed in a laboratory approximately 2~km from the telescope.

A schematic of the receiver and the digital spectrometer we used for this study is shown in Figure~\ref{fig:receiver}.
The telescope first feeds a corrugated horn.
An orthomode transducer (OMT) at the back of the horn splits the sky signals into two polarizations (polarization X and polarization Y).
The two outputs of the OMT are routed to a cryogenic stage that is cooled to approximately 15~K.
At this cryogenic stage, directional couplers are used to insert calibration signals from a noise diode.
These calibration signals were switched on and off during our observations to help monitor time-dependent gain variations.
The sky signals were then (i) amplified with a cryogenic low-noise amplifier (LNA), (ii) band-pass filtered, (iii) amplified a second time with a room-temperature amplifier, (iv) mixed down in frequency, and (v) routed to the laboratory via optical fibers.

%%%%%%%%%%%%%%%%%%%%%%%%%%%%%%%%%%%%%%%%%%%%%%%%%%%%%%%%%%%%%%%%%%%%%%%%%%%%%%%%

\begin{figure*}
\centering
\includegraphics[width=0.48\textwidth]{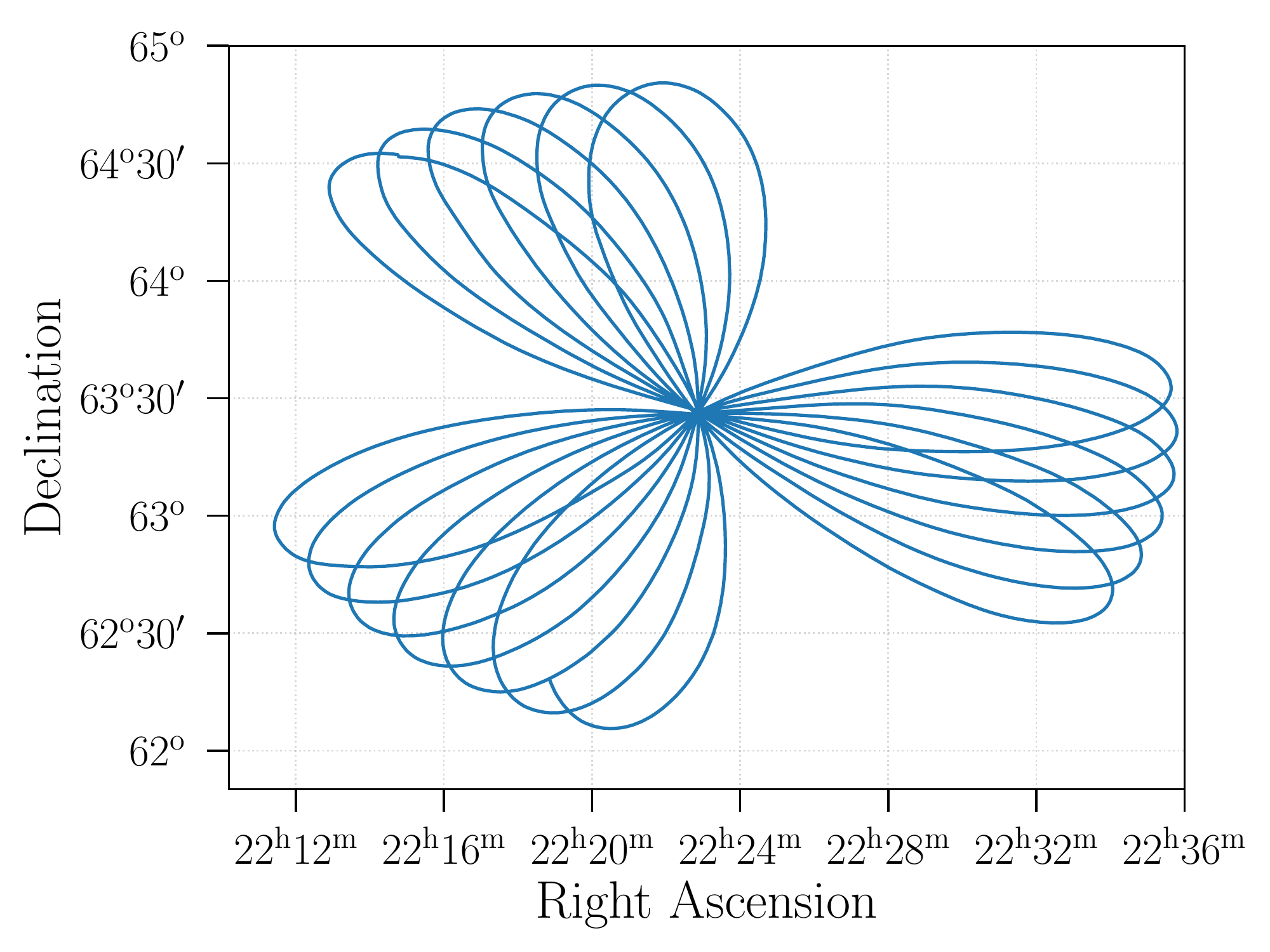}
\hspace{0.02\textwidth}
\includegraphics[width=0.48\textwidth]{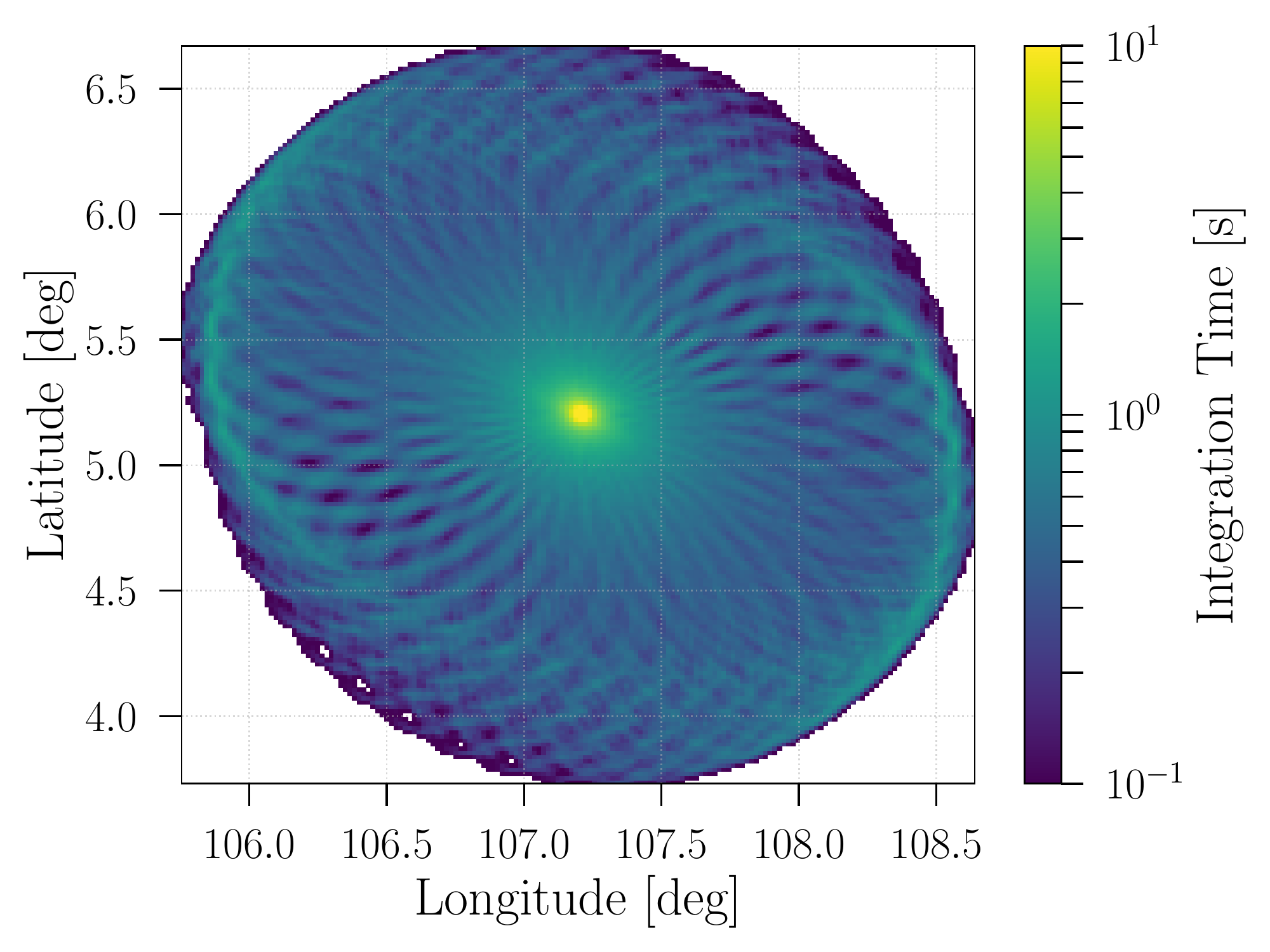}
\caption{
\textbf{Left:} The ``daisy'' scan pattern used for our observations.
Five minutes of pointing data are plotted as an example to show the three ``petals.''
A full cycle is completed every 25 minutes.
The scan strategy densely fills in a circle of radius $1.5^\circ$.
\textbf{Right:} Map of the total integration time per pixel for our observations.
The pixels are $1^{\prime} \times 1^{\prime}$, and the median integration time per pixel is 0.4~s.
Note that the color bar uses a log scale.
}
\label{fig:scan}
\end{figure*}

%%%%%%%%%%%%%%%%%%%%%%%%%%%%%%%%%%%%%%%%%%%%%%%%%%%%%%%%%%%%%%%%%%%%%%%%%%%%%%%%

In the laboratory, the signals were split into four banks: Bank A, B, C, and D.
Each bank used its own hardware chain to measure the spectrum of that bank.
For clarity, only one of the four spectrometer chains is shown in Figure~\ref{fig:receiver}.
The spectral band for each bank is determined by mixing up the signal in that bank using a tunable local oscillator (LO) and then band-pass filtering.
The chosen LO frequency ultimately defines the spectral band. 
The pass band of the filter is between 8.50 and 10.35~GHz.
The signals were then mixed down using a fixed 10.5~GHz LO.
At this stage, each bank has 1.85~GHz of bandwidth.
The signals were then (i) amplified, (ii) low-pass filtered to avoid aliasing (edge at 1.5~GHz), and (iii) sampled at 3~Gsps with an 8-bit analog to digital converter (ADC) that is connected to a Reconfigurable Open Architecture Computing Hardware 2 (ROACH2) board\footnote{https://casper.berkeley.edu/}.
The ROACH2 board uses a field-programmable gate array (FPGA) to compute the spectrum of the sampled data.
Each bank has 16,384 raw spectral channels that are each 91.552~kHz wide.
Spectra are integrated in the ROACH2, and one average spectrum is saved to disk every 40~ms.
In the following sections, we use the term ADC count to refer to the power measurement in each filter bank channel. 
The spectral banks are defined in Table~\ref{table:spectral_banks}.
%

%%%%%%%%%%%%%%%%%%%%%%%%%%%%%%%%%%%%%%%%%%%%%%%%%%%%%%%%%%%%%%%%%%%%%%%%%%%%%%%%

\subsection{Scan Strategy and Calibration}
\label{sec:scan_calibration}

%%%%%%%%%%%%%%%%%%%%%%%%%%%%%%%%%%%%%%%%%%%%%%%%%%%%%%%%%%%%%%%%%%%%%%%%%%%%%%%%

Our GBT observations were conducted in April and June of 2017. 
Ten total hours of mapping data were collected during observing sessions on April 5, April 10, and June 4.
Eight total hours of polarization calibration data were collected on April 3 and June 3.
We refer to these as Sessions 1 through 5 chronologically, so Sessions 1 and 4 are the polarization calibration sessions, and Sessions 2, 3, and 5 are the mapping sessions.
At the beginning of each session, the system temperature was measured.
For our five sessions, the mean system temperature was $19.5 \pm 1$~K.

We chose to use the ``daisy'' scan strategy available at GBT, which is typically used for MUSTANG mapping observations~\citep{korngut_2011}.
The daisy scan traces out three ``petals'' on the sky every 30 seconds (see Figure~\ref{fig:scan}).
Every 25 minutes, this scan strategy completes a full cycle densely covering both the innermost and the outermost portions of a nearly circular region. 
This approach works well with our map-making algorithm (see Section~\ref{sec:mapmaking}) because map pixels are revisited and sampled multiple times.
Given that we want a densely sampled map, we scanned GBT close to the speed and acceleration limits of the telescope\footnote{The maximum scan speed for GBT is 36~arcmin~s$^{-1}$ in azimuth and 18~arcmin~s$^{-1}$ in elevation.  The maximum acceleration is 3~arcmin~s$^{-2}$, and it is only possible to accelerate twice per minute.} and were able to observe a nearly circular region $3.0^\circ$ in diameter centered on G107.2+5.20.
Our maximum scan speed was 21.6~arcmin~s$^{-1}$, and the root mean squared (RMS) speed was 10~arcmin~s$^{-1}$.
The scan pattern is calculated in an astronomical coordinate system to ensure the center is always on G107.2+5.20.

To convert our measurements into flux units, we calibrated using observations of 3C295.
3C295 is an unpolarized radio galaxy that has a power-law-with-curvature spectrum~\citep{Perley2013, ott_1994}.
To mitigate the effects of any gain fluctuations, we switched the noise diode on and off at 25~Hz during all observations.
With this approach, every other spectrum output by VEGAS was a measurement of the noise-diode spectrum.
By comparing the noise-diode spectrum to the 3C295 spectrum, we calibrated the measured G107.2+5.20 spectra to the 3C295 calibration spectrum at every point in time during the observation session.
To calibrate the noise diode into flux units, at the beginning and end of each observation session we pointed the antenna directly at 3C295 and collected data for two minutes.
We then pointed 1~degree in RA away from 3C295 and collected two minutes of data.
These on-source/off-source measurements yielded the desired calibration spectrum, which was measured relative to the background.
Note that we assume the the on-source measurement includes signal from 3C295 plus the unknown background, while the nearby off-source measurement includes only the background signal.
%
%We also performed On/Off and Spider observations of the known sources 3C245, 3C273, 3C280, and 3C286 for polarization calibration.

%%%%%%%%%%%%%%%%%%%%%%%%%%%%%%%%%%%%%%%%%%%%%%%%%%%%%%%%%%%%%%%%%%%%%%%%%%%%%%%%

\subsection{Ancillary Data}
\label{sec:data}

%%%%%%%%%%%%%%%%%%%%%%%%%%%%%%%%%%%%%%%%%%%%%%%%%%%%%%%%%%%%%%%%%%%%%%%%%%%%%%%%

To measure the spectral flux density of the AME region G107.2+5.2 and to inspect its morphology at different frequencies we compiled data from a range of observatories.
A list of all the data sets used in our study is given in Table~\ref{tab:datasets}.
Data processing and unit conversions are required for each data set as described below. 

For the radio observations we used the Canadian Galactic Plane Survey (CGPS) data~\citep{cgps2003, cgps2017} at 408~MHz as well as the Reich all sky survey at 1.420~GHz~\citep{reich1982, reich1986, reich2001}.
The CGPS map was produced using Haslam data~\citep{haslam1981, haslam1982, haslamupdate, Remazeilles2016}, which is widely used to trace synchrotron and optically thin free-free emission on 1~degree angular scales.
The CGPS data\footnote{The CGPS data is available online at http://www.cadc-ccda.hia-iha.nrc-cnrc.gc.ca/en/cgps/} has arcminute resolution, which is useful for morphological comparisons.
To convert from thermodynamic units to flux units we used the Rayleigh Jeans approximation, 
\begin{equation}
I = \frac{2\nu^2k_B}{c^2}T_B \times \Omega_{p}\times 10^{26} \, ,
\end{equation}
where $\nu = 408$~MHz for the CGPS data and 1.420~GHz for the Reich data, and $\Omega_{p}$ is the solid angle of a pixel in steradians. 
This conversion brings the maps into spectral flux density units ($\rm{Jy~pixel^{-1}}$). 
The Reich data required a calibration correction factor of 1.55 to compensate for the full-beam to main-beam ratio, based on comparisons with bright calibrator sources~\citep{reich1988}.
We included an estimated $10\%$ calibration uncertainty on all the radio data. 

%%%%%%%%%%%%%%%%%%%%%%%%%%%%%%%%%%%%%%%%%%%%%%%%%%%%%%%%%%%%%%%%%%%%%%%%%%%%%%%%

We used Planck observations for measurements between 30 and 857~GHz~\citep{planck2015I}.
To convert Planck data from $K_{\rm{CMB}}$ to spectral radiance we used the Planck unit conversion and color correction code available on the Planck Legacy Archive\footnote{https://pla.esac.esa.int/pla/}.
Note that molecular CO lines have biased the 100 and 217~GHz Planck results, so these points are not included in the model fitting (see Section~\ref{sec:results}). 

%%%%%%%%%%%%%%%%%%%%%%%%%%%%%%%%%%%%%%%%%%%%%%%%%%%%%%%%%%%%%%%%%%%%%%%%%%%%%%%%

Far-infrared information was provided by IRIS (improved IRAS) and DIRBE data~\citep{iris2005, dirbe1998}.
For our spectrum analysis we only used the DIRBE data up to 3~THz because of complexities from dust grain absorption and emission lines at higher frequencies. 
The IRIS data was used for morphological comparisons only (see Appendix).
We applied color corrections to the DIRBE data according to the DIRBE explanatory supplement~\citep{dirbe1998}.
For this analysis we did not use Haslam or WMAP~\citep{wmap9results} data due to the low spatial resolution of those datasets. 
However we did check that our aperture photometry results using CGPS and Planck were consistent with the results using Haslam and WMAP.

%%%%%%%%%%%%%%%%%%%%%%%%%%%%%%%%%%%%%%%%%%%%%%%%%%%%%%%%%%%%%%%%%%%%%%%%%%%%%%%%

\section{GBT Data Analysis}
\label{sec:datareduction}

%%%%%%%%%%%%%%%%%%%%%%%%%%%%%%%%%%%%%%%%%%%%%%%%%%%%%%%%%%%%%%%%%%%%%%%%%%%%%%%%

The data processing algorithm consists of five steps: (i) data selection, (ii) noise diode calibration, (iii) data calibration, (iv) map making, and (v) aperture photometry.
Each of these steps is described in the subsections below.
The time-ordered data from each mapping session are processed with steps (i), (ii), (iii), and (iv).
Data from Sessions~3~and~5 are processed with step (v).
The results presented in this paper come from data collected during Session~5, which was 4.5 hours long.
The data from Sessions~2~and~3 are used for jackknife tests.
The mapping observations are stored in files containing 25 minutes of data arranged in 5 minute long segments.
Some of the steps in the data processing algorithm operate on these 5 minute long segments.

%%%%%%%%%%%%%%%%%%%%%%%%%%%%%%%%%%%%%%%%%%%%%%%%%%%%%%%%%%%%%%%%%%%%%%%%%%%%%%%%

\subsection{Data Selection}
\label{sec:rfi}

%%%%%%%%%%%%%%%%%%%%%%%%%%%%%%%%%%%%%%%%%%%%%%%%%%%%%%%%%%%%%%%%%%%%%%%%%%%%%%%%

Parts of the data sets are corrupted by radio frequency interference (RFI), transient signals, and 
instrumental artifacts.
These spurious signals need to be removed before making maps.
The transient signals and instrumental artifacts are excised by hand after inspection.
To find RFI corrupted spectral channels we search for high noise levels and non-Gaussianity using two statistics: the coefficient of variation and the spectral kurtosis~\citep{nita2010}.
The RFI removal techniques based on these statistics are described below.

The subscript $\nu$ denotes the frequency channel index and $t$ denotes the time index.  
For example, $\xi_{\nu, t}$ is data in ADC counts in frequency channel $\nu$ at time $t$. 

For each 5 minute long data segment, we calculated the coefficient of variation in each spectral channel, which is the the inverse signal-to-noise ratio ($NSR_{\nu}$). 
This statistic finds spectral channels with persistently high noise levels.
We define the mean and the standard deviation in time per channel as
\begin{eqnarray}
&\mu_{\nu} = \langle \xi_{\nu, t} \rangle_t \\
&\sigma_{\nu} = \sqrt{\big\langle(\xi_{\nu, t} - \mu_{\nu} )^2 \big\rangle_t},
\end{eqnarray}
therefore
\begin{equation}
NSR_{\nu} = \frac{\sigma_{\nu}}{\mu_{\nu}}.
\end{equation}
We masked channels with $NSR_{\nu}$ greater than 7.5 times the median absolute deviation of the $NSR_{\nu}$.
We empirically chose this cutoff level because it corresponds to approximately $5\sigma$ and effectively detects outliers.
In addition, we calculated the spectral kurtosis (or the fourth standardized moment), 
\begin{equation}
K_{\nu} = \frac{\big\langle ( \xi_{\nu, t} - \mu_{\nu} )^4 \big\rangle_t}{\big\langle ( \xi_{\nu, t} - \mu_{\nu} )^2 \big\rangle_t^2}.
\end{equation}
This statistic finds channels with non-Gaussian noise properties.
Again we mask spectral channels with $K_{\nu}$ greater than 7.5 times the median absolute deviation of $K_{\nu}$. 

Finally, we only used the selected bandwidth that is listed in Table~\ref{table:spectral_banks} for each bank because at the spectral bank edges the band-pass filters in the receiver (see Figure~\ref{fig:receiver}) attenuate the sky signals and the gain is low.
In total, for Banks A, B, and C in Session~5, 0.7\% of the bandwidth-selected data was excised because of RFI contamination, 2\% was excised because of transient signals, and 7\% was excised because of instrumental artifacts.
The signal-to-noise ratio (SNR) for Bank~D was low so the data in this bank was ultimately unusable.

%%%%%%%%%%%%%%%%%%%%%%%%%%%%%%%%%%%%%%%%%%%%%%%%%%%%%%%%%%%%%%%%%%%%%%%%%%%%%%%%

\subsection{Calibration}
\label{sec:calibration}

%%%%%%%%%%%%%%%%%%%%%%%%%%%%%%%%%%%%%%%%%%%%%%%%%%%%%%%%%%%%%%%%%%%%%%%%%%%%%%%%

To convert the mapping data from ADC units to spectral radiance (Jy sr$^{-1}$), we first calibrate the noise diode using the point source 3C295 and then calibrate the mapping data using the noise diode (see Section~\ref{sec:scan_calibration}).
The point source observations take place at the beginning and the end of the observing sessions, and they allow us to convert the data to Janskys.
The noise diode is flashed at 25~Hz during both the point source and the mapping observations, so the noise diode is used as a calibration signal to track gain stability.
The assumptions are the noise diode spectrum is stable in time and the gain is linear as a function of signal brightness over the observing session. 

%%%%%%%%%%%%%%%%%%%%%%%%%%%%%%%%%%%%%%%%%%%%%%%%%%%%%%%%%%%%%%%%%%%%%%%%%%%%%%%%

\begin{figure}
\centering
\includegraphics[width=\columnwidth]{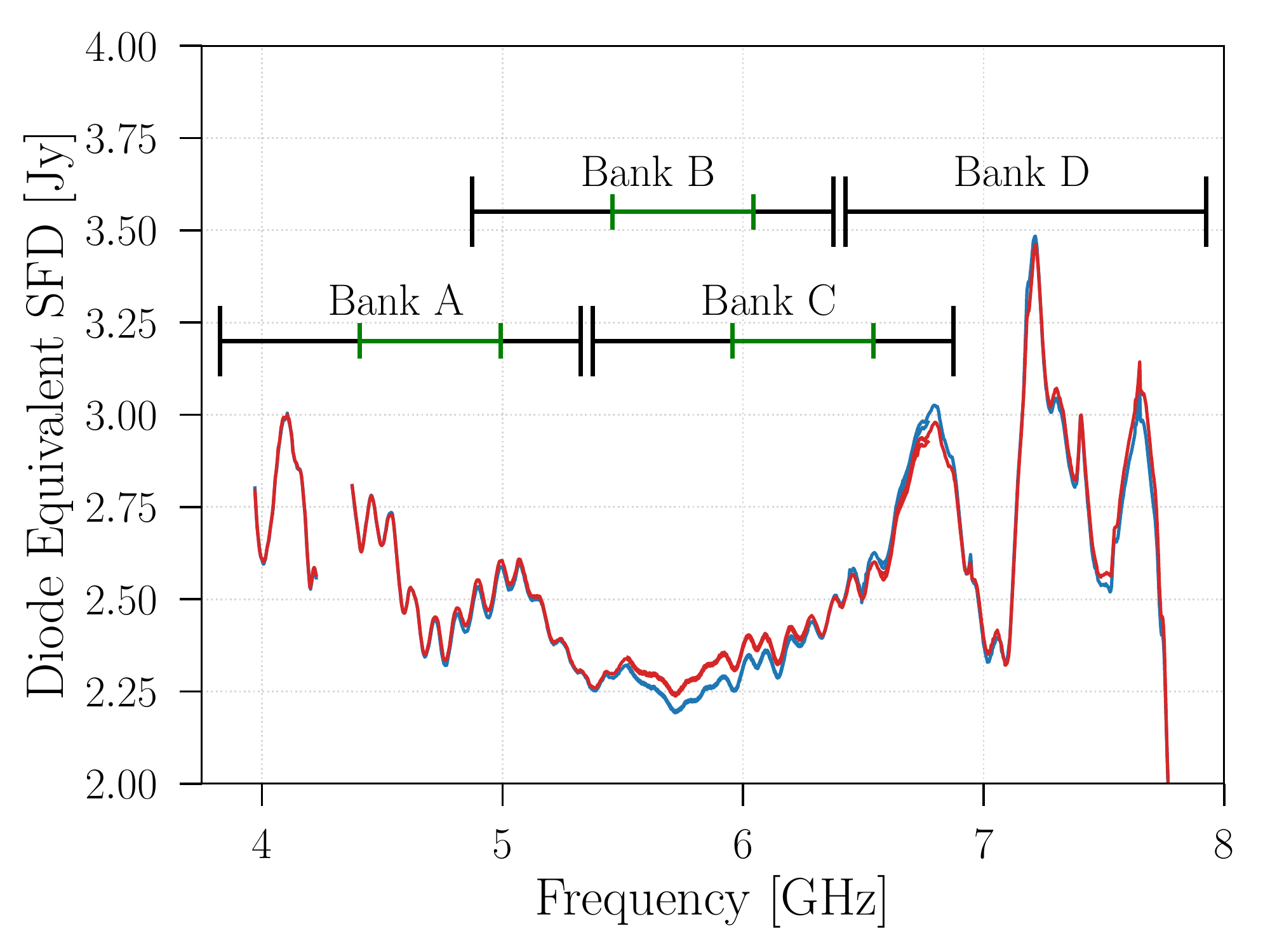}
\caption{
Equivalent spectral flux density (SFD) of the noise diode.
The noise diode brightness was calibrated on 3C295 at the beginning (red) and end (blue) of Session~5.
The amplitude of the noise-diode spectrum used for calibration is very stable in time throughout the observations.
The RMS of the difference between the two calibrations during Session~5 over banks A to C is 30~mJy, approximately a 1\% difference.
The full bandwidth of Bank~A through D is shown in black, and the selected bandwidth for Bank A, B, and C is shown in green (see Table~\ref{table:spectral_banks}).
}
\label{fig:calibration}
\end{figure}

%%%%%%%%%%%%%%%%%%%%%%%%%%%%%%%%%%%%%%%%%%%%%%%%%%%%%%%%%%%%%%%%%%%%%%%%%%%%%%%%

In this subsection, we now define $x$ as calibration data while pointing away from 3C295 (off-source), $y$ as calibration data while pointing at 3C295 (on-source), and $z$ as mapping data, scanning G107.2+5.2. 
We use the superscripts $on$ or $off$ to denote whether the noise diode is on or off.
For example, $x^{on}_{\nu,t}$ is off-source calibration data at time $t$ for channel $\nu$ while the noise diode is on. 
We calculated the average noise diode level in a spectral channel as 
\begin{equation}
D_{\nu} = \langle \, x^{on}_{\nu,t} - x^{off}_{\nu,t} \, \rangle_t \, ,
\end{equation}
which has units of ADC counts.
We computed the average source level in a spectral channel as
\begin{equation}
S_{\nu} = \langle \, y^{off}_{\nu,t} - x^{off}_{\nu,t} \, \rangle_t \, ,
\end{equation}
which also has units of ADC counts.
Both $D_{\nu}$ and $S_{\nu}$ were averaged over two minutes, which was the total duration of the point source calibration observations.
The noise diode signal was calibrated using the known spectral flux density of 3C295~\citep{Perley2013} in the following way:
\begin{equation}
P_{\nu} = \frac{I_{\nu}}{S_{\nu}} D_{\nu}.
\end{equation}
Here $P_{\nu}$ is the calibrated noise diode signal in units of Janskys (see Figure~\ref{fig:calibration}) and $I_{\nu}$ is the spectral flux density of 3C295.
We then used $P_{\nu}$ to calibrate the mapping data $z$ into Janskys. 

%%%%%%%%%%%%%%%%%%%%%%%%%%%%%%%%%%%%%%%%%%%%%%%%%%%%%%%%%%%%%%%%%%%%%%%%%%%%%%%%

\begin{figure*}
\centering
\includegraphics[scale=0.4]{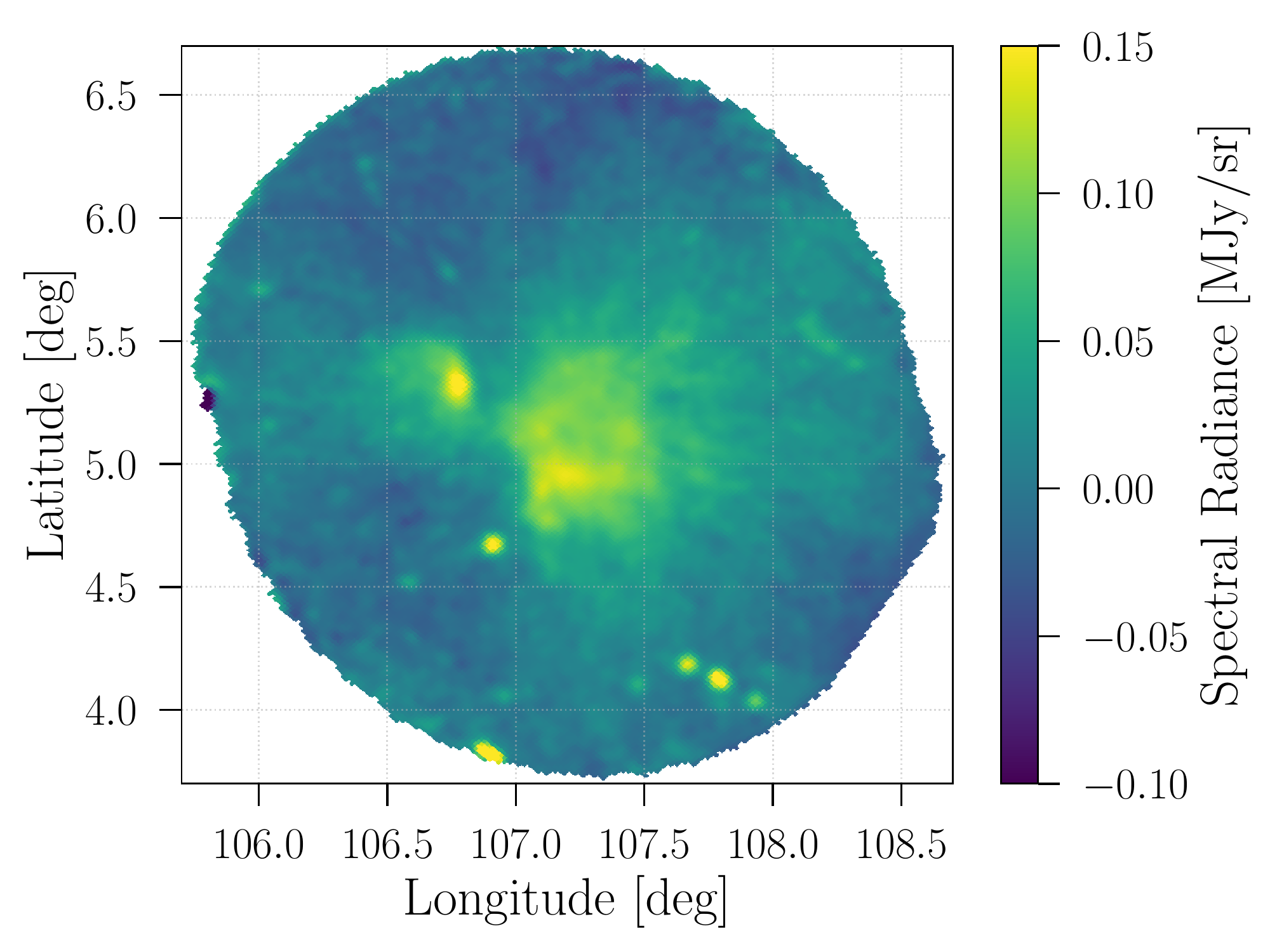}
\hspace{0.1in}
\includegraphics[scale=0.4]{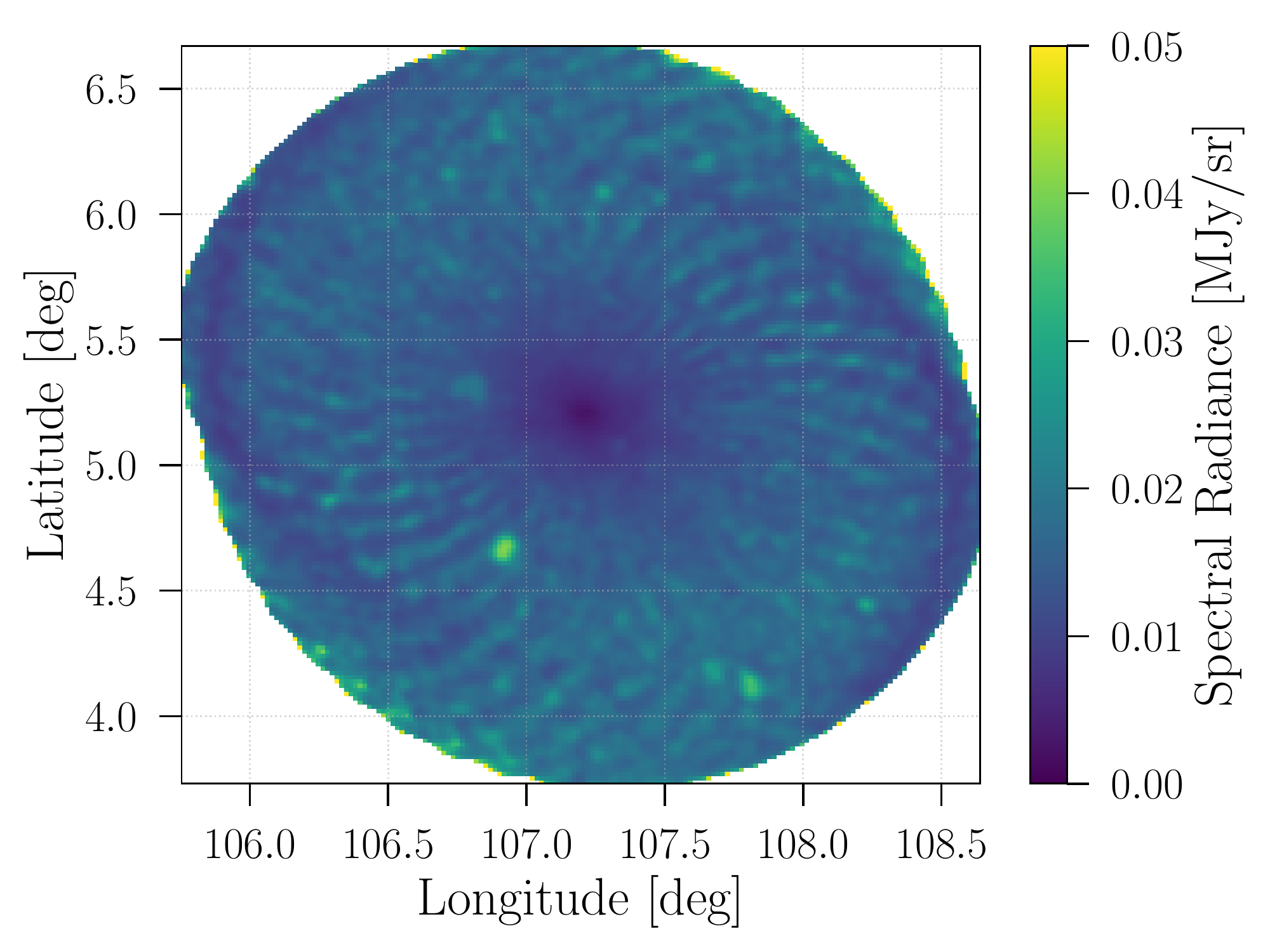}
\includegraphics[scale=0.4]{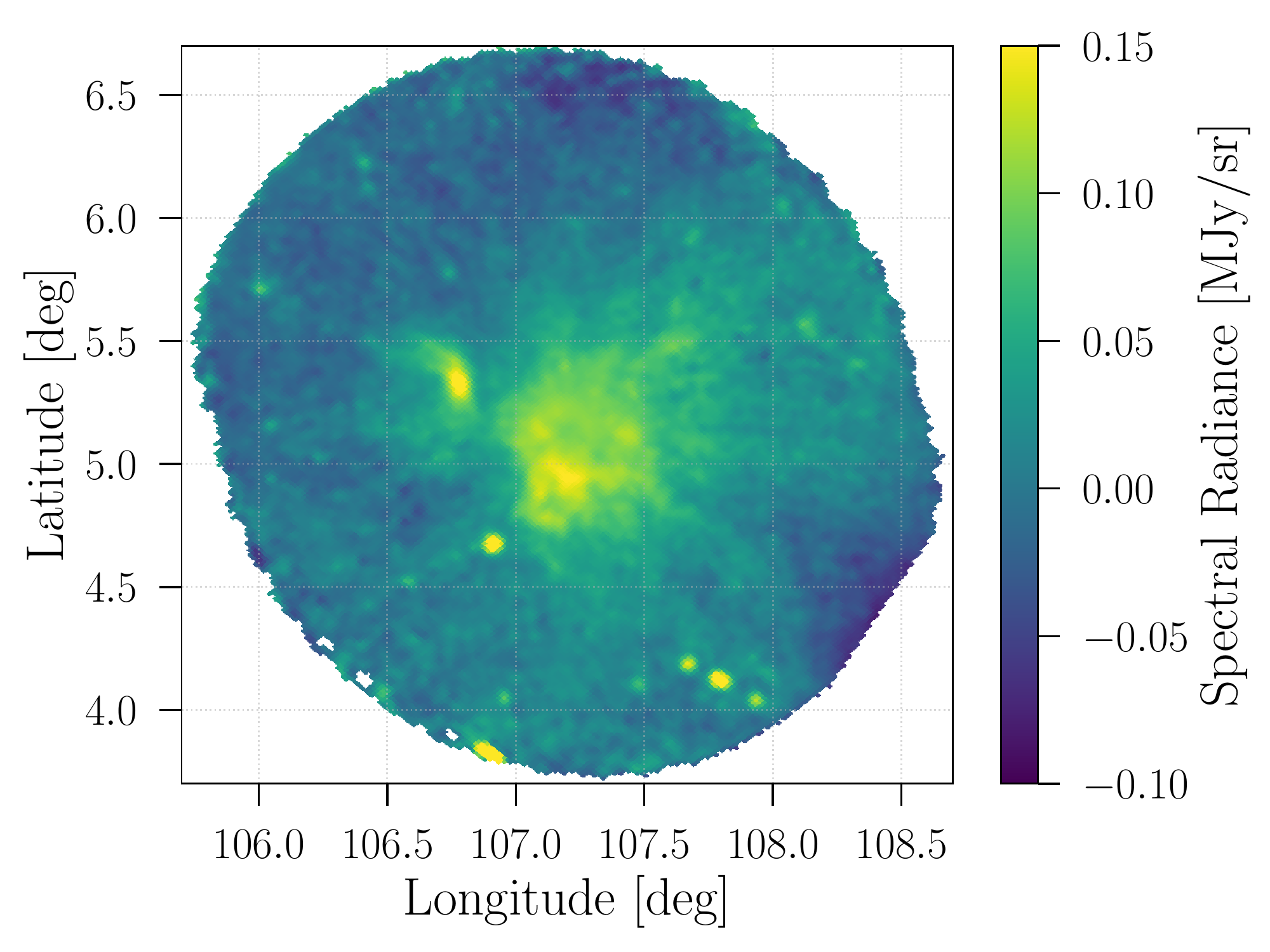}
\hspace{0.1in}
\includegraphics[scale=0.4]{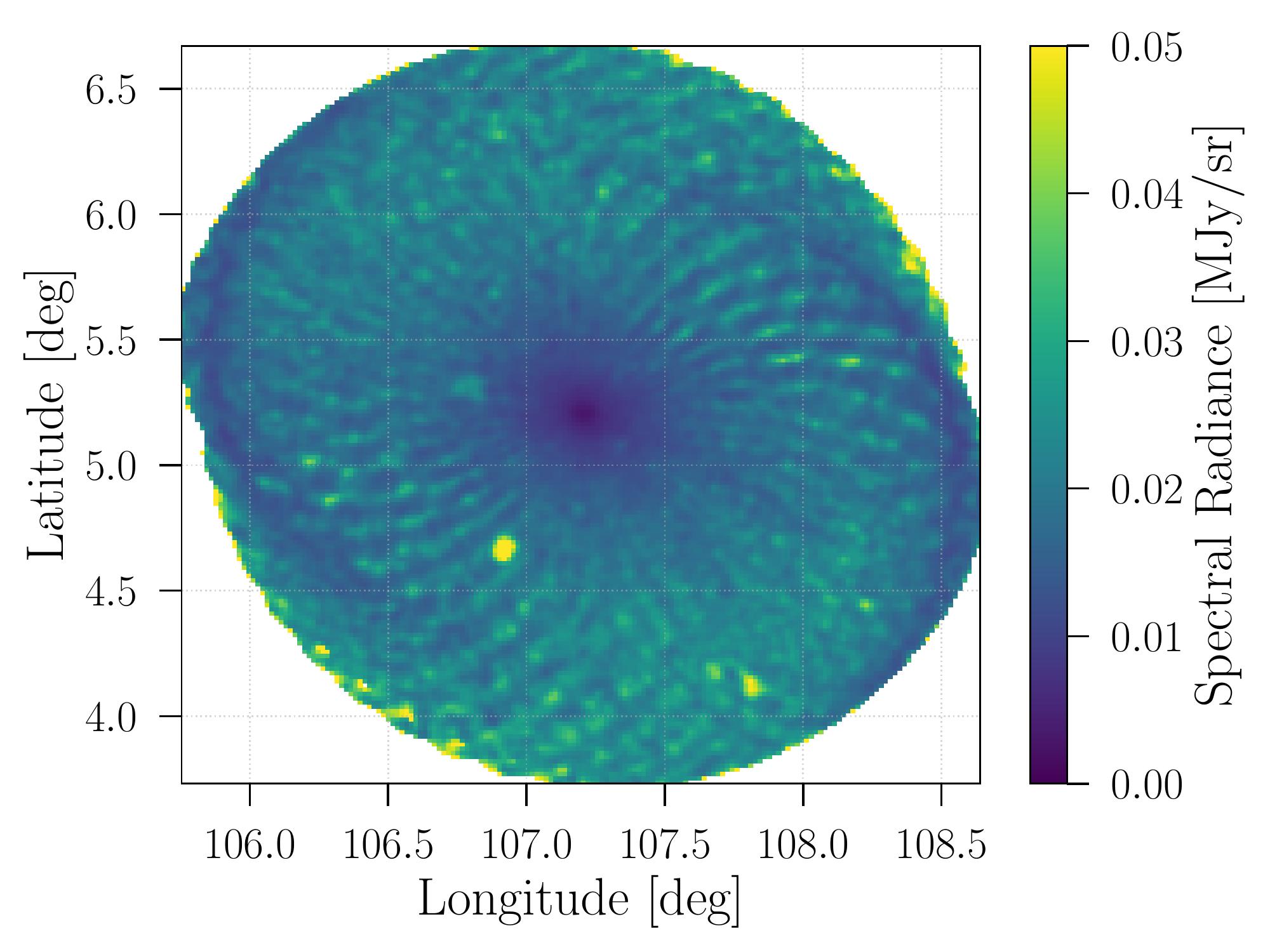}
\hspace{0.1in}
\includegraphics[scale=0.4]{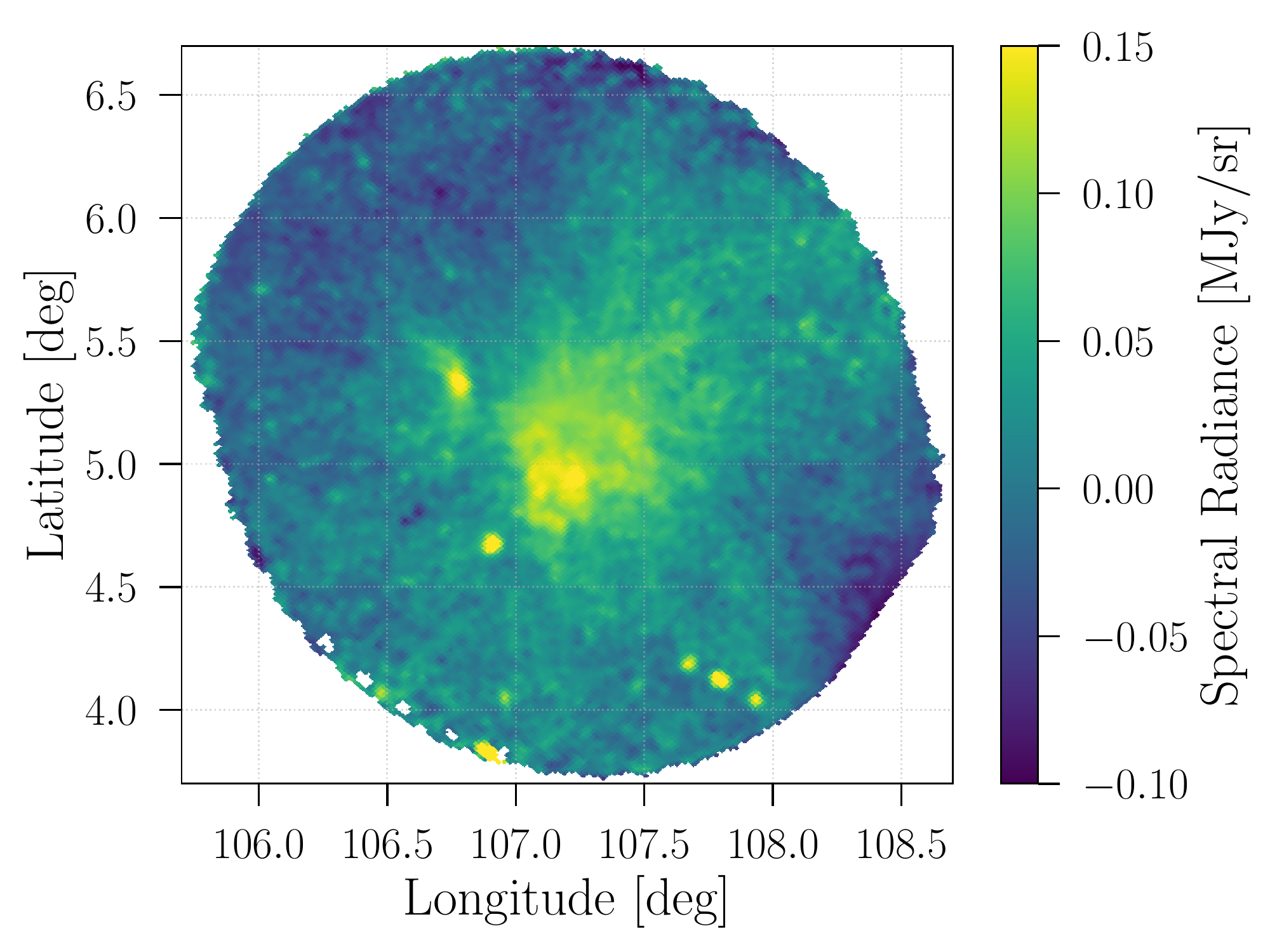}
\hspace{0.1in}
\includegraphics[scale=0.4]{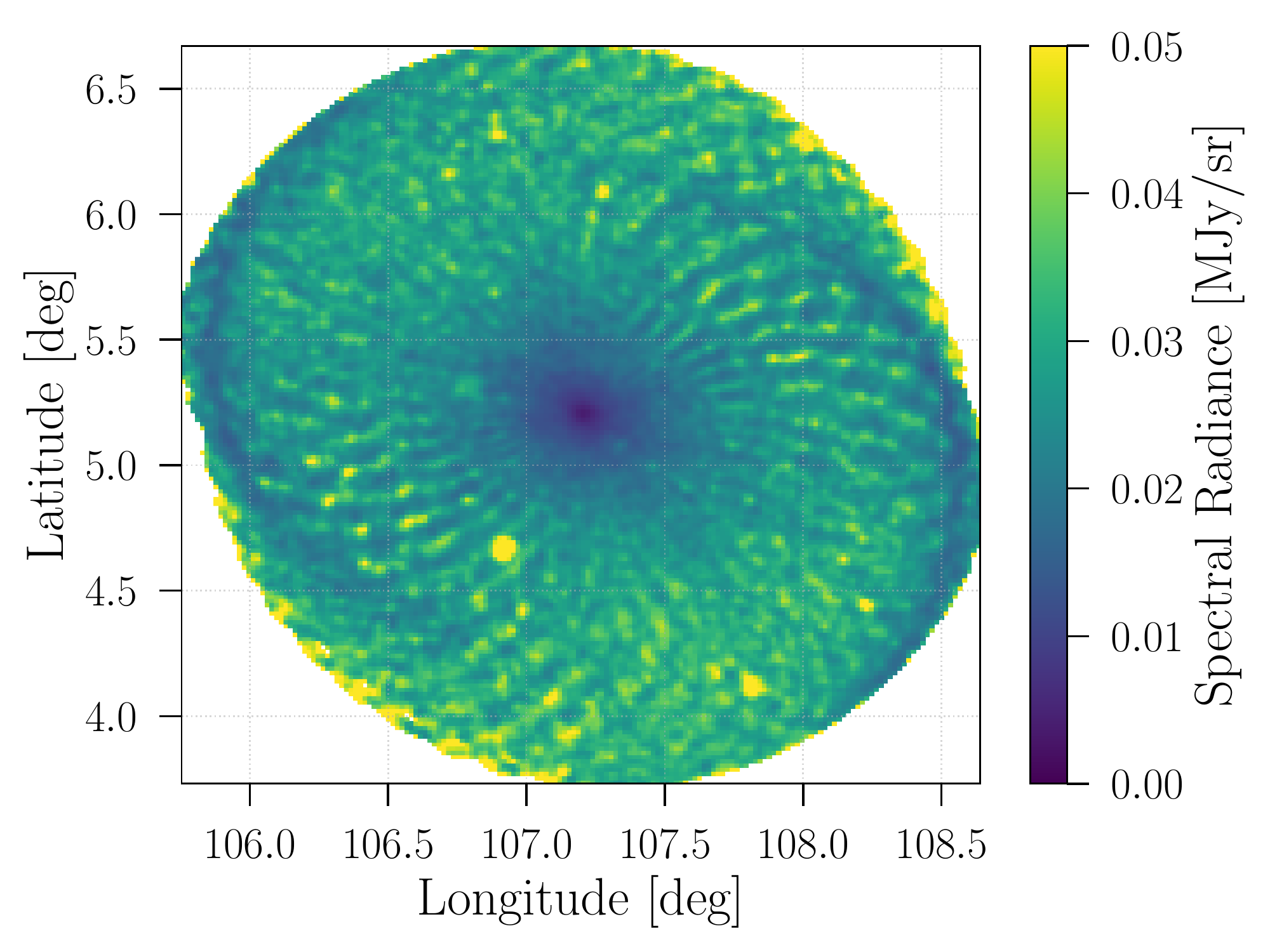}
\caption{
Maps of the AME region centered on G107.2+5.2.
The left column shows the destriped maps (see Section~\ref{sec:mapmaking}) and the right column shows the estimated uncertainty map.
The top, middle, and bottom rows correspond to Bank~A, B, and C, respectively.
The star-forming region, S140, is the bright feature located at $(l, b) = (106.80^\circ, +5.31^\circ$), and G107.2+5.20 is the center of each map.
The stripes in the uncertainty maps come from the visible stripes in the integration time map (see Figure~\ref{fig:scan}), and they are not map-making artifacts.
The peak SNR is 40, 36, and 26, and the median SNR inside the photometry aperture is 3.2, 2.6, and 1.9 for Bank~A, B, and C, respectively.
}
\label{fig:bankABmaps}
\end{figure*}

%%%%%%%%%%%%%%%%%%%%%%%%%%%%%%%%%%%%%%%%%%%%%%%%%%%%%%%%%%%%%%%%%%%%%%%%%%%%%%%%

Let $z^{on}_{\nu,t}$ and $z^{off}_{\nu,t}$ be the mapping data at time $t$ and frequency channel $\nu$ with the noise diode on and off, respectively. 
We calculated the inverse receiver gain
\begin{equation}
G_{\nu} = \frac{P_{\nu}}{\langle \, z^{on}_{\nu,t} - z^{off}_{\nu,t} \, \rangle_t} \, ,
\end{equation}
which has units of Janskys per ADC count.
$G_{\nu}$ was calculated for every 5 minute long data segment. 
When making maps of diffuse sky signals we divide the data by the beam solid angle,
\begin{equation}
\Omega_{\nu} = \frac{\pi}{4 \log{2}} \, \mathrm{FWHM}^2_{\nu}.
\end{equation}
Here, $\rm{FWHM}_{\nu}$ is the beam full-width at half-maximum at the frequency channel $\nu$ and we assume a Gaussian beam profile.   
The FWHM values for the center frequencies of the four Banks are given in Table~\ref{table:spectral_banks}.
The calibrated time-ordered mapping data were then calculated as
\begin{equation}
d_{t} = \left\langle \frac{G_{\nu} \, z^{off}_{\nu,t}}{\Omega_{\nu}} \right\rangle_{\nu} \, ,
\label{eq:tod}
\end{equation}
which have units of spectral radiance (Jy sr$^{-1}$).
The average is taken over the selected bandwidth in a given spectral bank (see Table~\ref{table:spectral_banks}) after data selection (see Section~\ref{sec:rfi}).

%%%%%%%%%%%%%%%%%%%%%%%%%%%%%%%%%%%%%%%%%%%%%%%%%%%%%%%%%%%%%%%%%%%%%%%%%%%%%%%%

\subsection{Map Making}
\label{sec:mapmaking}

%%%%%%%%%%%%%%%%%%%%%%%%%%%%%%%%%%%%%%%%%%%%%%%%%%%%%%%%%%%%%%%%%%%%%%%%%%%%%%%%

\begin{figure}
\centering
\includegraphics[width=\columnwidth]{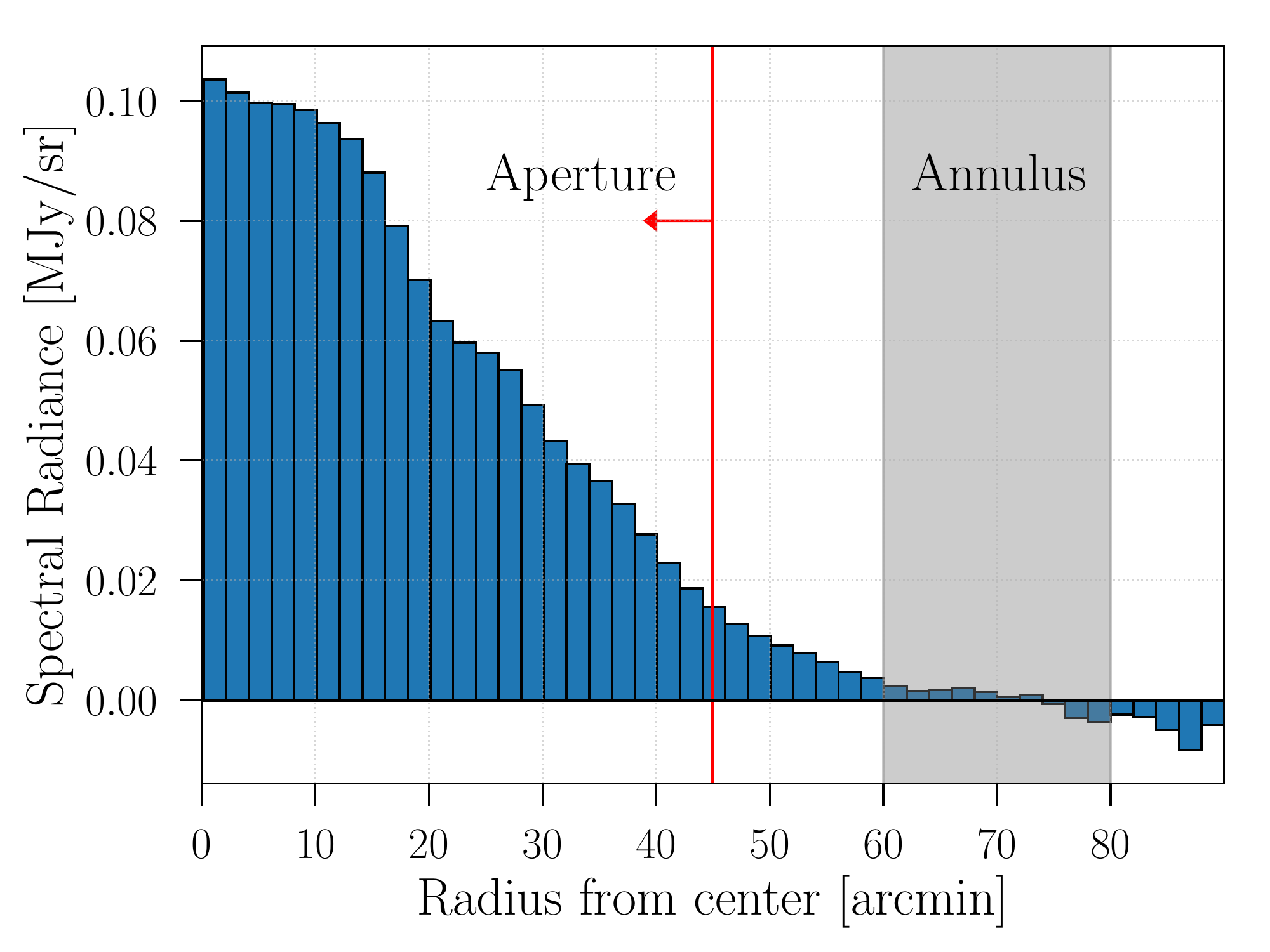}
\caption{
Mean spectral radiance per pixel as a function of radial distance from G107.2+5.20 for the Bank~A map shown in Figure~\ref{fig:bankABmaps}.
The histogram bins are annuli 2$^{\prime}$ wide centered on G107.2+5.20.
The zero-point annulus and the aperture radius are highlighted.
}
\label{fig:fluxradius}
\end{figure}

%%%%%%%%%%%%%%%%%%%%%%%%%%%%%%%%%%%%%%%%%%%%%%%%%%%%%%%%%%%%%%%%%%%%%%%%%%%%%%%%

Variations in the gain and system temperature of a receiver result in a form of correlated noise that is often referred to as $1/f$ noise.
To separate the sky signal from the $1/f$ noise we implemented a form of the destriping map-making method as described in \citet{Delabrouille1998, Sutton2009, Sutton2010}.
The aim of the destriping map-making method is to solve for the $1/f$ noise in the time-ordered data as a series of linear offsets.
To do this the time-ordered data from a receiver system is defined as
\begin{equation}
\vdata = \mP \, \vmap + \mF \, \vamp + \mathbf{n}_w ,
\end{equation}
where the $\vmap$ is a the map vector of the true sky signal, $\mP$ is the pointing matrix that transforms pixel locations on the sky into time positions in the data stream, $\mF\vamp$ describes the $1/f$ noise linear offsets and $\mathbf{n}_w$ is the white noise vector.
For our GBT data, the $\vdata$ is populated with $d_t$, which is the calibrated time-ordered data for a spectral bank given in Equation~\ref{eq:tod}.

Solving for the amplitudes of the $1/f$ noise linear offsets ($\vamp$) requires minimizing
\begin{equation}
\chi^2 = \left( \vdata - \mP \, \vmap - \mF \, \vamp \right)^\mathrm{T} \mathbf{N}^{-1} \left( \vdata - \mP \, \vmap -\mF \, \vamp \right) ,
\label{eqn:minimise}
\end{equation}
where $\mathbf{N}$ is a diagonal matrix describing the receiver white noise.
By minimizing derivatives of Equation~\ref{eqn:minimise} with respect to the sky signal $\vmap$ and $1/f$ noise amplitude $\vamp$, it is possible to derive the following maximum-likelihood estimate for the amplitudes
\begin{equation}
\hat{\vamp} = \left( \ \mF\T \, \mZ\T \, \ninv \, \mZ \, \mF \, \right)^{-1} \mF\T \, \mZ\T \, \ninv \, \mZ \, \vdata.
\end{equation}
Here, we have made the substitution 
\begin{equation}
\mZ = \mathbf{I} - \mP \left( \, \mP\T \, \ninv \, \mP \, \right)^{-1} \mP \, \ninv.
\end{equation}
Once the $1/f$ noise amplitudes have been computed the, $1/f$ noise can be subtracted in the time domain, and the sky map becomes
\begin{equation}
\hat{\vmap} = \left( \, \mP\T \, \ninv \, \mP \, \right)^{-1} \mP\T \, \ninv \, \left( \, \vdata - \mF \, \vamp \, \right) ,
\end{equation}
which is a noise weighted histogram of the data. % and $1/f$ noise residual.

For our GBT observations a linear offset length of 1~second was chosen, which removes $1/f$ noise on scales larger than 10$^\prime$.
The noise weights for each data point were calculated by subtracting neighboring pairs of data and taking the running RMS within 2-second chunks of the auto-subtracted data.
%
% The improvement in the noise level between the destriped and naively binned maps is of the order 10\,per\,cent at pixel scales of 1\,arcmin for banks A to C, but critically non-Gaussian $1/f$ noise structures on large-scales are significantly reduced.
%
The destriped sky maps and the associated uncertainty-per-pixel maps for Bank~A, B, and C are shown in Figure~\ref{fig:bankABmaps}.

%%%%%%%%%%%%%%%%%%%%%%%%%%%%%%%%%%%%%%%%%%%%%%%%%%%%%%%%%%%%%%%%%%%%%%%%%%%%%%%%

\begin{table}
\centering
\caption{
Measured spectral flux density in an aperture 45$^\prime$ in radius centered on G107.2+5.20.
Here, $\sigma_r$ is the random error from noise in the measurement, $\sigma_s$ is the systematic error from uncertainty in the calibration, and $\sigma_t$ is the total uncertainty.
These points are plotted in Figure~\ref{fig:spectrum} and \ref{fig:amefit}.
}
\label{table:our_data_points}
\begin{tabular}{cccccc}
\hline 
Bank & $\nu_c$   & Aperture SFD & $\sigma_r$  & $\sigma_s$    & $\sigma_t$  \\ [-1mm]
     & [\,GHz\,] & [\,Jy\,]     & [\,Jy\,]    &  [\,Jy\,]     & [\,Jy\,]    \\ [1mm] \hline
A    & 4.575     & 18.09        & 0.08       &  1.81         & 1.81        \\
B    & 5.625     & 17.51        & 0.10       &  1.75         & 1.75        \\
C    & 6.125     & 17.75        & 0.15       &  1.78         & 1.79        \\ 
D    & 7.175     & 32.39        & 0.77       &  3.24         & 3.33        \\ [1mm] \hline
\end{tabular}
\end{table}

%%%%%%%%%%%%%%%%%%%%%%%%%%%%%%%%%%%%%%%%%%%%%%%%%%%%%%%%%%%%%%%%%%%%%%%%%%%%%%%%

\subsection{Aperture Photometry}
\label{sec:photometry}

%%%%%%%%%%%%%%%%%%%%%%%%%%%%%%%%%%%%%%%%%%%%%%%%%%%%%%%%%%%%%%%%%%%%%%%%%%%%%%%%

\begin{figure*}
\centering
\includegraphics[width=0.9\textwidth]{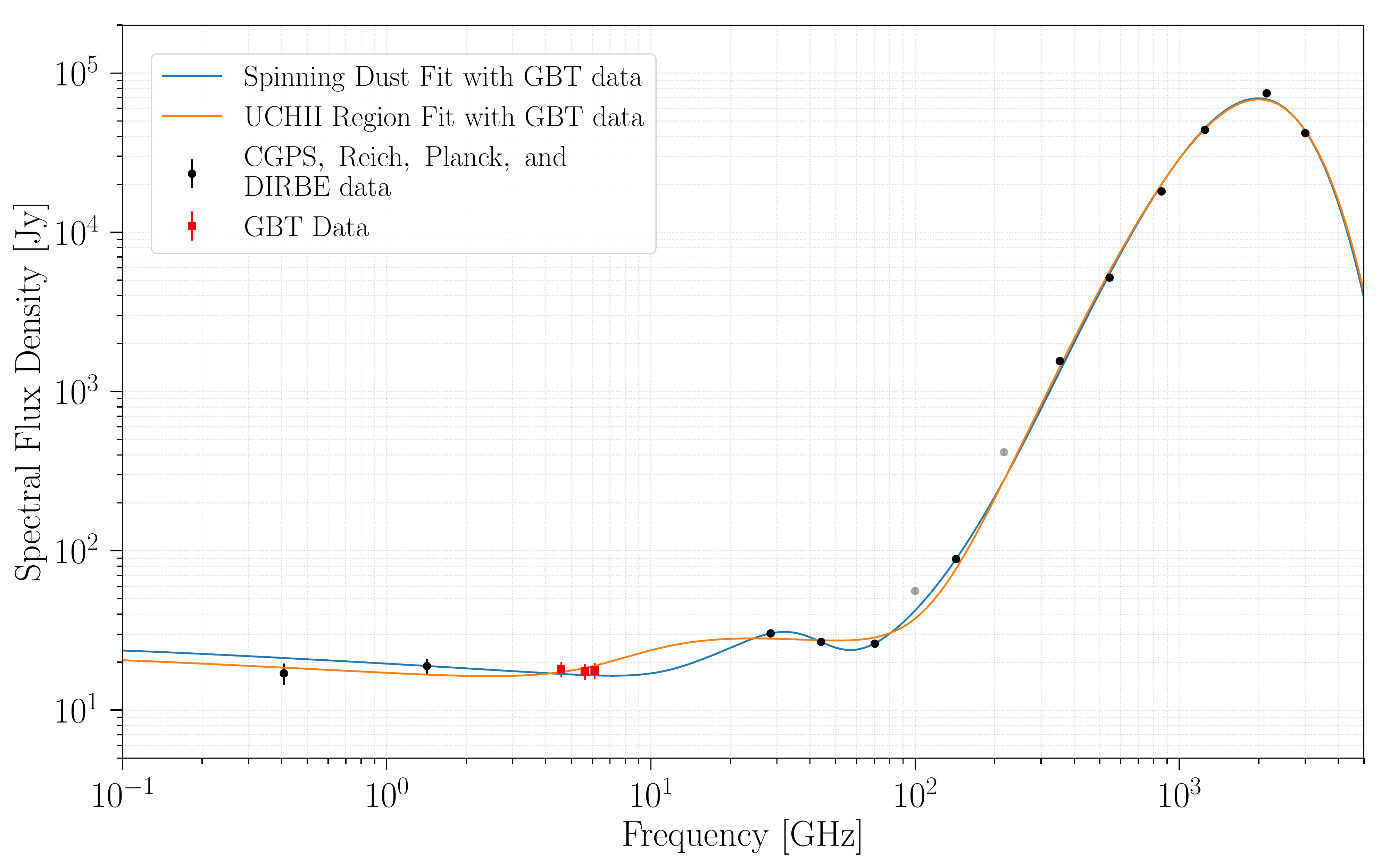}
\caption{
Spectral flux density for G107.2+5.2.
The aperture radius used for each point in the spectrum is 45$^{\prime}$ with the zero-point annulus extending from 60$^{\prime}$ to 80$^{\prime}$ (see Section~\ref{sec:photometry}).
The data points in black come from CGPS, Reich, Planck, and DIRBE (see references in Table~\ref{tab:datasets}).
Our new data points from this GBT study are shown in red.
The gray points are from Planck (100 and 217~GHz), but contain known CO contamination and are not used in the fit.
The solid curves correspond to the best-fit foreground models.
These models include optically thin free-free emission, thermal dust emission, the CMB, and one AME component.
If the included AME component is spinning dust, then the best-fit model is the blue curve.
If the included AME component is UCHII free-free, then the best-fit model is the orange curve.
The foreground models are given in Table~\ref{tab:models}, and the best-fit model parameters are given in Table~\ref{tab:parameters}.
A close-up view of the result between 300~MHz and 200~GHz is shown in Figure~\ref{fig:amefit}.
}
\label{fig:spectrum}
\end{figure*}

%%%%%%%%%%%%%%%%%%%%%%%%%%%%%%%%%%%%%%%%%%%%%%%%%%%%%%%%%%%%%%%%%%%%%%%%%%%%%%%%

We used aperture photometry~\citep{PlanckXV, Genova-Santos2017} to measure the spectral flux density (Jy) of the AME region centered on G107.2+5.2. 
This analysis used seventeen total maps including our GBT maps and maps from CGPS, Reich, Planck, and DIRBE (see Table~\ref{tab:datasets}).
This aperture photometry procedure involves five key steps.
First, we removed a spatial gradient and smoothed all the maps in the study to a common resolution of $40^\prime$ by convolving the maps with a two-dimensional Gaussian with 
\begin{equation}
\mathrm{FWHM} = \sqrt{(40^\prime)^2 - \Theta^2} \, ,  
\end{equation}
where $\Theta$ is the beam FWHM of each data set.
The common $40^\prime$ resolution is set by DIRBE, which has the largest beam of all of the data sets used in this study.
The beam sizes are given in Table~\ref{tab:datasets}.
Second, we integrated the spectral radiance over the solid angle of a map pixel $\Omega_p$ to convert the units to Jy pixel$^{-1}$.
For our GBT maps, 
\begin{equation}
\Omega_p = s^2_p \, , 
\end{equation}
where $s_p = 1^{\prime}$ is the length of the side of each square pixel in the map. 
Third, the map offsets needed to be subtracted because the aperture photometry technique references a common zero point among all the maps.
We determined this zero point by calculating the median value of all the pixels in an annulus with an inner radius of 60$^{\prime}$ and an outer radius of 80$^{\prime}$ centered on G107.2+5.2.
We found the results do not strongly depend on the precise annulus dimensions as long as it is away from the aperture and within the boundaries of our maps (see Figure~\ref{fig:fluxradius}). 
Fourth, we summed all the pixels inside a circular aperture with a radius of 45$^{\prime}$ centered on G107.2+5.2 to get the spectral flux density of the AME region.
The aperture radius we chose is well matched to the map resolution after smoothing.
Fifth, we estimated the uncertainty in the aperture spectral flux density by computing the standard deviation of the pixel values in the annulus and propagating this uncertainty through to each pixel within the aperture. 
See Equations 4 and 5 in \citet{Genova-Santos2015}.
A breakdown of the statistical and systematic uncertainty of the spectral flux density measurements from our GBT maps is listed in Table~\ref{table:our_data_points}.
The spectral flux density values from all maps computed with this aperture photometry technique are listed in Table\ref{tab:datasets} and plotted in Figures~\ref{fig:spectrum}~and~\ref{fig:amefit}.
All of the maps and the smoothed versions of the maps are shown in the Appendix in Figures~\ref{fig:contours1}, \ref{fig:contours2}, and \ref{fig:contours3}.

%%%%%%%%%%%%%%%%%%%%%%%%%%%%%%%%%%%%%%%%%%%%%%%%%%%%%%%%%%%%%%%%%%%%%%%%%%%%%%%%

%%%%%%%%%%%%%%%%%%%%%%%%%%%%%%%%%%%%%%%%%%%%%%%%%%%%%%%%%%%%%%%%%%%%%%%%%%%%%%%%

\begin{table*}
\centering
\caption{Foreground models.}
\label{tab:models}
\begin{tabular}{lccc}
\hline \hline \\[-3mm]
Foreground &
Spectral Radiance Model [\,Jy/sr\,] &
Free Parameters &
Additional Information 
\\[2mm] \hline \hline \\[-3mm]
Conversion &
$\displaystyle{I_{RJ} = \frac{2 k_B\nu^2}{c} \times 10^{26}}$ &
&
from K to Jy/sr
\\[3mm] \hline \\[-2mm]
Thermal Dust &
$\displaystyle{ x = \frac{h \nu}{k T_{\rm D}} }$ &
$\displaystyle{ A_{\rm D} }$~[\,K\,] & 
\\[4mm]
&
$\displaystyle{ I_{\rm D}(\nu) = A_{\rm D}\,\left(\frac{x}{x_0}\right)^{\beta_{\rm D}+1}\, \frac{\expf{x_0} - 1}{\expf{x}-1} \times I_{RJ} }$ &
$\displaystyle{ \beta_{\rm D} }$ &
$\displaystyle{ \nu_0 = 545 }$~GHz
\\[4mm]
& 
&
$T_{\rm D}$~[\,K\,] & 
\\[2mm] \hline \\[-2mm]
Free-Free &
$\displaystyle{ g_{\rm FF} = \log\left(\frac{0.04955}{(\nu/10^9)}\right) + 1.5\,\log(T_{\rm e}) }$ &
$\displaystyle{ {\rm EM} }$~[\,$\rm{cm}^{-6} $pc\,] &
$\displaystyle{ T_{\rm e} = 8000 }$~K
\\[4mm]
& 
$\displaystyle{ T_{\rm FF} = 0.0314 \, \frac{T^{-.15}_{\rm e}}{(\nu/10^9)^{2}} \, {\rm EM} \, g_{\rm FF} }$ & 
& 
\\[4mm]
& 
$\displaystyle{ I_{\rm FF}(\nu) = T_{\rm e} \, \left( 1 - e^{-T_{\rm FF}} \right) \times I_{RJ} }$ & 
& 
\\[3mm] \hline \\[-2mm]
Spinning Dust &
$\displaystyle{ \Theta_{\rm SD}(\nu) = \textnormal{SD template}(\nu) }$ &
$\displaystyle{ A_{\rm SD} }$~[\,K\,] &
$\displaystyle{ \nu_0 = 22.8 }$~GHz
\\[4mm]
& 
$\displaystyle{ \Delta I_{\rm SD}(\nu) = A_{\rm SD} \left(\frac{\nu_0}{\nu}\right)^2\frac{\Theta_{\rm SD}(\nu\nu_{p_0}/\nu_p)}{\Theta_{\rm SD}(\nu_0\nu_{p_0}/\nu_p)}\times I_{RJ} }$ & $\displaystyle{ \nu_p }$~[\,GHz\,] &
$\displaystyle{\nu_{p_0} = 30.0 }$~GHz
\\[5mm] \hline \\[-2mm]
CMB &
$\displaystyle{ X = \frac{h \nu}{k_B T_{\rm CMB}} }$ &
$\displaystyle{ A_{\rm CMB} }$~[\,K\,] &
$T_{CMB}$ = 2.7255~K
\\[4mm]
& 
$\displaystyle{ g_f = \frac{(e^X - 1)^2}{X^2 e^X} }$ & 
& 
\\[4mm]
&
$\displaystyle{ I_{\rm CMB} = A_{\rm CMB} / g_f \times I_{RJ} }$ & 
&
\\[2mm] \hline
\end{tabular}
\end{table*}

%%%%%%%%%%%%%%%%%%%%%%%%%%%%%%%%%%%%%%%%%%%%%%%%%%%%%%%%%%%%%%%%%%%%%%%%%%%%%%%%

\section{Results}
\label{sec:results}

To understand the emission mechanism in the G107.2+5.2 AME region, we fit model spectra to the data points from our aperture photometry analysis.
These models are composed of CMB, thermal dust emission, optically thin free-free emission, and one AME component. 
The AME component is either spinning dust emission or optically thick free-free emission. 
These component models are the same used in \citet{PlanckXV}.
We fit the models to the data using the affine invariant Markov chain Monte Carlo (MCMC) ensemble sampler from the {\tt emcee} package~\citep{Foreman2012}, which gives model parameter values and parameter posterior probability distributions.
The maximum-likelihood parameter values are given in Table~\ref{tab:parameters} and the marginalized posteriors are plotted in the Appendix in Figures~\ref{fig:spinningdust_fit}~and~\ref{fig:freefree_fit}.
A physical description of the model components is given below in Section~\ref{sec:spectrum}, and the functional form of each model component is given in Table~\ref{tab:models}. 
We also compare the angular morphology of all the maps, which are also plotted in the Appendix in Figures~\ref{fig:contours1}~to~\ref{fig:contours5}.
Our interpretation of the results is given in Section~\ref{sec:ame}. \\

%%%%%%%%%%%%%%%%%%%%%%%%%%%%%%%%%%%%%%%%%%%%%%%%%%%%%%%%%%%%%%%%%%%%%%%%%%%%%%%%

\subsection{Emission Mechanisms}
\label{sec:spectrum}

%%%%%%%%%%%%%%%%%%%%%%%%%%%%%%%%%%%%%%%%%%%%%%%%%%%%%%%%%%%%%%%%%%%%%%%%%%%%%%%%

\begin{table*}
\centering
\caption{
Best-fit AME parameter values for an aperture region 45$^{\prime}$ in radius.
The associated models are given in Table~\ref{tab:models}, and the posteriors are plotted in Figures~\ref{fig:spinningdust_fit} and \ref{fig:freefree_fit}.
}
\label{tab:parameters}
\begin{tabular}{cccccccc}
\hline
                     & $\mathrm{EM_{Diffuse}}$ & $\mathrm{EM_{UCHII}}$             & $\mathrm{\theta_{UCHII}}$ & $A_{\rm D}$        & $\beta_{\rm D}$          & $T_{\rm D}$            & $A_{\rm CMB}$       \\
                     & [\,$\rm{cm^{-6} pc}$\,] & [\,$\rm{cm^{-6} pc}$\,]           & [\,arcsec\,]              & [\,$\mu$K\,]       & [\,$-$\,]                & [\,K\,]                & [\,$\mu$K\,]        \\
 UCHII Model         & $300^{+22}_{-24}$       & $5.27^{+2.5}_{-1.5}\times 10^{8}$ & $2.49^{+0.47}_{-0.44}$    & $1160^{+27}_{-27}$ & $1.83^{+0.057}_{-0.056}$ & $20.0^{+0.35}_{-0.34}$ & $-20.9^{+27}_{-28}$ \\[1mm] 
\hline
                     & $\mathrm{EM_{Diffuse}}$ & $A_{\rm SD}$                      & $\nu_p$                   & $A_{\rm D}$        & $\beta_{\rm D}$          & $T_{\rm D}$            & $A_{\rm CMB}$       \\
                     & [\,$\rm{cm^{-6} pc}$\,] & [\,$\mu$K\,]                      & [\,GHz\,]                 & [\,$\mu$K\,]       & [\,$-$\,]                & [\,K\,]                & [\,$\mu$K\,]        \\
 Spinning Dust Model & $339^{+16}_{-16}$       & $1380^{+160}_{-150}$              & $30.9^{+1.4}_{-1.4}$      & $1110^{+27}_{-27}$ & $1.94^{+0.057}_{-0.056}$ & $19.4^{+0.32}_{-0.31}$ & $142^{+20}_{-20}$   \\[1mm]
\hline
\end{tabular}
\end{table*}

%%%%%%%%%%%%%%%%%%%%%%%%%%%%%%%%%%%%%%%%%%%%%%%%%%%%%%%%%%%%%%%%%%%%%%%%%%%%%%%%

\subsubsection{Free-Free}

Free-free emission is electron-ion collision radiation in our Galaxy, typically in HII regions. 
The model we used in this study was derived by \citet{Draine2011}. 
We used the same model for optically thin and optically thick free-free emission.
The optically thin free-free emission is the diffuse signal commonly considered in CMB foreground analyses, while the optically thick free-free emission, which could be the source of the AME signal, has a much higher emission measure and is spatially compact.
In both cases we found the spectrum is very weakly dependent on the electron temperature, and therefore we set it to the commonly used value of 8000~K. 
Since the optically thick signal is compact, we do not resolve it, and an additional solid angle parameter is added to the model to account for the size of the compact region.
HII regions of the size and density we are considering are typically classified as ultra compact, so in this paper we commonly call the optically thick free-free emission UCHII.
The difference between the optically thin and the optically thick spectra is shown in the right panel of Figure~\ref{fig:amefit}.

%%%%%%%%%%%%%%%%%%%%%%%%%%%%%%%%%%%%%%%%%%%%%%%%%%%%%%%%%%%%%%%%%%%%%%%%%%%%%%%%

\subsubsection{Thermal Dust}

Thermal dust emission is the dominant radiation source above approximately 100~GHz.
The model we used is a modified blackbody spectrum with a power law emissivity. 
The dust grain properties can widely vary, which is accounted for by the emissivity power law.
This results in the 3-parameter modified black body spectrum we used. 
In principle, several different grain populations at different temperatures may be present in the G107.2+5.2 region and could be described by the inclusion of several modified blackbody spectra with different parameters. 
The presence of the star forming region S140 as well as the surrounding diffuse emission indeed could harbor grains at different temperatures, but the high-frequency data does not allow us to constrain multiple modified blackbody models. 
Additionally, the DIRBE beam size is 40$^\prime$ which does not allow us to spatially identify different regions within the beam.  

%%%%%%%%%%%%%%%%%%%%%%%%%%%%%%%%%%%%%%%%%%%%%%%%%%%%%%%%%%%%%%%%%%%%%%%%%%%%%%%%

\subsubsection{CMB}

The temperature of the CMB varies between our annulus and aperture because of the angular anisotropy. 
To account for this fact we included a CMB spectrum in our fit described by the first derivative of a blackbody with respect to the temperature.  
The amplitude of this derivative spectrum is a free parameter. 

%%%%%%%%%%%%%%%%%%%%%%%%%%%%%%%%%%%%%%%%%%%%%%%%%%%%%%%%%%%%%%%%%%%%%%%%%%%%%%%%

\subsubsection{Spinning Dust}

We used the spinning dust template from the Planck analysis~\citep{PlanckXV}, that is derived from the SPDust code~\citep{Yacine2009, silsbee2011} using the warm ionized medium (WIM) spinning-dust parameters. 
The free parameters in the model are the amplitude and peak frequency. 
Spinning dust emission is typically correlated with thermal dust emission because the two signals are produced by the same dust grains. 
We searched for radio/infrared map-domain correlations (see Section~\ref{sec:maps}), but this study was limited by the comparatively low resolution of the 28~GHz Planck data. 

%%%%%%%%%%%%%%%%%%%%%%%%%%%%%%%%%%%%%%%%%%%%%%%%%%%%%%%%%%%%%%%%%%%%%%%%%%%%%%%%

\subsubsection{Other}

We considered but ruled out other AME models including hard synchrotron radiation and thermal magnetic dust emission.
Hard synchrotron radiation has a falling spectral flux density, which was ruled out because the spectrum would have to be increasing instead to produce the observed excess near 30~GHz.
Note that we did not include conventional synchrotron radiation in our analysis for two reasons.
First, synchrotron radiation is not expected to vary appreciably on scales less than 1 degree, so it would appear as an offset in the map and should not effect the detected signal morphology.
Second, the shallowness of the measured spectrum below approximately 10~GHz is not consistent with the common $\beta \approx -1$ spectral index in Jansky units~\citep{Wehus2016}, so if there is any background synchrotron radiation, then it has to be negligible. 
Thermal magnetic dust is a possible AME source, but this signal is expected to have a spectrum that peaks near 70~GHz~\citep{Dickinson2018, Draine1999}, so it can not produce the observed excess near 30~GHz. 

%%%%%%%%%%%%%%%%%%%%%%%%%%%%%%%%%%%%%%%%%%%%%%%%%%%%%%%%%%%%%%%%%%%%%%%%%%%%%%%%

\subsection{Maps and Spatial Morphology}
\label{sec:maps}

%%%%%%%%%%%%%%%%%%%%%%%%%%%%%%%%%%%%%%%%%%%%%%%%%%%%%%%%%%%%%%%%%%%%%%

% description of GBT maps

Our GBT maps show diffuse emission inside the photometry aperture, which extends out to approximately 45$^\prime$ away from G107.2+5.2 (see Figure~\ref{fig:bankABmaps}). 
This diffuse emission spatially correlates very well visually with the high resolution CGPS data at 408~MHz.
Inside the photometry aperture we see the star forming region S140, a diffuse cloud centered on G107.2+5.2 (hereafter the cloud), and one bright radio point source. 
Outside the photometry aperture, we also detected three additional bright radio point sources and several other point sources with low SNR.

% description of common signals and spatial correlation 

\begin{figure*}
\centering
\includegraphics[width=0.48\textwidth]{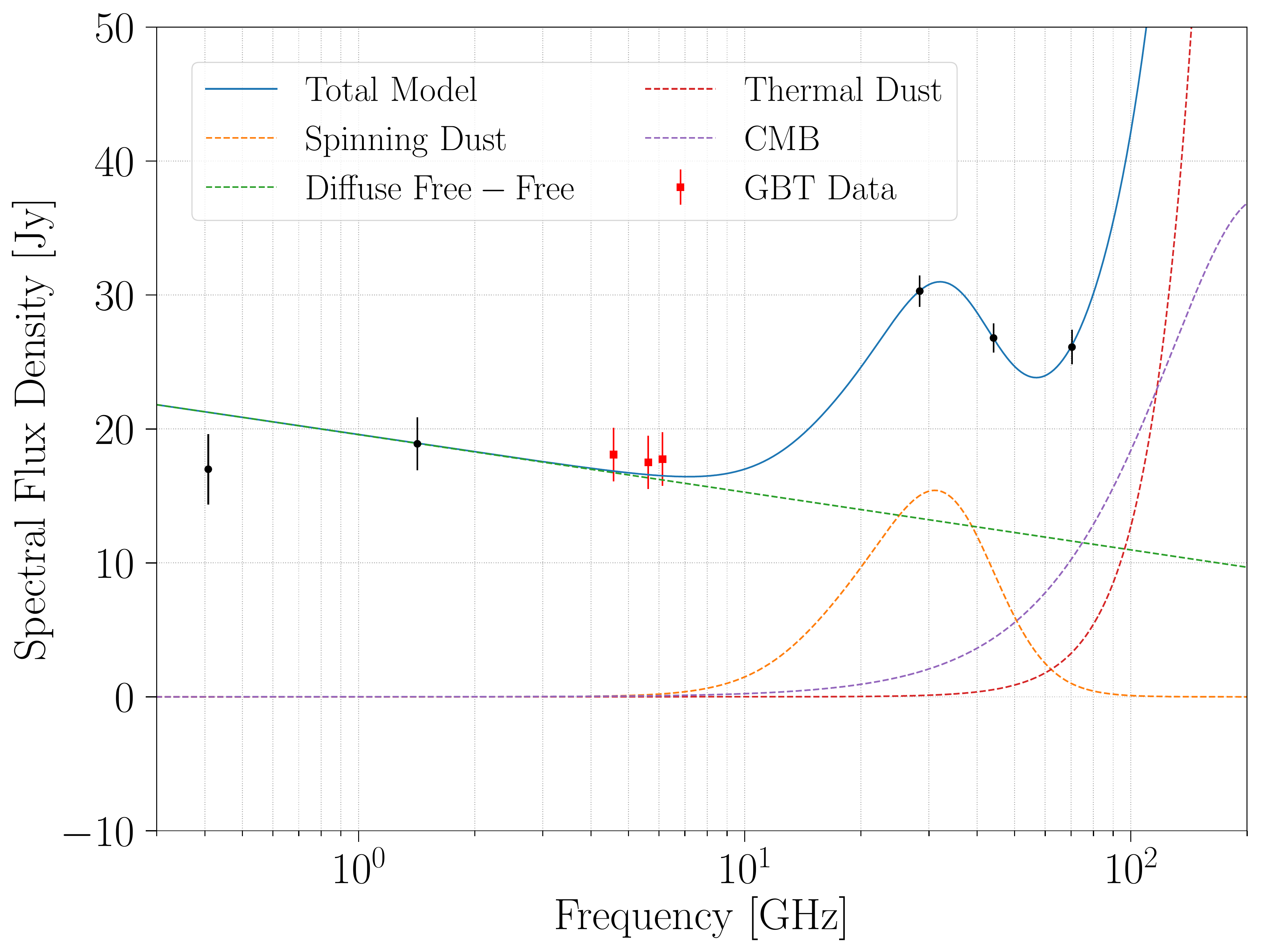}
\hspace{0.02\textwidth}
\includegraphics[width=0.48\textwidth]{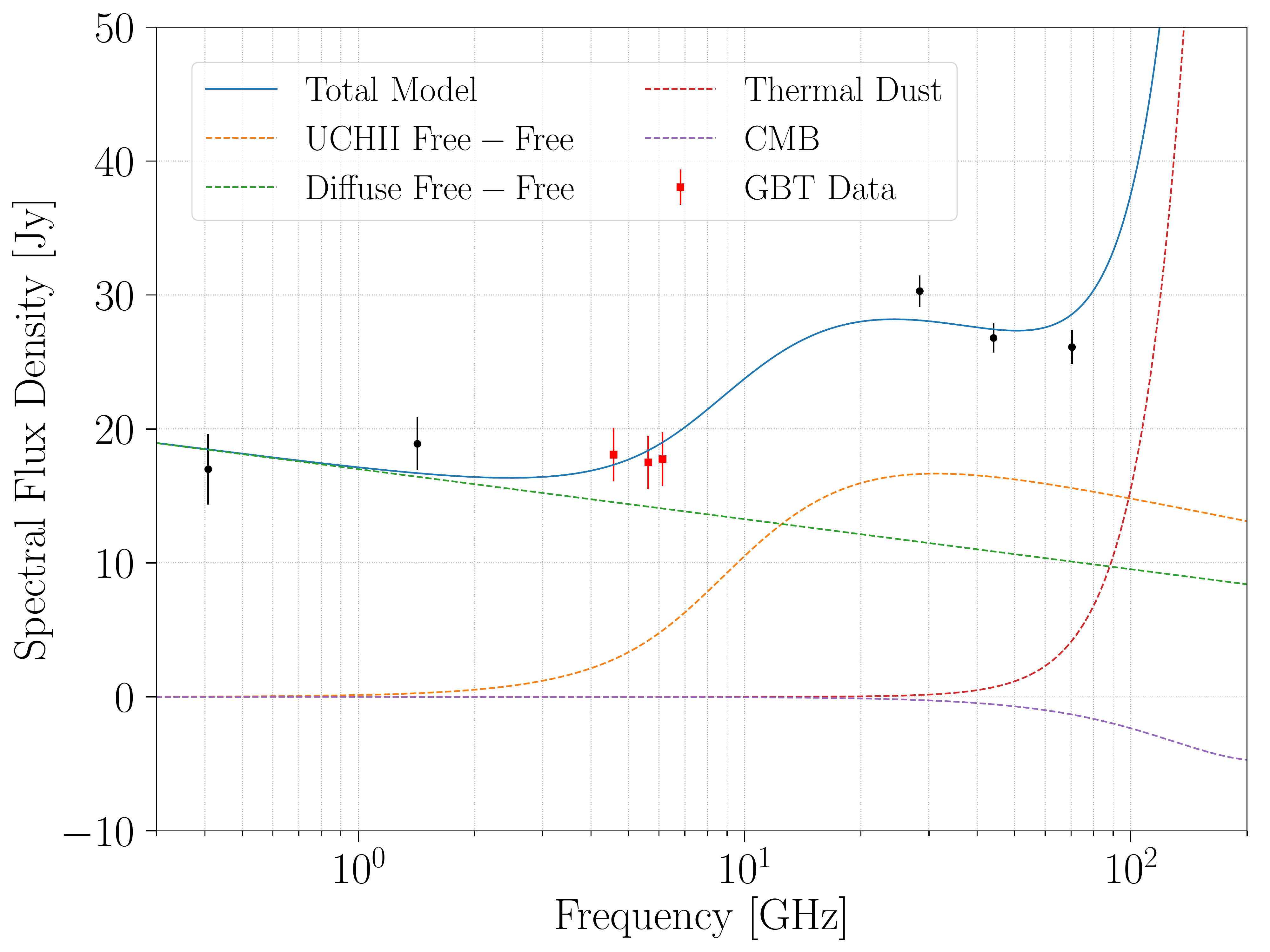}
\caption{
Data and best-fit models plotted between 300~MHz and 200~GHz.
As in Figure~\ref{fig:spectrum}, the models include diffuse free-free emission, thermal dust emission, the CMB, and one AME component -- either spinning dust (left) or UCHII free-free (right).
The foreground models are given in Table~\ref{tab:models}, the best-fit model parameters are given in Table~\ref{tab:parameters}, and the posteriors are plotted in Figures~\ref{fig:spinningdust_fit} and \ref{fig:freefree_fit}.
}
\label{fig:amefit}
\end{figure*}

The diffuse emission centered on G107.2+5.2 appears in all of the maps from 408~MHz up to 100~GHz.
This seems to indicated that this emission is diffuse free-free plus possibly AME near 30~GHz.
Above 100~GHz the diffuse signal in this region is faint when compared with the signal from S140.
This seems to indicate S140 contributes the majority of the thermal dust emission that appears in the measured spectrum.
Since S140 appears all the way down to 408~MHz, this seems to indicated that it contains a range of signals because thermal dust emission should be negligible below 70~GHz and effectively zero below 10~GHz (see Figure~\ref{fig:amefit}).

\subsection{Interpretation of Results}
\label{sec:ame}

%%%%%%%%%%%%%%%%%%%%%%%%%%%%%%%%%%%%%%%%%%%%%%%%%%%%%%%%%%%%%%%%%%%%%%%%%%%%%%%%

The spectrum shows a clear deviation from a simple model consisting of only optically thin free-free emission and thermal dust emission near 30~GHz indicating there is AME somewhere in the region defined by our photometry aperture.
The AME could be either in S140 or in the cloud or both.
Given the varied angular resolutions of all of the data in this study -- in particular the coarse resolution at 28~GHz -- it is difficult to say which case is correct.
Our GBT measurements near 5~GHz suggest the signal from the cloud is predominantly optically thin free-free emission.
Therefore, viable AME models must rapidly rise above approximately 5~GHz, peak near 30~GHz, and then remain sub dominant to thermal dust emission above 100~GHz.
Models based on both the spinning dust signal and the UCHII signal match this description.
However, this new information puts a tighter constraint on the angular size and emission measure of viable UCHII AME scenarios.

Fitting the combined model with the UCHII AME component to the observed spectrum results in a best-fit emission measure of $5.27^{+2.5}_{-1.5}\times 10^8$~$\rm{cm^{-6}pc}$ and an angular size of $2.49^{+0.47}_{-0.44}$~arcseconds. 
Given that S140 is 910~pc away, the angular size from the fit corresponds to an HII region with a physical extent of $1.01^{+0.21}_{-0.20} \times 10^{-2}$~pc.
Note that an UCHII region of this size and emission measure might be better classified as a hyper-compact HII region~\citep{murphy2010}.
High-resolution, interferometric measurements of S140 at 15~GHz from AMI did indeed reveal a rising spectrum but did not conclusively resolve any UCHII regions and the AMI collaboration concluded the AME signal is likely from spinning dust~\citep{perrott2013}.
%
% This information when combined with our fit parameters suggests that any UCHII signal originating from within the region defined by our photometry aperture comes from the G107.2+5.2 region, but more investigation is required.
%
Our spectrum fit suggests that, if it is present, we have enough sensitivity to see the UCHII signal in our GBT maps, however our maps do not conclusively show compact discrete sources in the the cloud.
Therefore, if the AME signal is from UCHII emission in the cloud, then it seems there must be multiple UCHII sources that together look like the single diffuse region we detected.

The combined model with the spinning dust AME component also explains the AME excess. 
The best-fit model gives a spinning-dust peak frequency of $30.9\pm{1.4}$~GHz, with a peak amplitude of $15.2^{+1.8}_{-1.7}$~Jy. 
Spinning dust should correlate well with thermal dust emission.
The spinning-dust AME signal could be from S140, where there is obviously a significant amount of thermal dust emission, or it could be from the cloud or both.
However, the Planck maps show that any thermal dust emission in the cloud is small.
Therefore, if the AME is from spinning dust it seems likely that it is coming from S140.
The resulting emission at 28~GHz is then both spinning dust emission emanating from around S140 and optically thin free-free emission from the diffuse cloud present at 28~GHz.
The comparatively low angular resolution of the 28~GHz map results in the bright region over both S140 and the cloud as seen in Figure~\ref{fig:contours1}.

%%%%%%%%%%%%%%%%%%%%%%%%%%%%%%%%%%%%%%%%%%%%%%%%%%%%%%%%%%%%%%%%%%%%%%%%%%%%%%%%

\section{Discussion}
\label{sec:discussion}

%%%%%%%%%%%%%%%%%%%%%%%%%%%%%%%%%%%%%%%%%%%%%%%%%%%%%%%%%%%%%%%%%%%%%%%%%%%%%%%%

The goal for this study was to determine the AME mechanism in the G107.2+5.2 region.
Our measurements are consistent with and further support the spinning dust scenario, and they conclusively ruled out some of parameter space for the UCHII scenario.
Additional measurements are needed to concretely determine the emission mechanism.

High angular resolution measurements near 30~GHz are ideal.
Ku-band (12.0 to 15.4~GHz) observations at GBT, for example, would provide valuable spectral information where the AME signal rises.
If the AME signal is in fact from spinning dust, then polarization measurements in Ku band could convincingly reveal the polarization fraction of this spinning-dust signal.
Additionally, the angular resolution in Ku band would be higher, providing a better view of the morphology of the region.
Our original project proposal requested both C-band and Ku-band observations.
Unfortunately, the Ku-band receiver was not available in the 17A semester at GBT when we observed.
Therefore, we are planning a follow-up observing proposal for these Ku-band observations.

High resolution H-alpha measurements would also help because H-alpha is a tracer of free-free emission.
We investigated the Finkbeiner composite H-alpha map that uses data from the Wisconsin $\rm{H-\alpha}$ Survey and Virginia Tech Spectral Lines Survey~\citep{finkbeiner2003_halpha}.
However, in the G107.2+5.2 region the resolution of the survey is approximately 1 degree, which makes spatial comparisons difficult, and significant dust extinction is present. \vspace{10mm}

%%%%%%%%%%%%%%%%%%%%%%%%%%%%%%%%%%%%%%%%%%%%%%%%%%%%%%%%%%%%%%%%%%%%%%%%%%%%%%%%

%%%%%%%%%%%%%%%%%%%%%%%%%%%%%%%%%%%%%%%%%%%%%%%%%%%%%%%%%%%%%%%%%%%%%%%%%%%%%%%%

\section{Conclusions}
\label{sec:conclusions}

%%%%%%%%%%%%%%%%%%%%%%%%%%%%%%%%%%%%%%%%%%%%%%%%%%%%%%%%%%%%%%%%%%%%%%%%%%%%%%%%

In this study, we performed follow-up C-band observations of the region G107.2+5.2 and fit two potential AME models to the resulting spectra to explain the excess microwave emission at 30~GHz. 
We find that spinning dust emission or optically thick free-free emission can explain the AME in this region. 
Additional studies including higher spatial resolution data at 30~GHz as well as high resolution H-alpha data are necessary disentangle the two emission mechanics. 
Our analysis of the C-band polarization data are ongoing.
We also plan to look at radio recombination lines between 4-8 GHz, using the high spectral resolution of the GBT data. \vfill

%%%%%%%%%%%%%%%%%%%%%%%%%%%%%%%%%%%%%%%%%%%%%%%%%%%%%%%%%%%%%%%%%%%%%%%%%%%%%%%%

\section{Acknowledgements}
\label{sec:achnowledgements}

%%%%%%%%%%%%%%%%%%%%%%%%%%%%%%%%%%%%%%%%%%%%%%%%%%%%%%%%%%%%%%%%%%%%%%%%%%%%%%%%

We would like to thank the National Radio Astronomy Observatory (NRAO) and the Green Bank Observatory for allowing us to use GBT for this study. 
Our observation award was GBT17A-259.
In particular, we would like to thank David Frayer who provided invaluable project support.
This research was supported in part by a grant to BRJ from the Research Initiatives for Science and Engineering (RISE) program at Columbia University.
CD and SH acknowledge support from an STFC Consolidated Grant (ST/P000649/1) and an ERC Consolidator Grant (no.~307209) under the FP7.
We also thank Roland Kothes for providing 408~MHz Canadian Galactic Plane Survey data of the region.
The Green Bank Observatory is a facility of the National Science Foundation operated under cooperative agreement by Associated Universities, Inc.

%%%%%%%%%%%%%%%%%%%%%%%%%%%%%%%%%%%%%%%%%%%%%%%%%%%%%%%%%%%%%%%%%%%%%%%%%%%%%%%%

\begin{appendix}

\begin{figure*}
\centering
\includegraphics[width=0.8\textwidth]{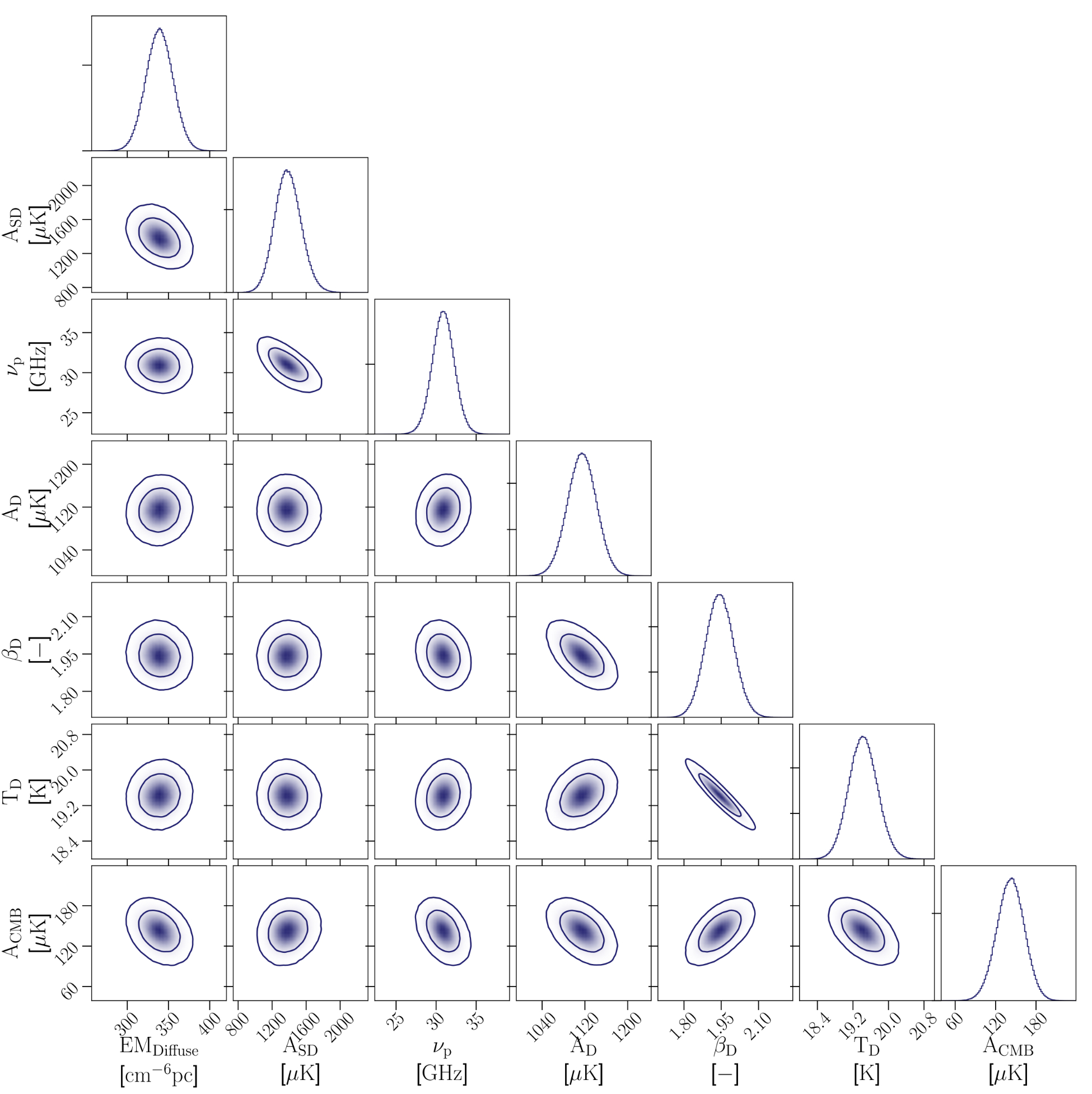}
\caption{
Posterior from fit using the spinning-dust model.
The foreground models are given in Table~\ref{tab:models}, and the best-fit model parameters are given in Table~\ref{tab:parameters}.
}
\label{fig:spinningdust_fit}
\end{figure*}

\begin{figure*}
\centering
\includegraphics[width=0.8\textwidth]{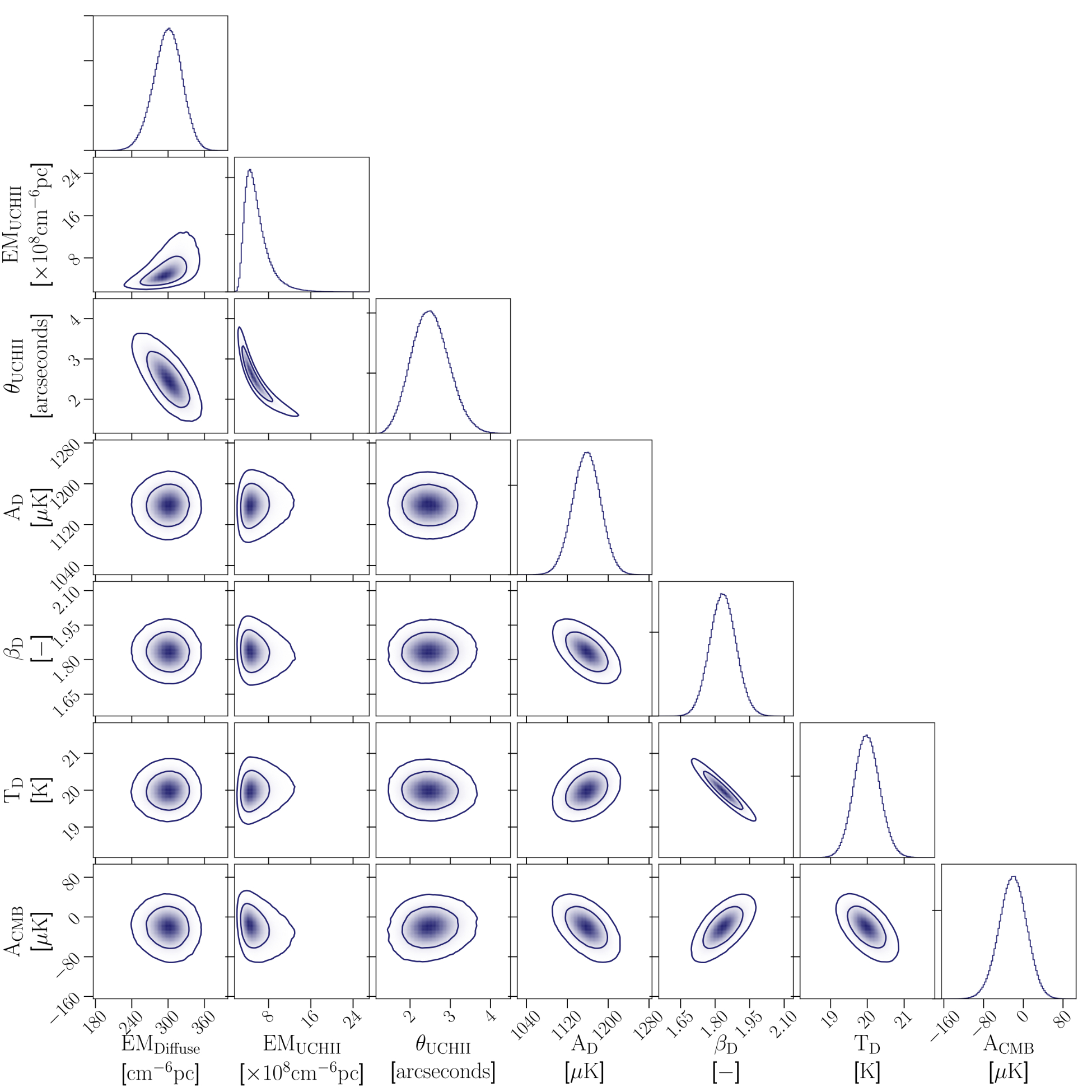}
\caption{
Posterior from fit using the UCHII model.
The foreground models are given in Table~\ref{tab:models}, and the best-fit model parameters are given in Table~\ref{tab:parameters}.
}
\label{fig:freefree_fit}
\end{figure*}

\begin{figure*}
\centering
\includegraphics[scale=0.4]{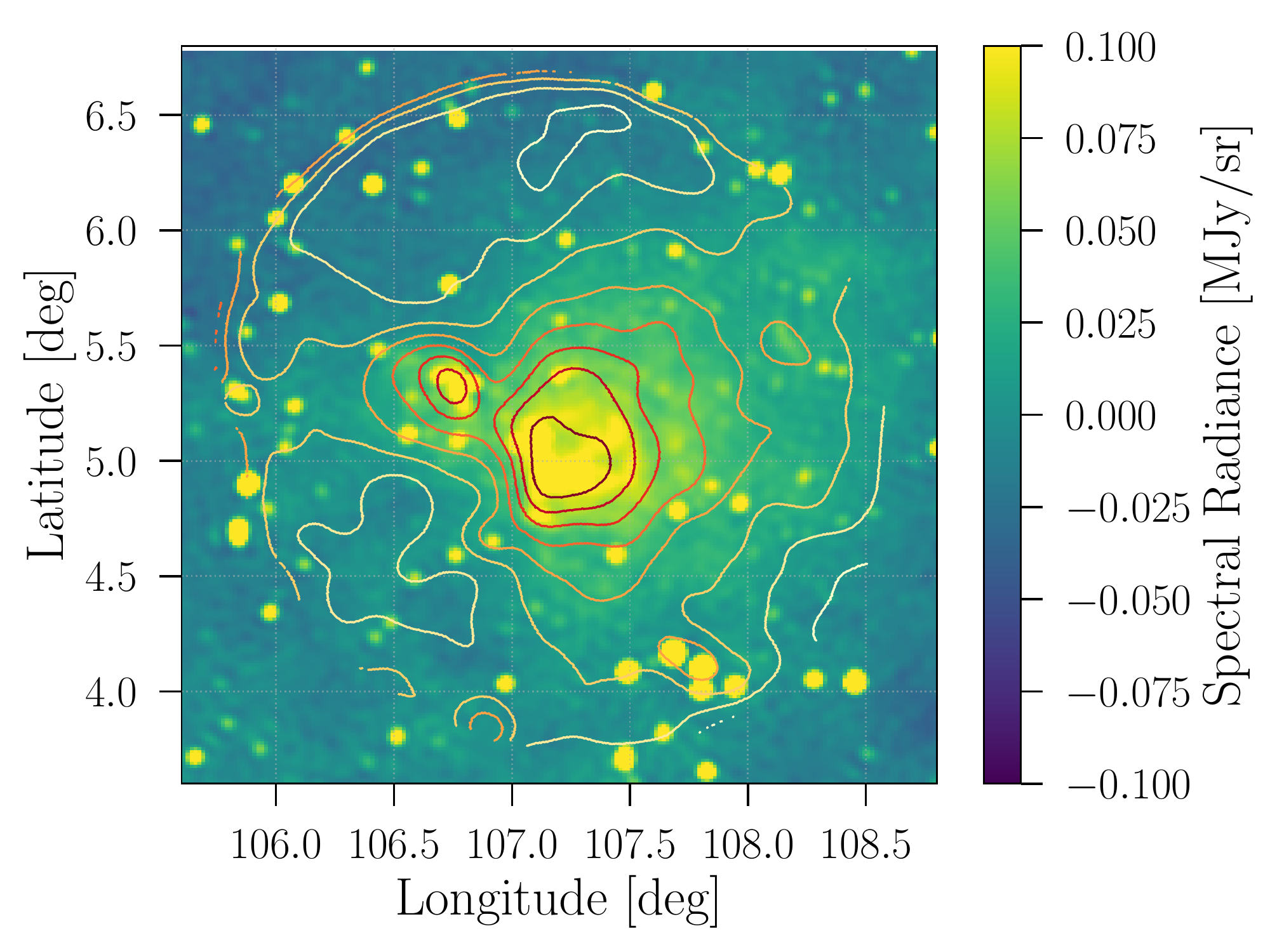}
\hspace{0.1in}
\includegraphics[scale=0.4]{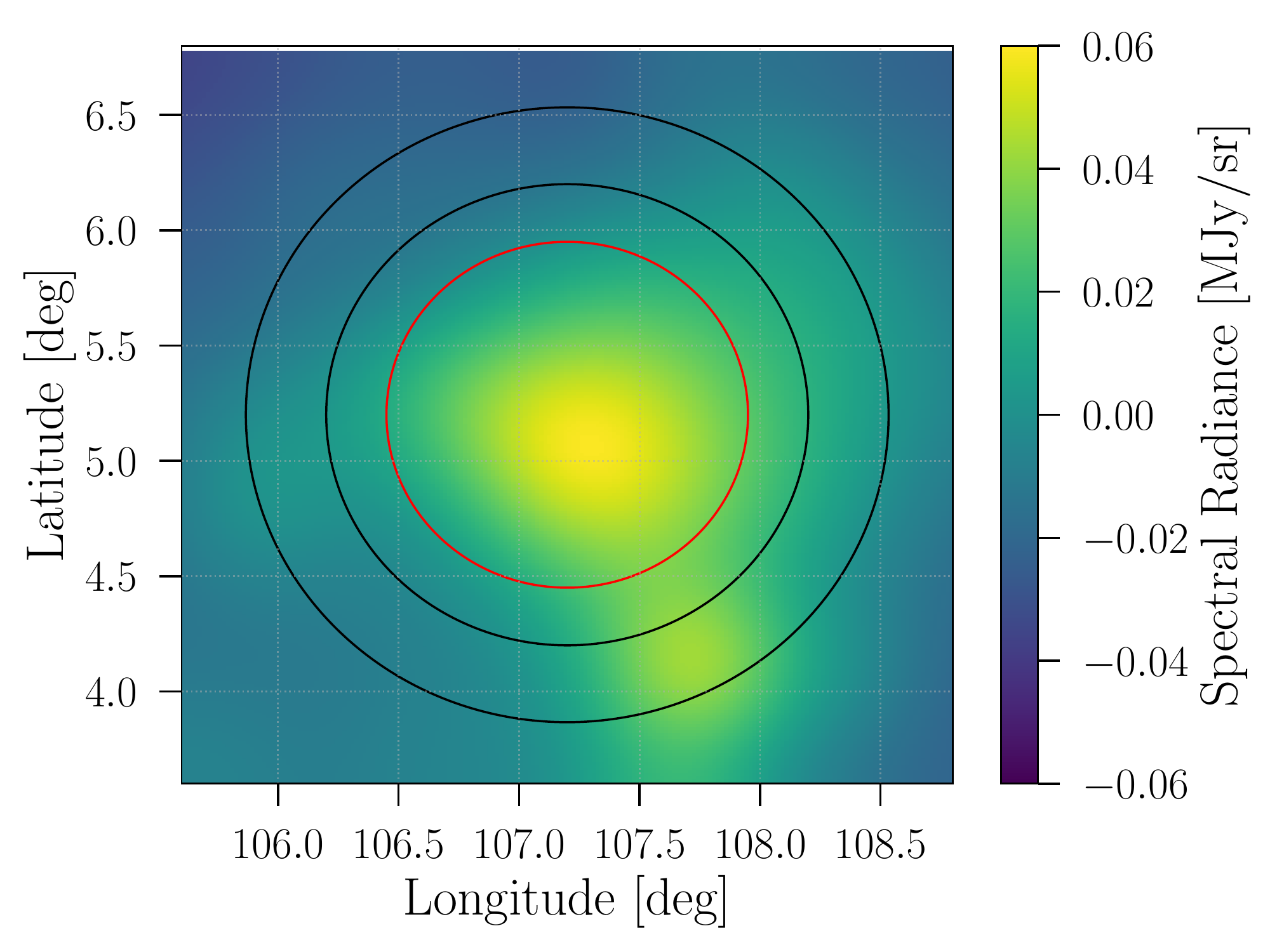}
\includegraphics[scale=0.4]{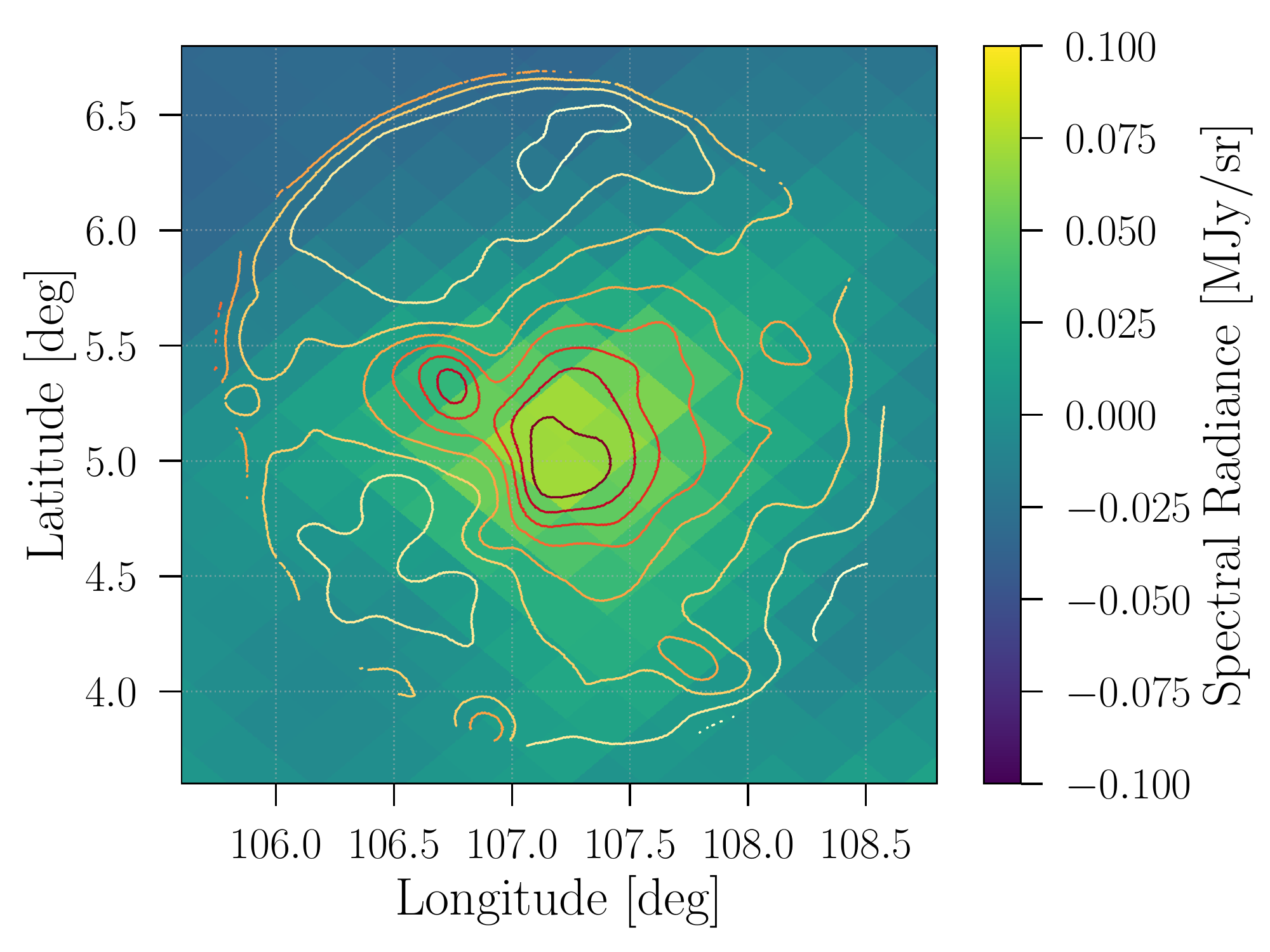}
\hspace{0.1in}
\includegraphics[scale=0.4]{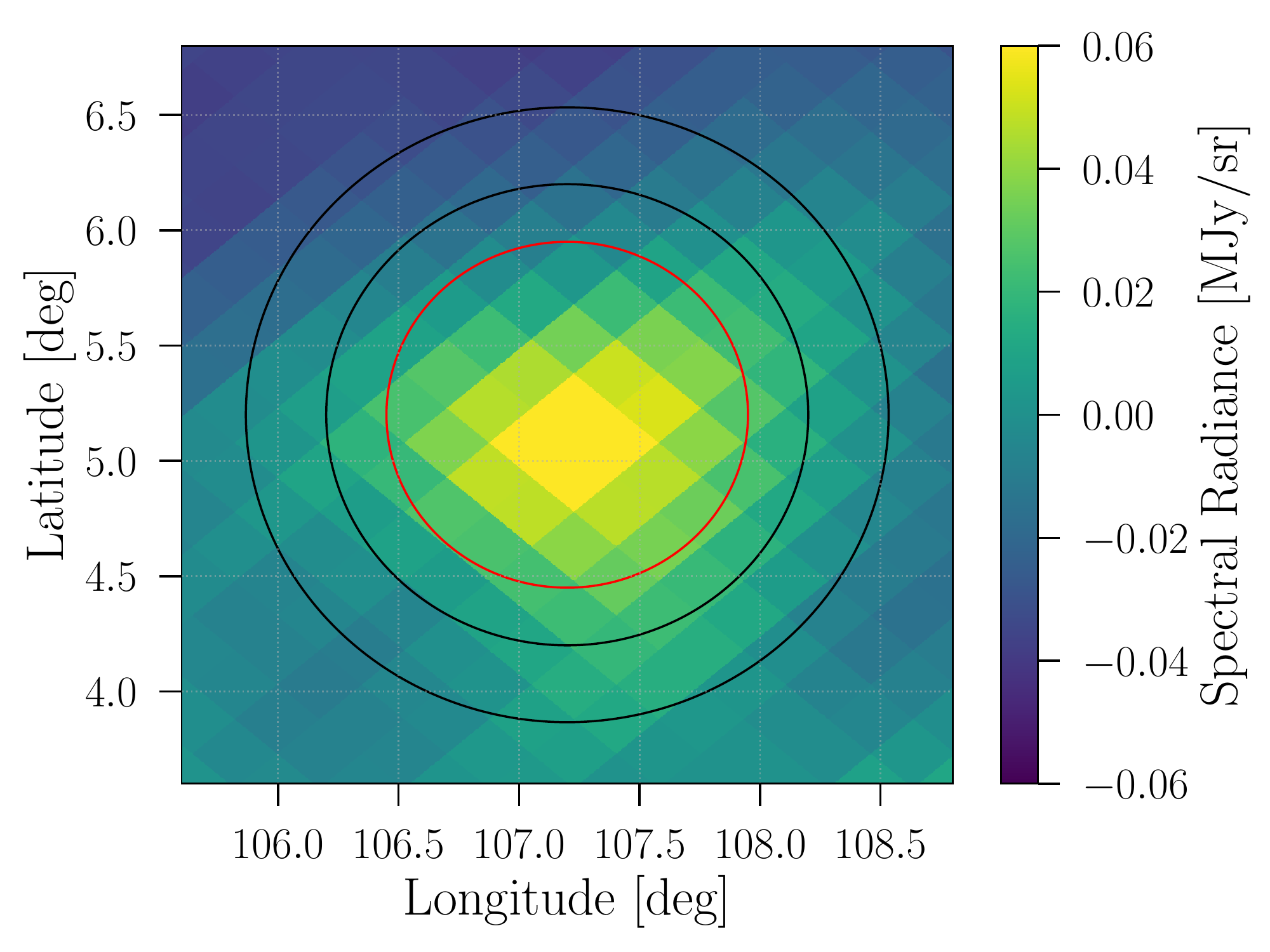}
\includegraphics[scale=0.4]{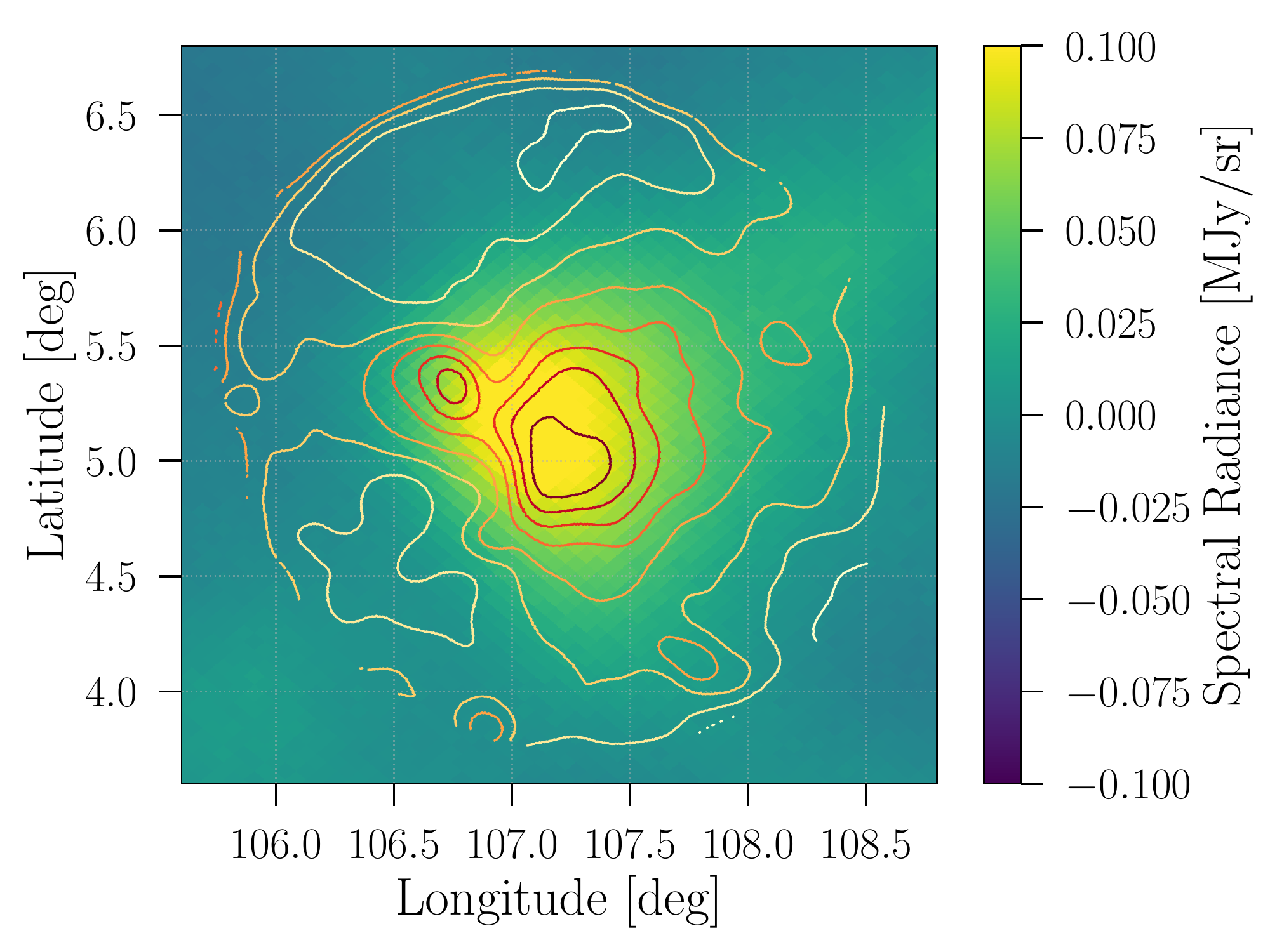}
\hspace{0.1in}
\includegraphics[scale=0.4]{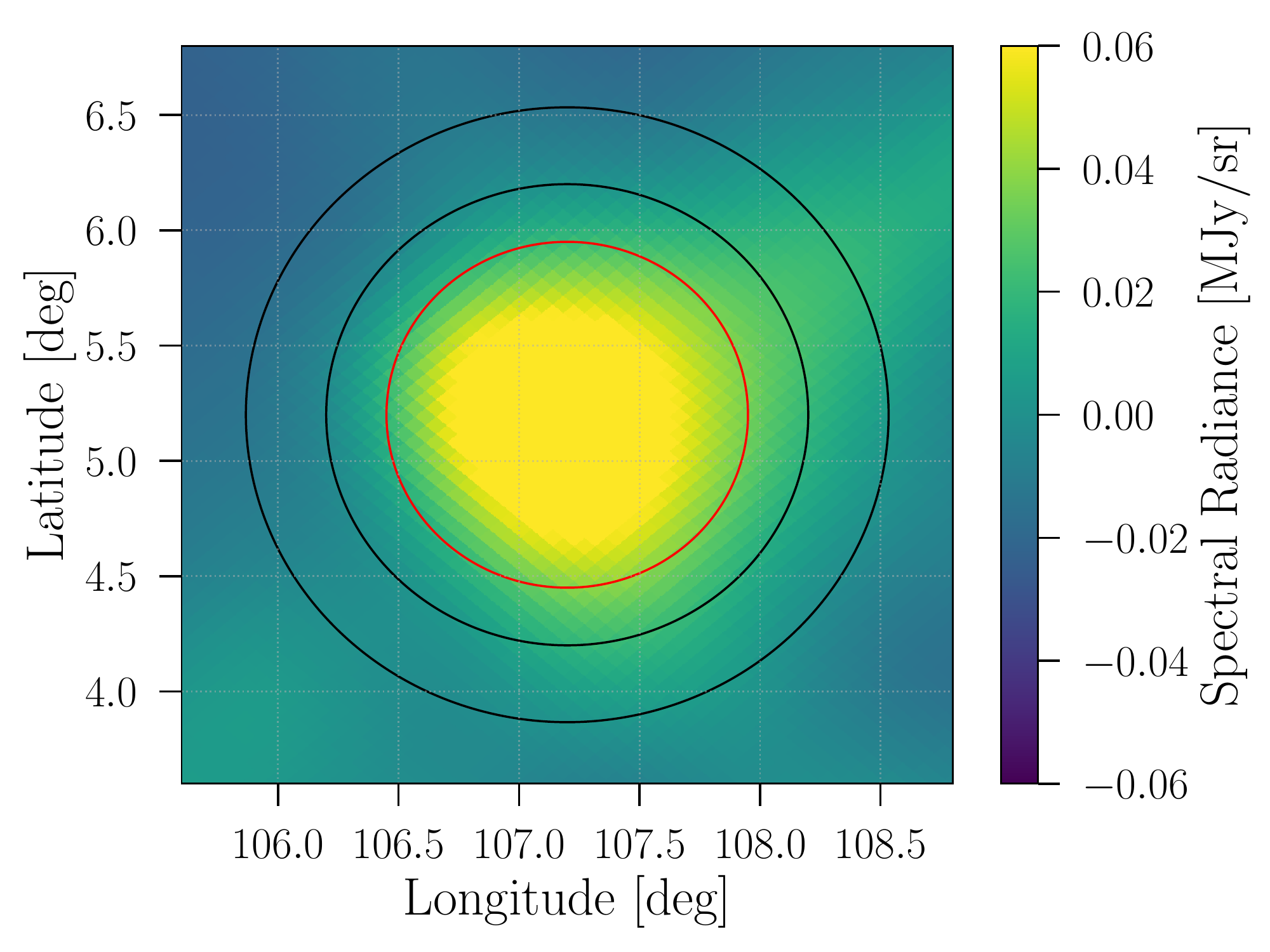}
\caption{
Maps used in this study.
The left column shows the map with contours from our Bank~A (4.575~GHz) map oveplotted for comparison.
The right column shows the map on the left smoothed to 40$^{\prime}$ resolution.
Here, the aperture (circle 45$^{\prime}$ radius) and the zero-point annulus ( 60$^{\prime}$ to 80$^{\prime}$) are overplotted.
From top to bottom, the rows are CGPS (0.408~GHz), Stocker (1.42~GHz), and Planck (28~GHz).
Maps in each column are plotted with the same color scale for straightforward comparison.
References are given in Table~\ref{tab:datasets}.
}
\label{fig:contours1}
\end{figure*}

\begin{figure*}
\centering
\includegraphics[scale=0.4]{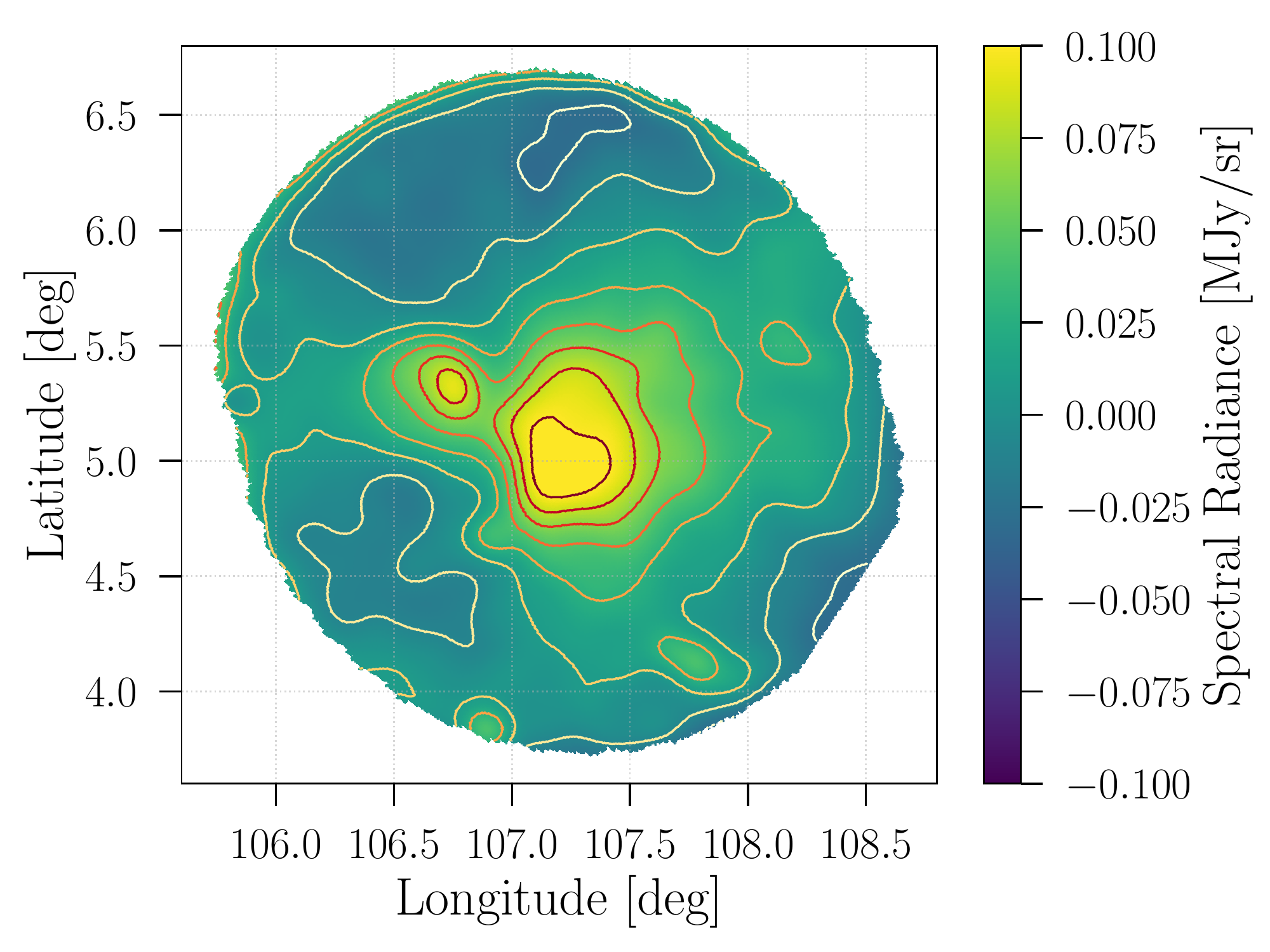}
\hspace{0.1in}
\includegraphics[scale=0.4]{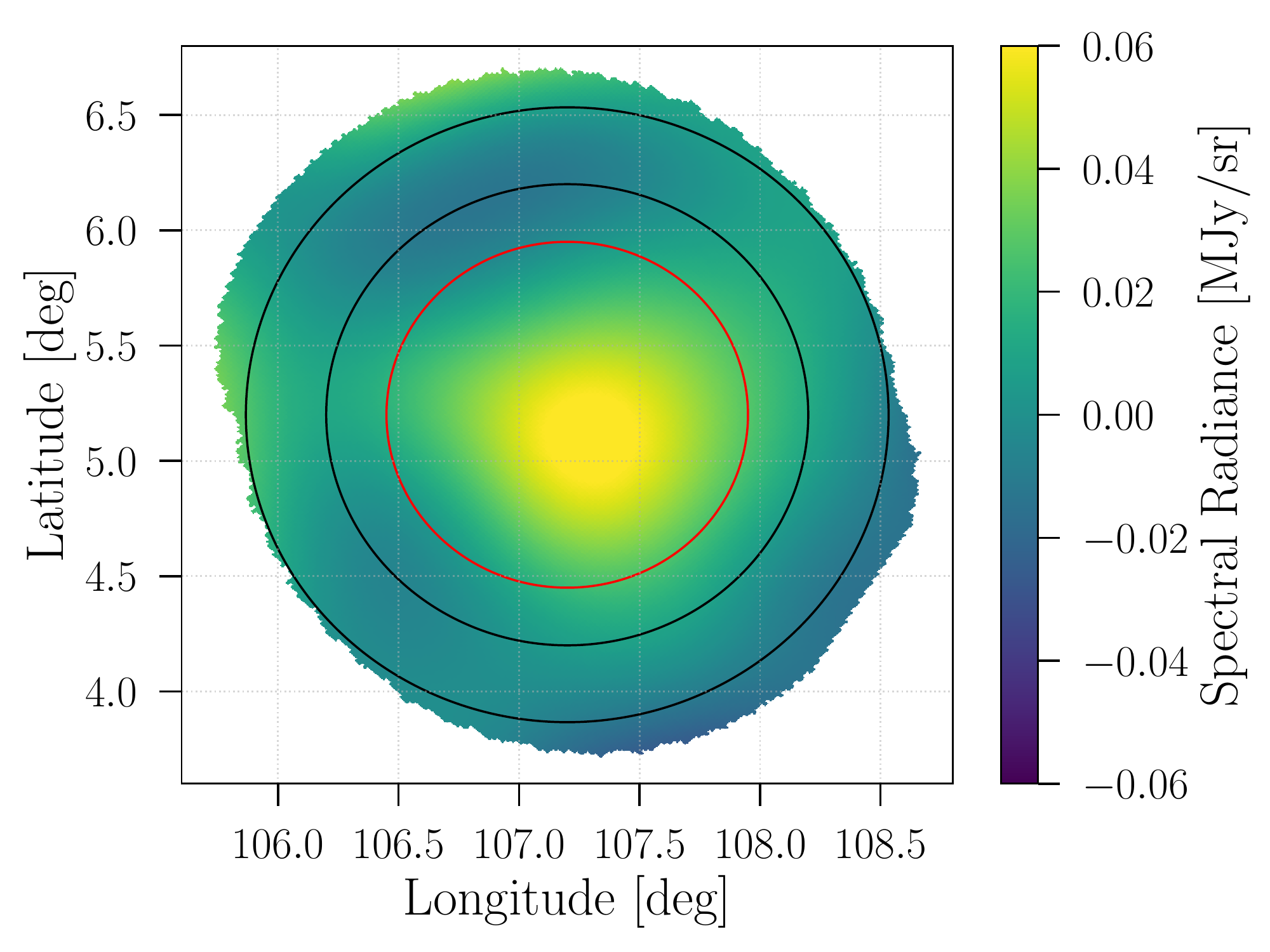}
\includegraphics[scale=0.4]{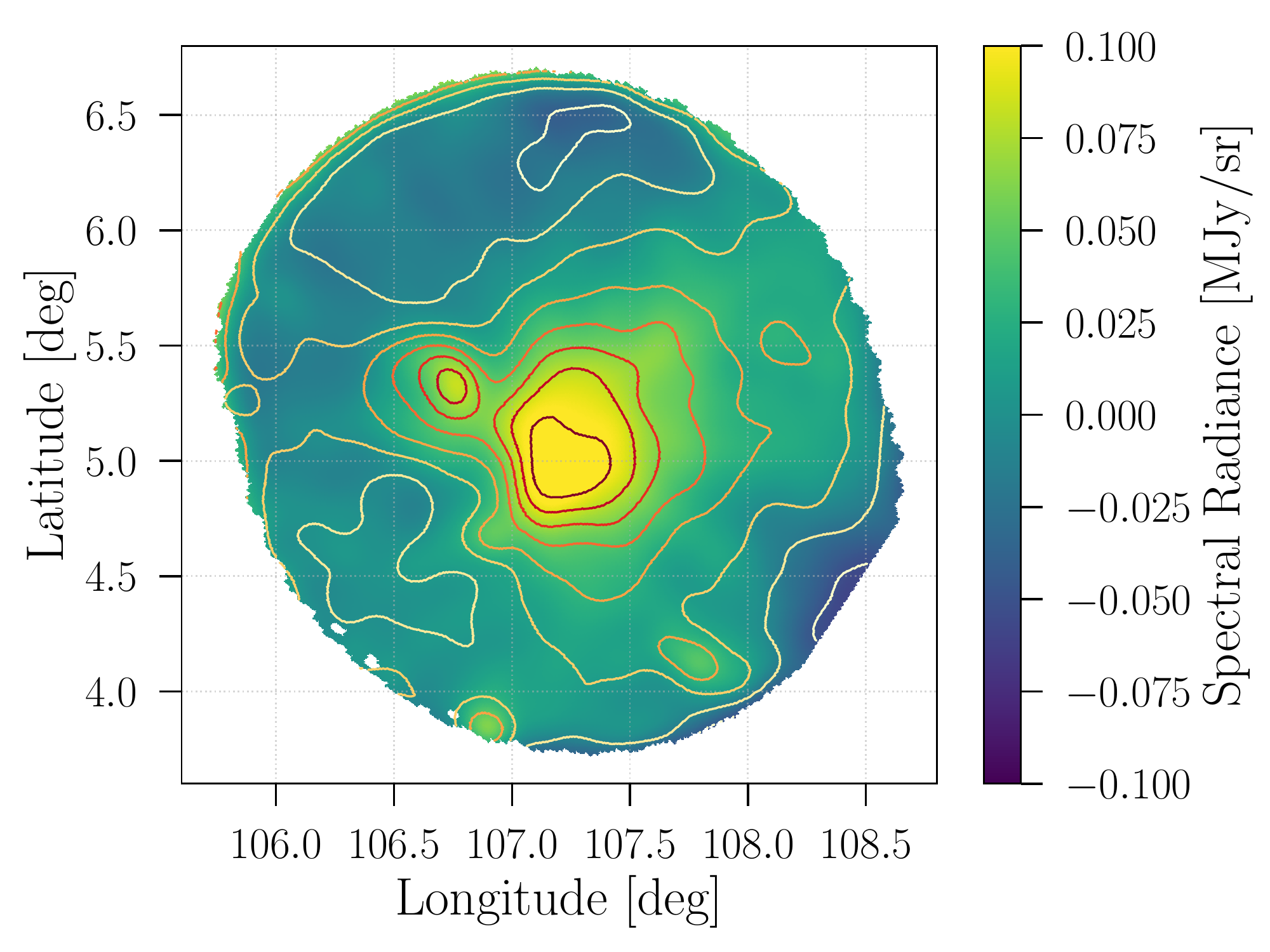}
\hspace{0.1in}
\includegraphics[scale=0.4]{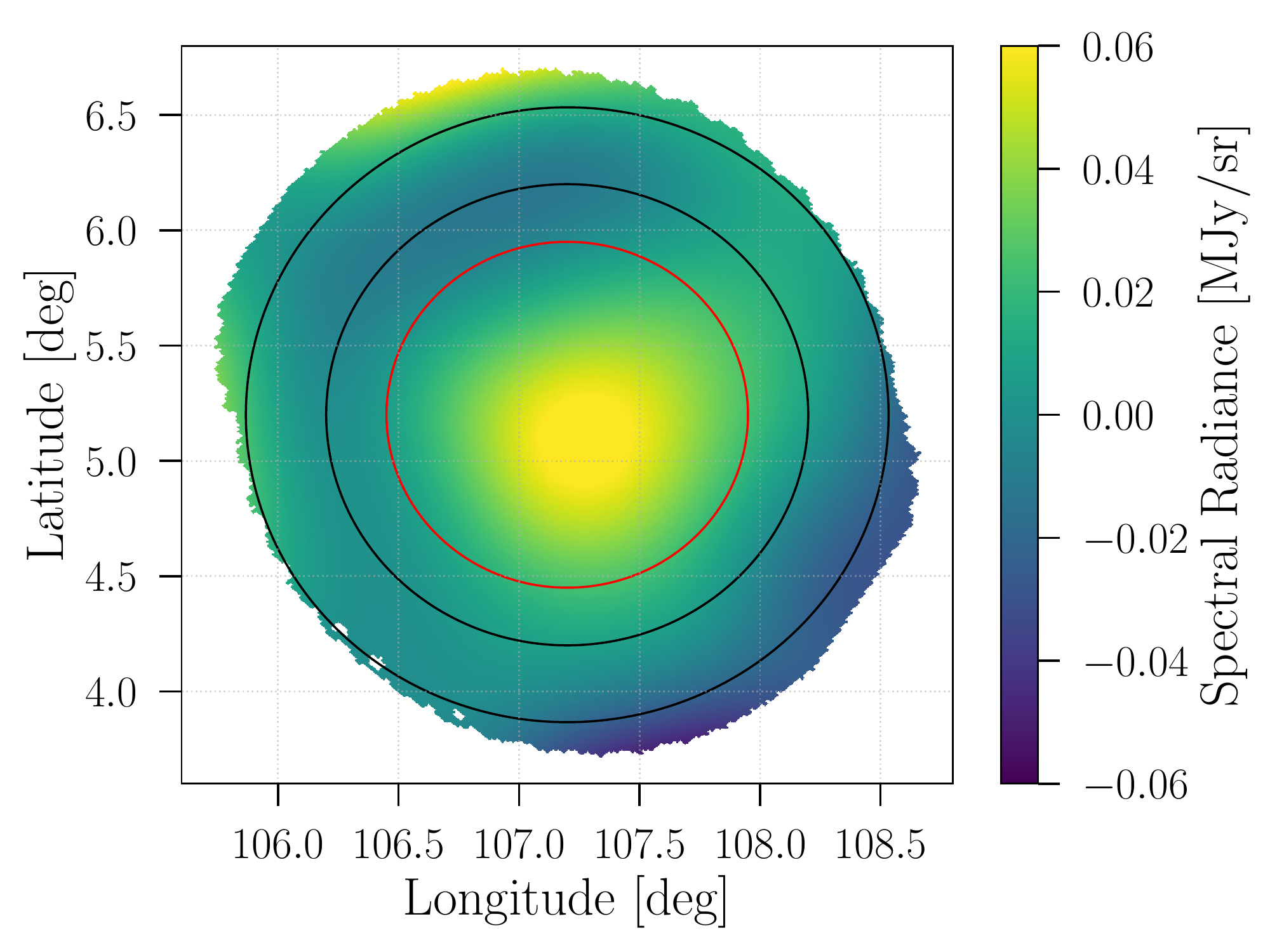}
\includegraphics[scale=0.4]{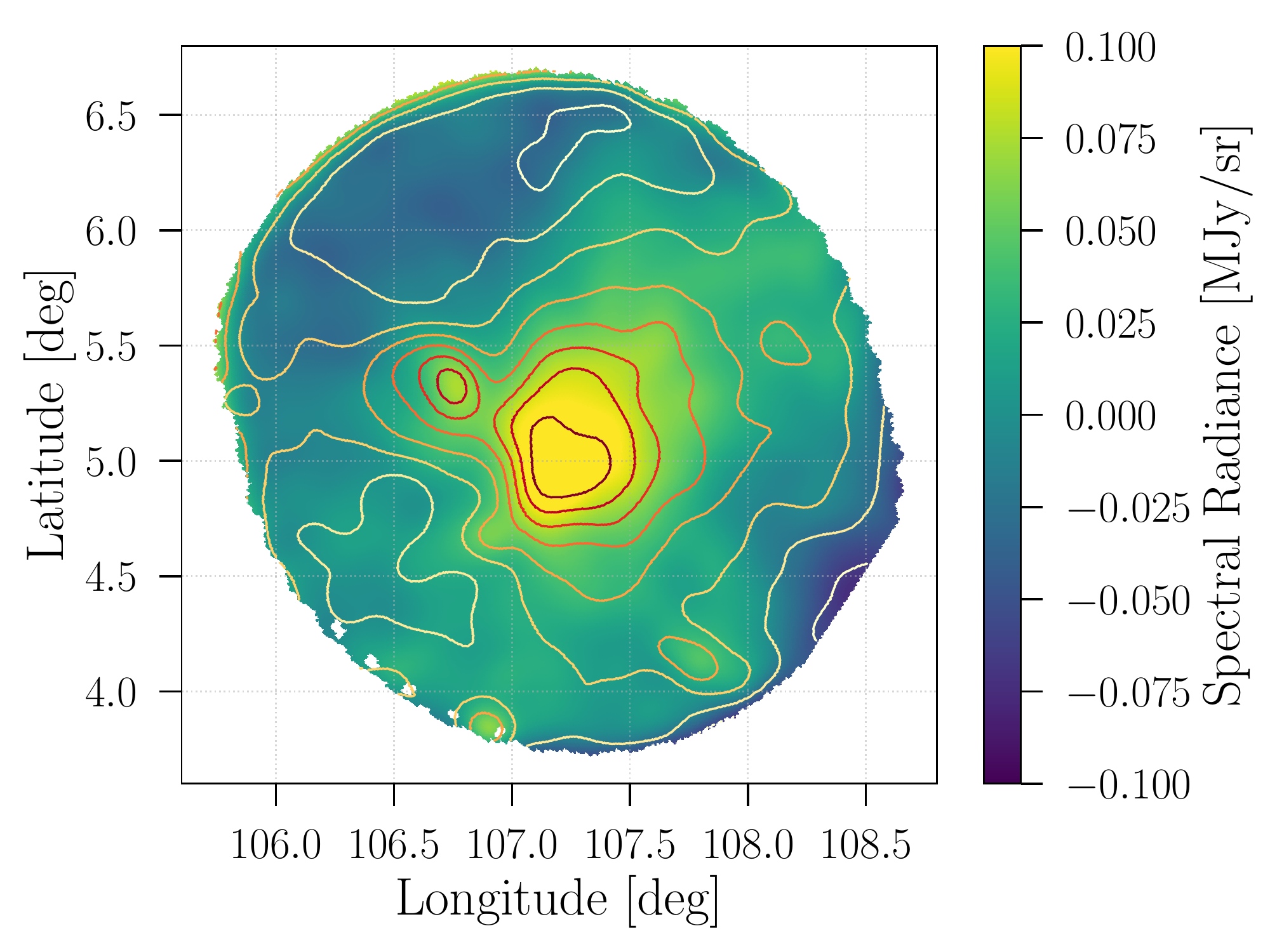}
\hspace{0.1in}
\includegraphics[scale=0.4]{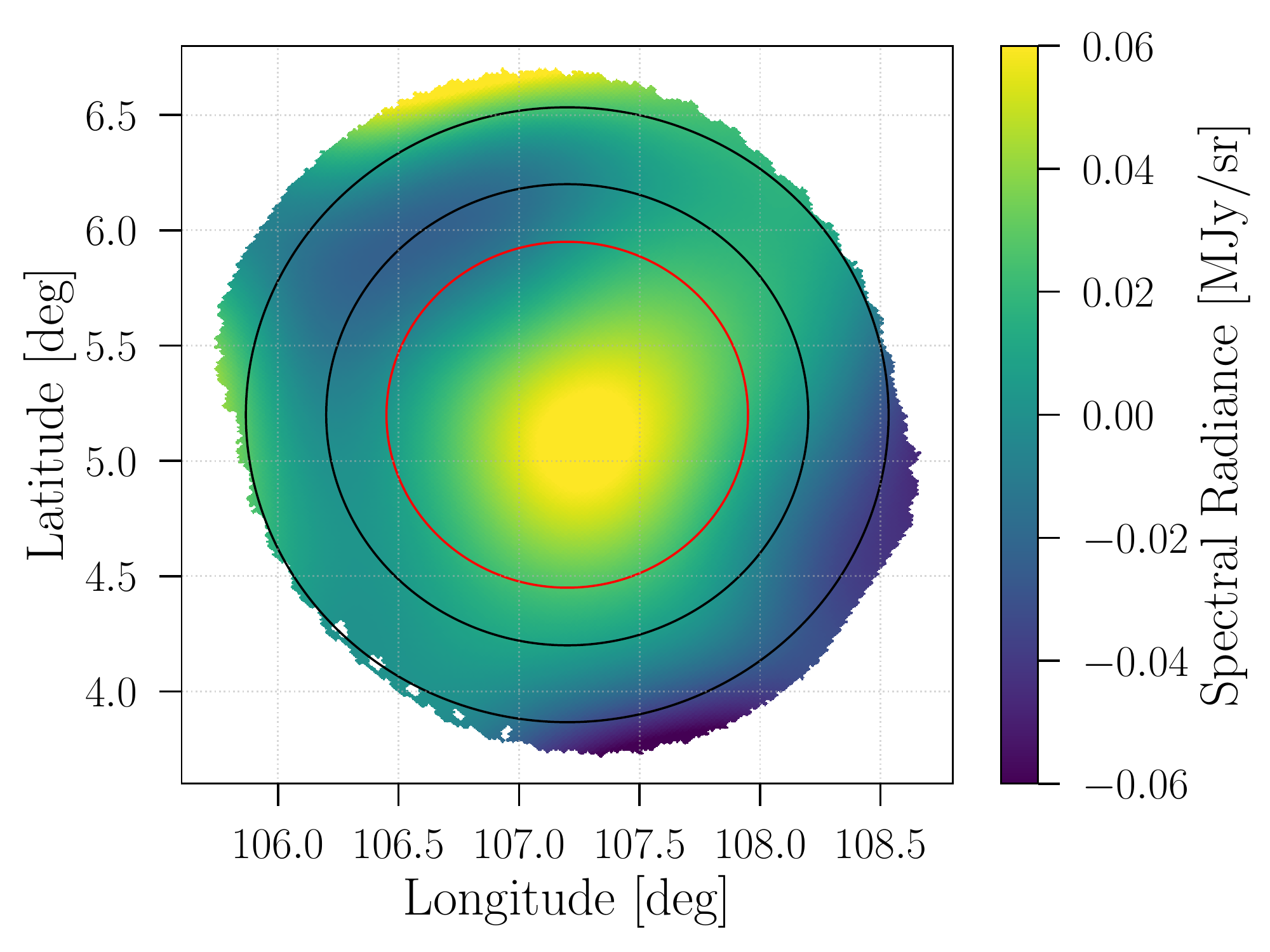}
\caption{
Maps used in this study.
The left column shows the map with contours from our Bank~A (4.575~GHz) map oveplotted for comparison.
The right column shows the map on the left convolved with a 40$^{\prime}$ Gaussian.
Here, the aperture (circle 45$^{\prime}$ radius) and the zero-point annulus ( 60$^{\prime}$ to 80$^{\prime}$) are overplotted.
From top to bottom, the rows are GBT Bank A (4.575~GHz), GBT Bank B (5.625~GHz), and GBT Bank C (6.125~GHz).
Maps in each column are plotted with the same color scale for straightforward comparison.
References are given in Table~\ref{tab:datasets}.
}
\label{fig:contours2}
\end{figure*}

\begin{figure*}
\centering
\includegraphics[scale=0.4]{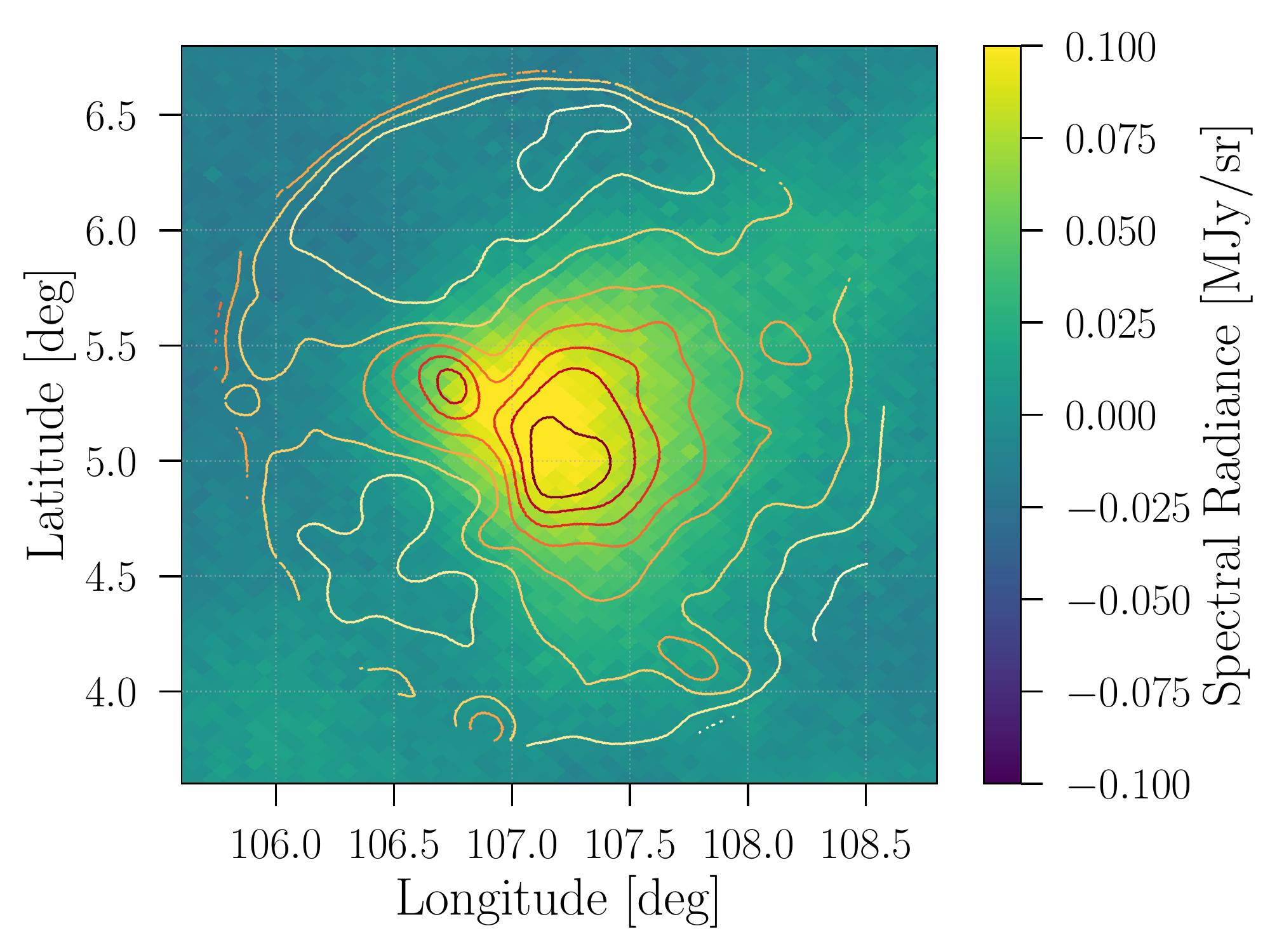}
\hspace{0.1in}
\includegraphics[scale=0.4]{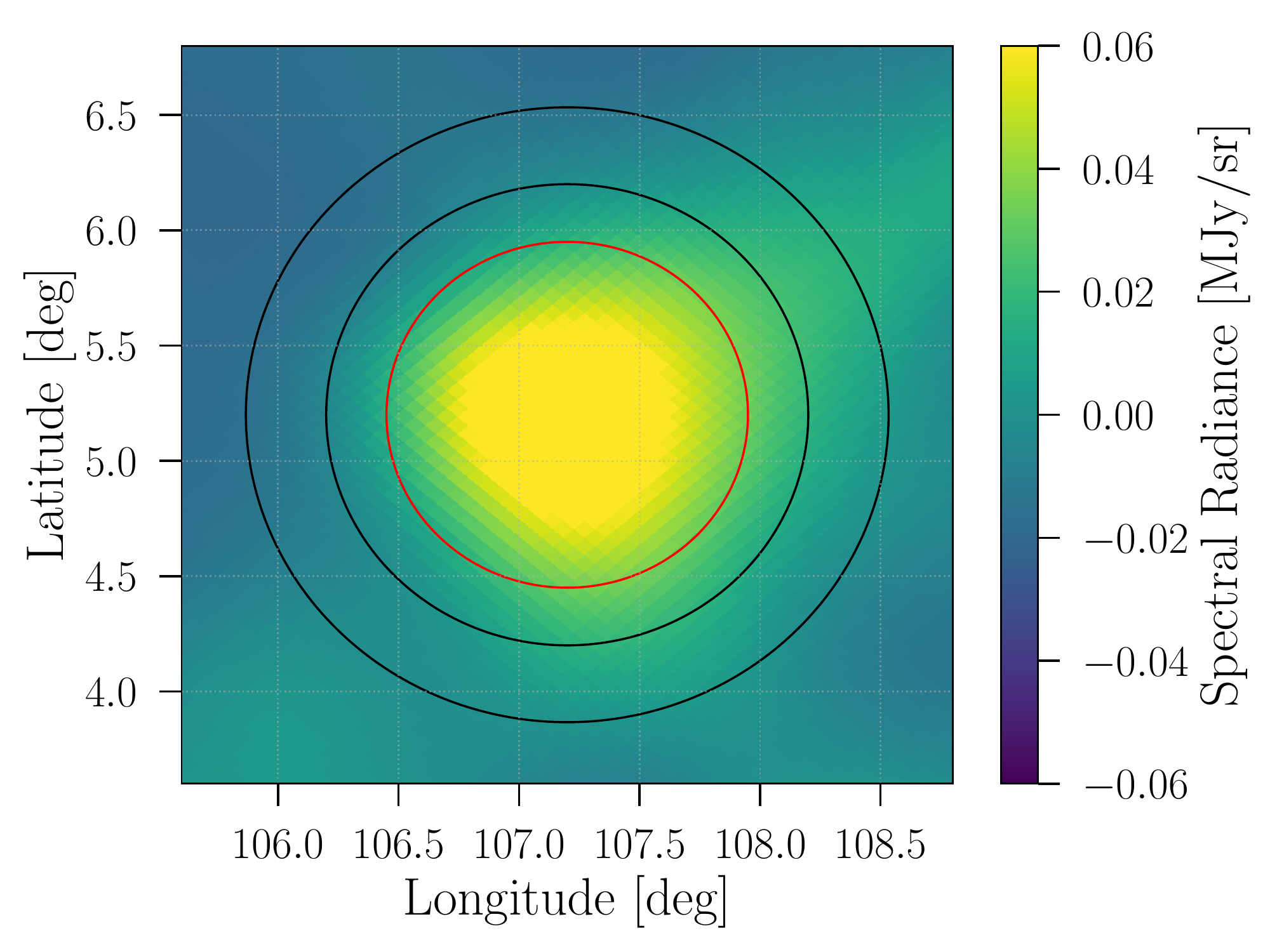}
\includegraphics[scale=0.4]{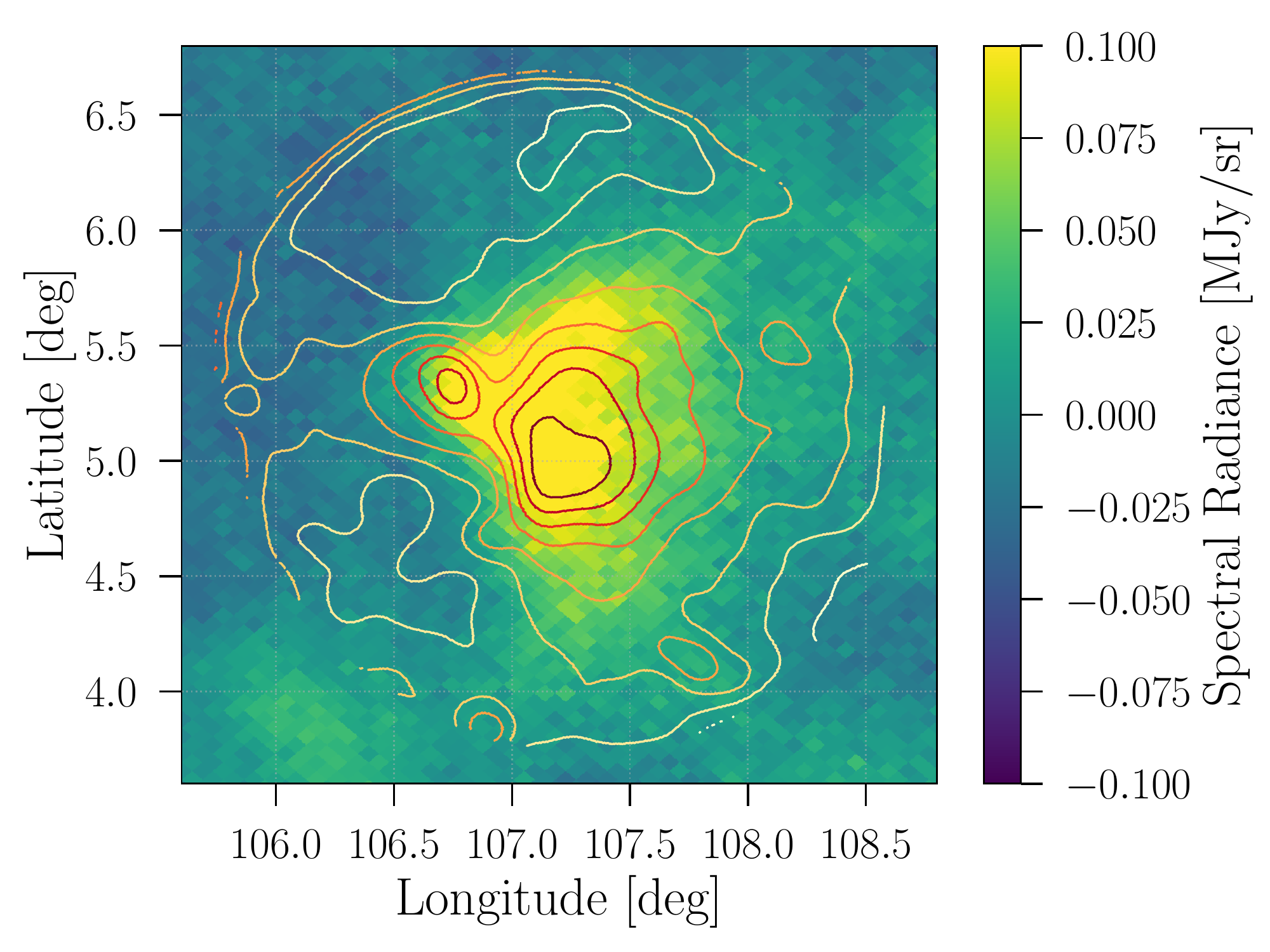}
\hspace{0.1in}
\includegraphics[scale=0.4]{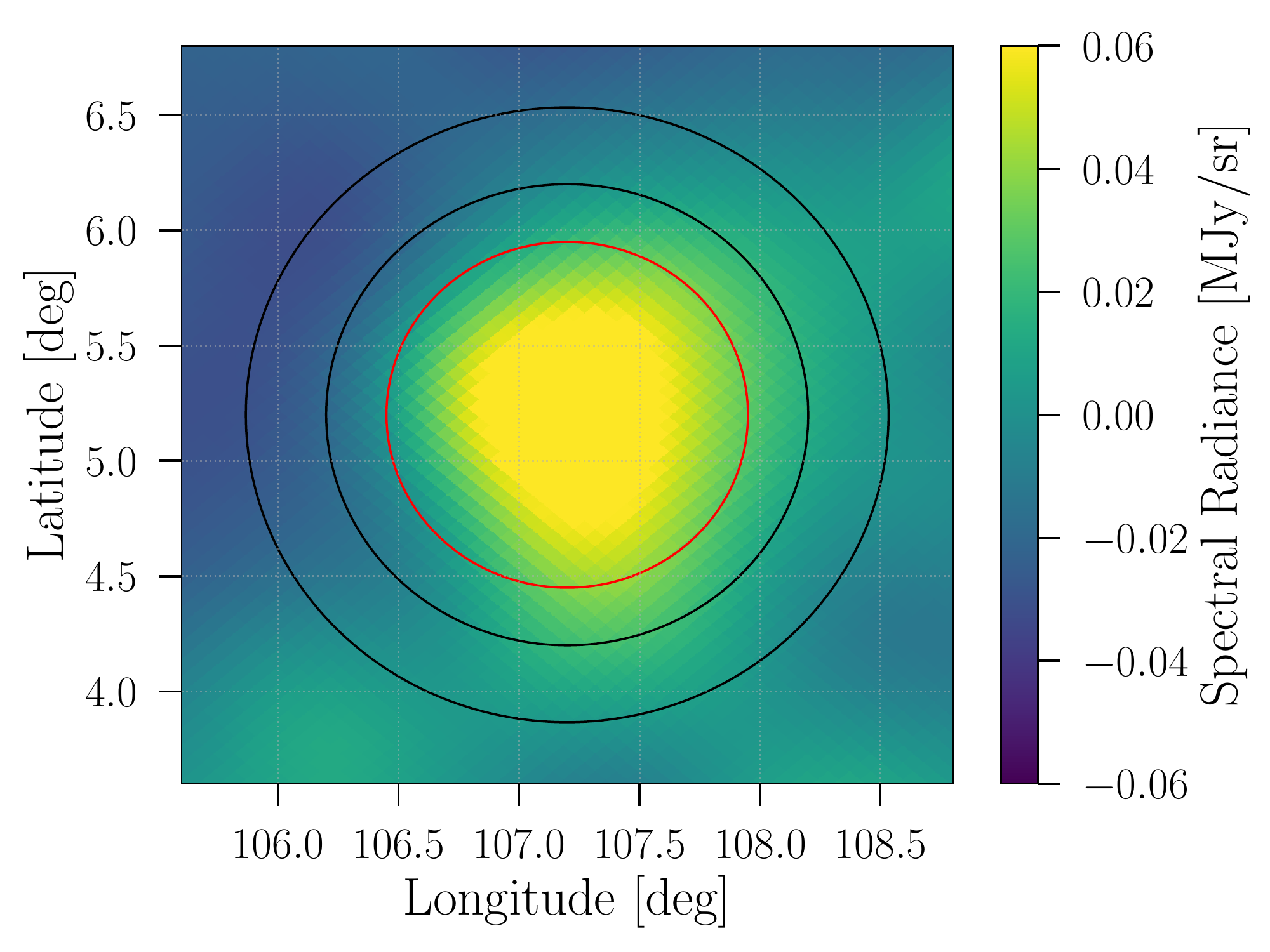}
\includegraphics[scale=0.4]{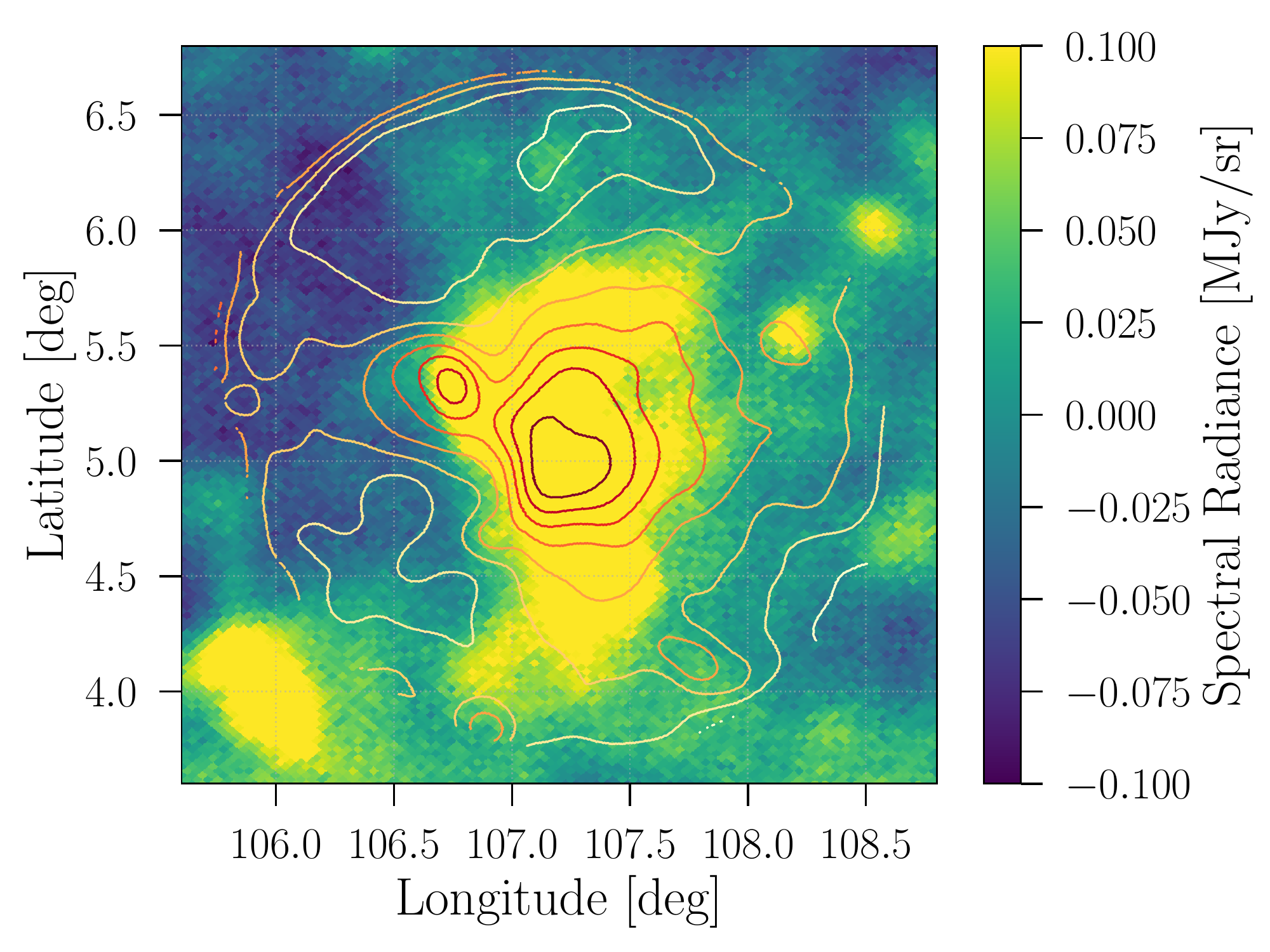}
\hspace{0.1in}
\includegraphics[scale=0.4]{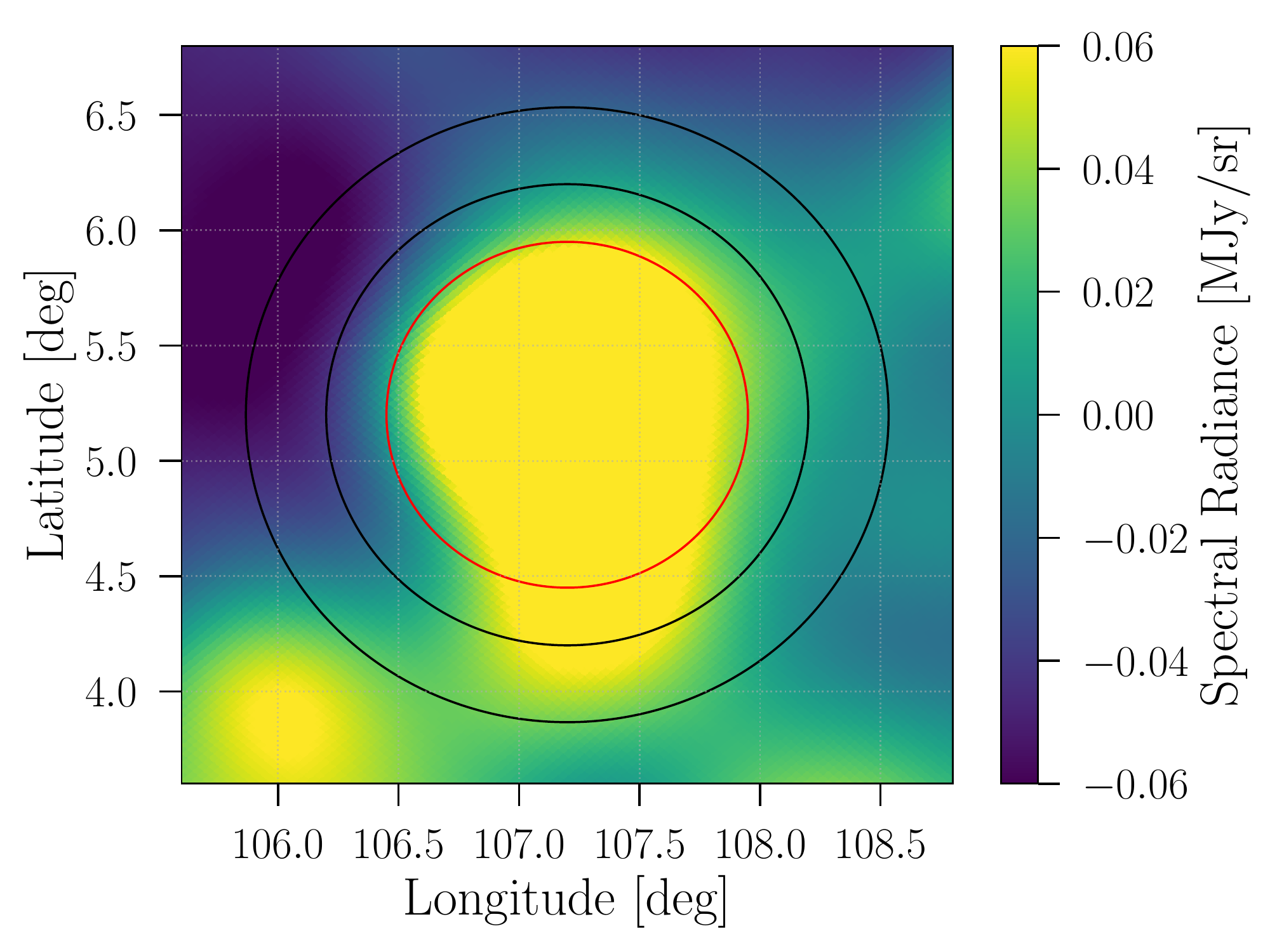}
\caption{
Maps used in this study.
As in Figure~\ref{fig:contours1}, the left column shows the given map with our GBT contours oveplotted for comparison, while the right column shows the map on the left convolved with a 40$^{\prime}$ Gaussian.
Again, the aperture (45$^{\prime}$ radius) and the zero-point annulus ( 60$^{\prime}$ to 80$^{\prime}$) are overplotted.
From top to bottom, the rows are Planck 44~GHz, Planck 70~GHz, and Planck 100~GHz.
Maps in each column are plotted with the same color scale for straightforward comparison.
References are given in Table~\ref{tab:datasets}.
}
\label{fig:contours3}
\end{figure*}

\begin{figure*}
\centering
\includegraphics[scale=0.4]{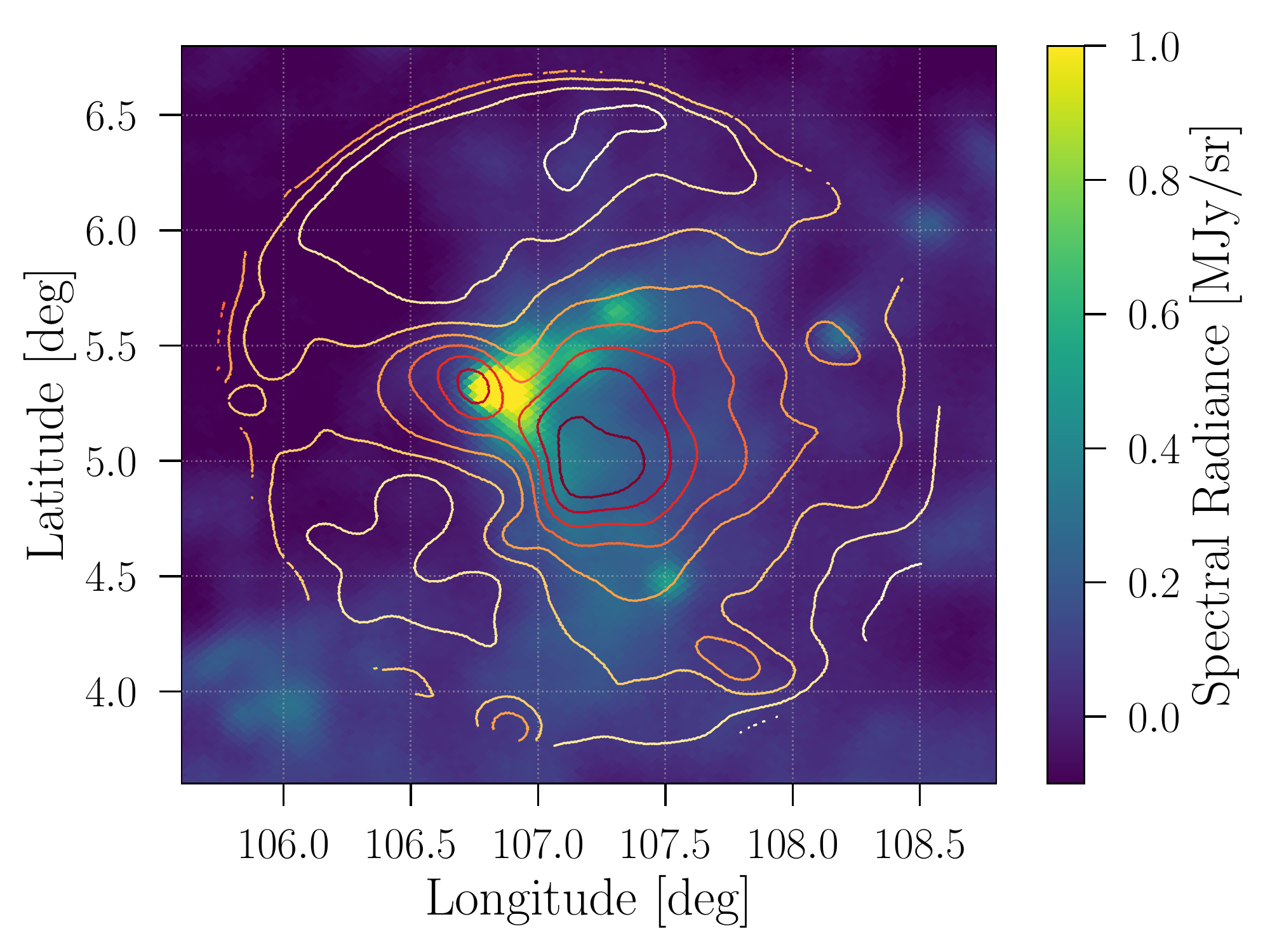}
\hspace{0.1in}
\includegraphics[scale=0.4]{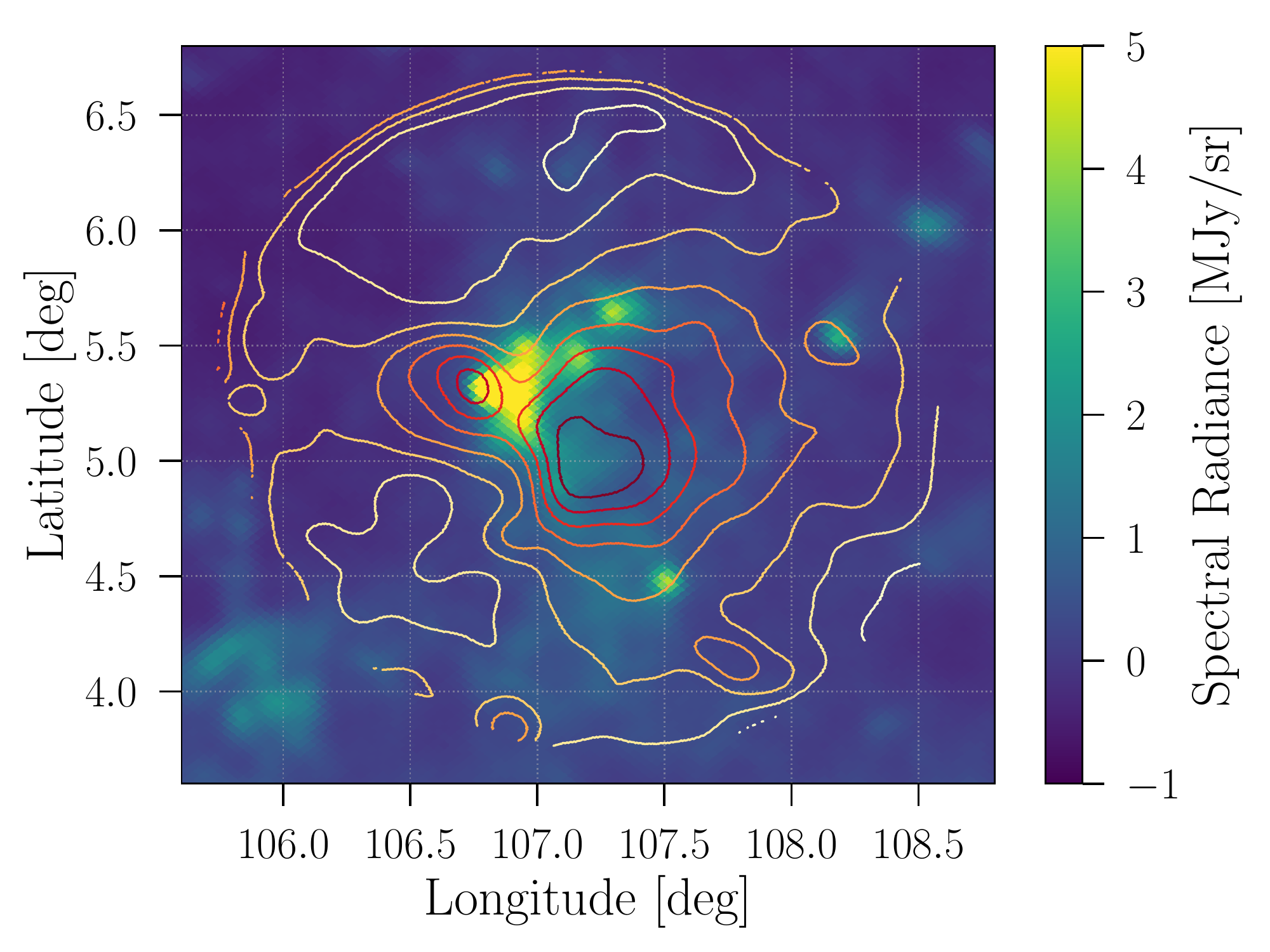}
\includegraphics[scale=0.4]{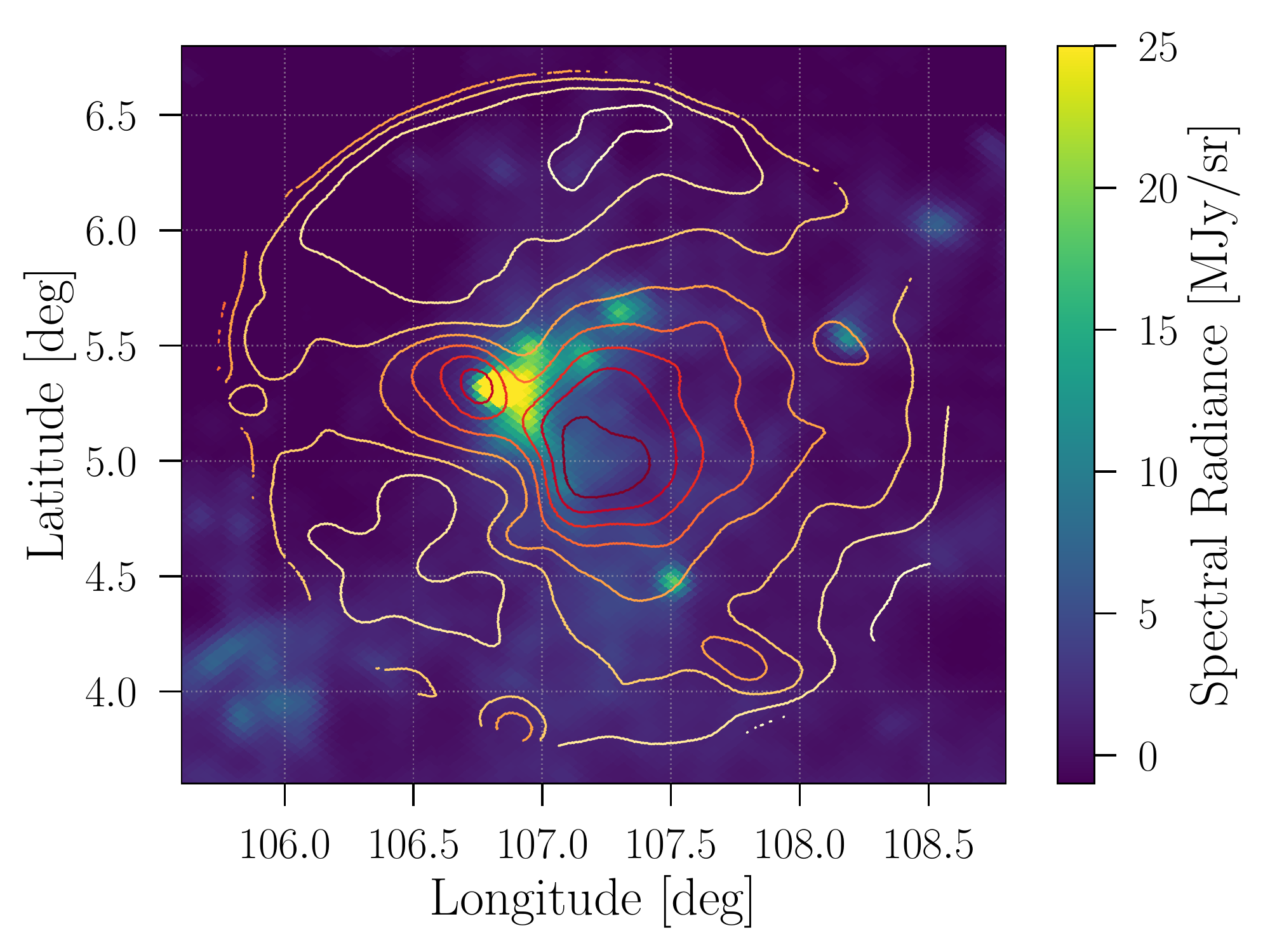}
\hspace{0.1in}
\includegraphics[scale=0.4]{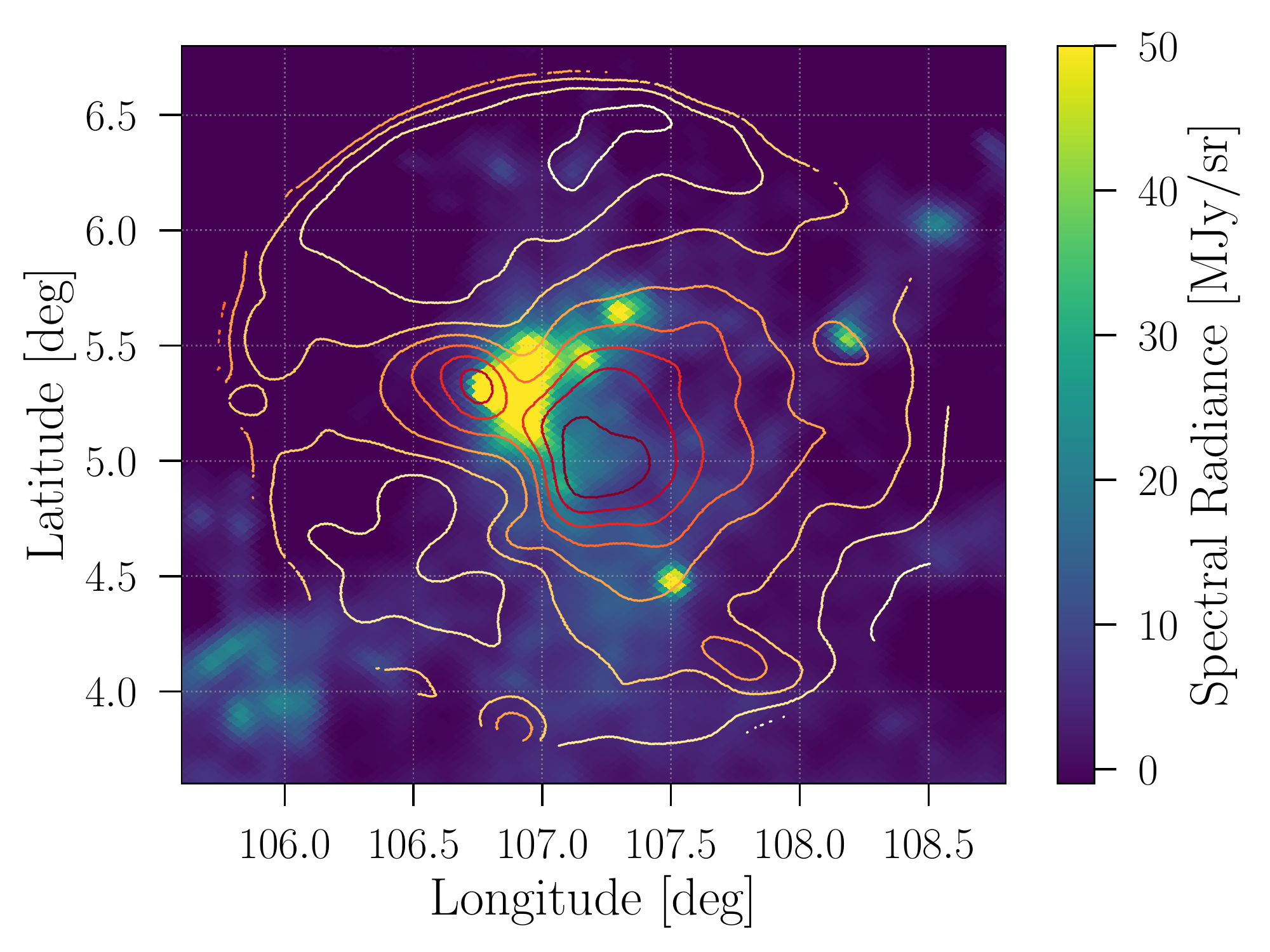}
\includegraphics[scale=0.4]{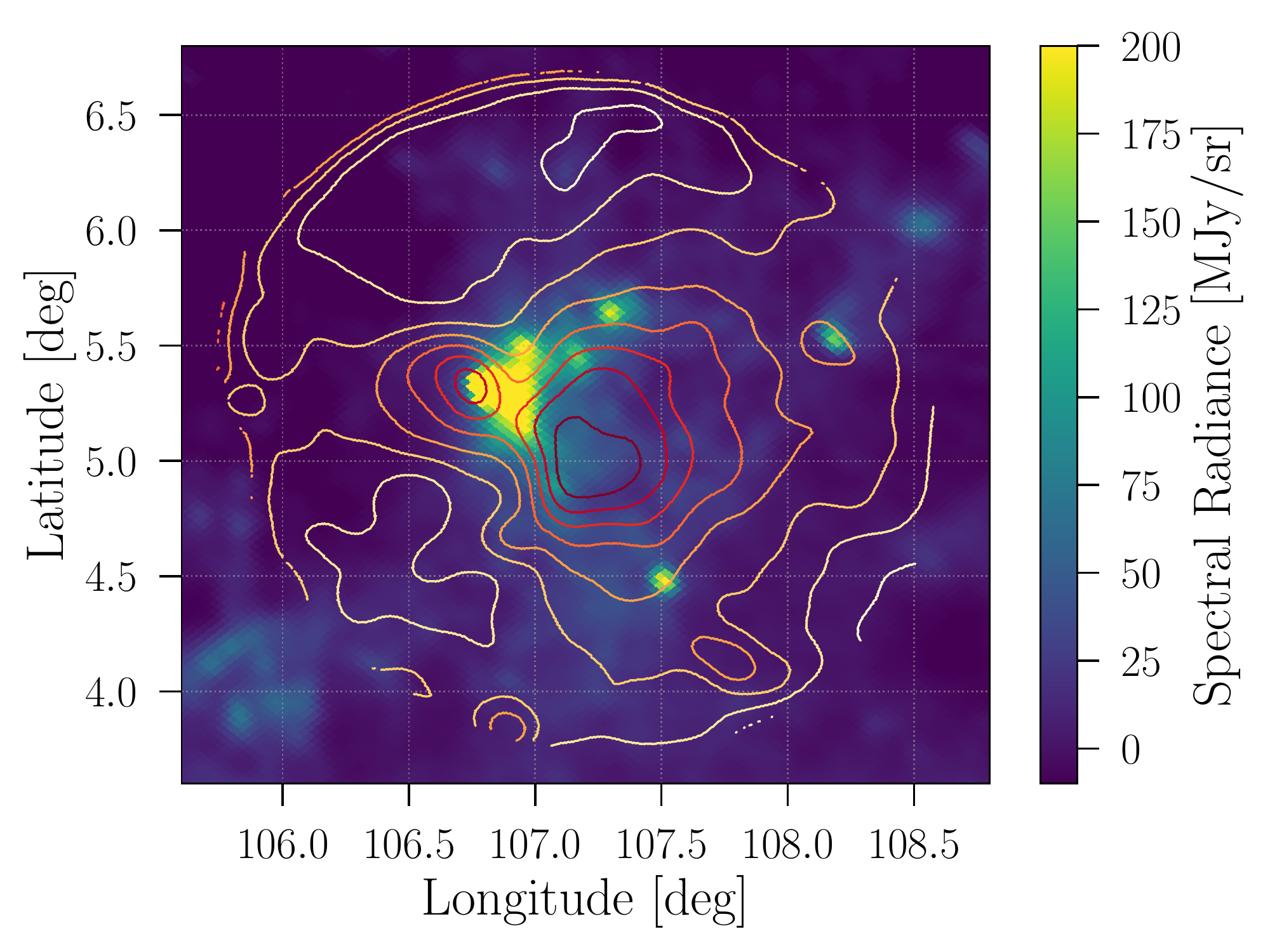}
\caption{
Morphological comparison between our GBT maps and the Planck maps between 143~GHz and 857~GHz.
\textbf{Top Row:} Planck 143~GHz and 217~GHz maps.
\textbf{Middle Row:} Planck 353~GHz and 545~GHz maps.
\textbf{Bottom Row:} Planck 857~GHz map.
The overplotted contours come from our Bank A (4.575~GHz) GBT map.
References are given in Table~\ref{tab:datasets}.
}
\label{fig:contours4}
\end{figure*}

\begin{figure*}
\centering
\includegraphics[scale=0.4]{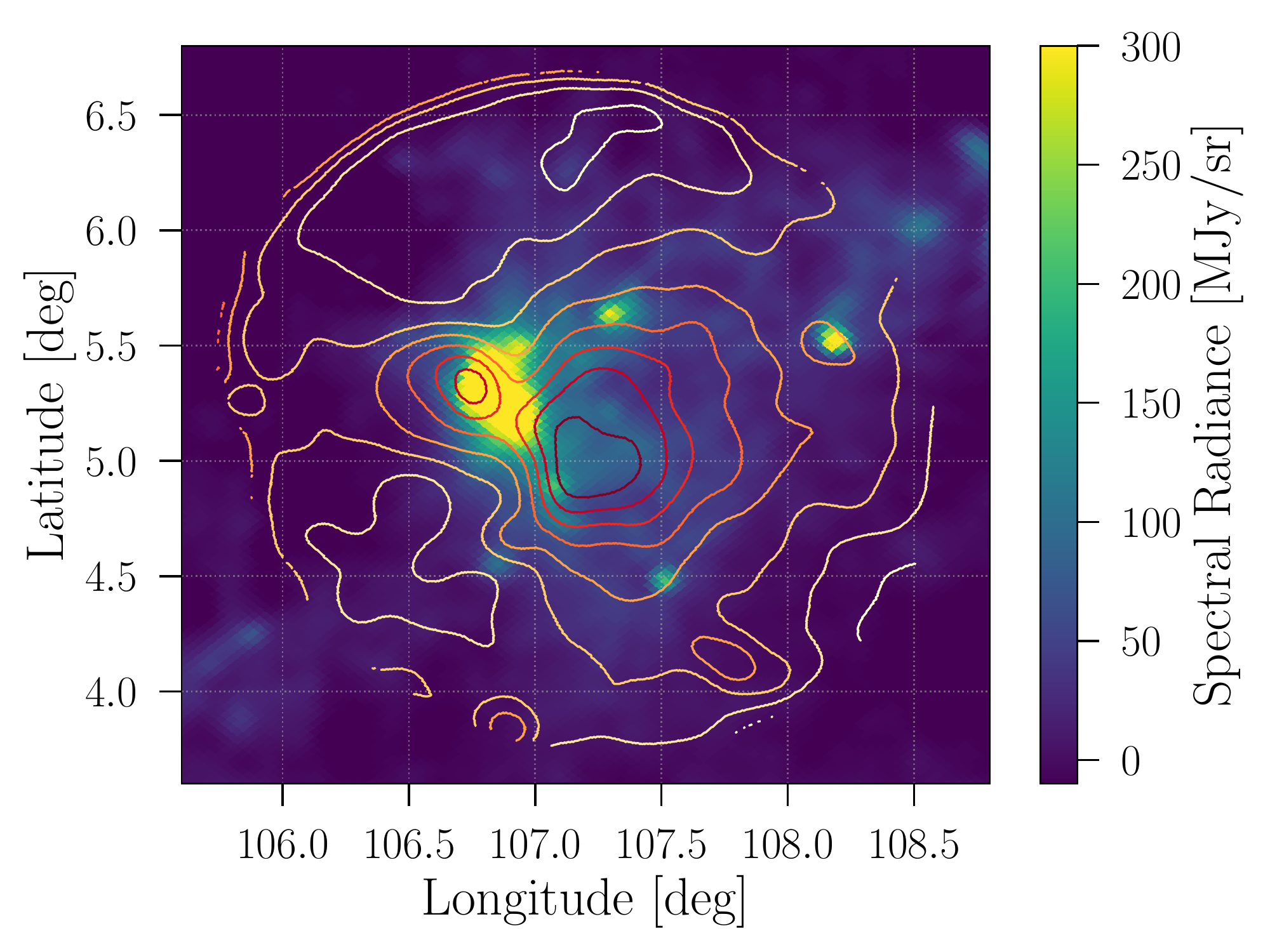}
\hspace{0.1in}
\includegraphics[scale=0.4]{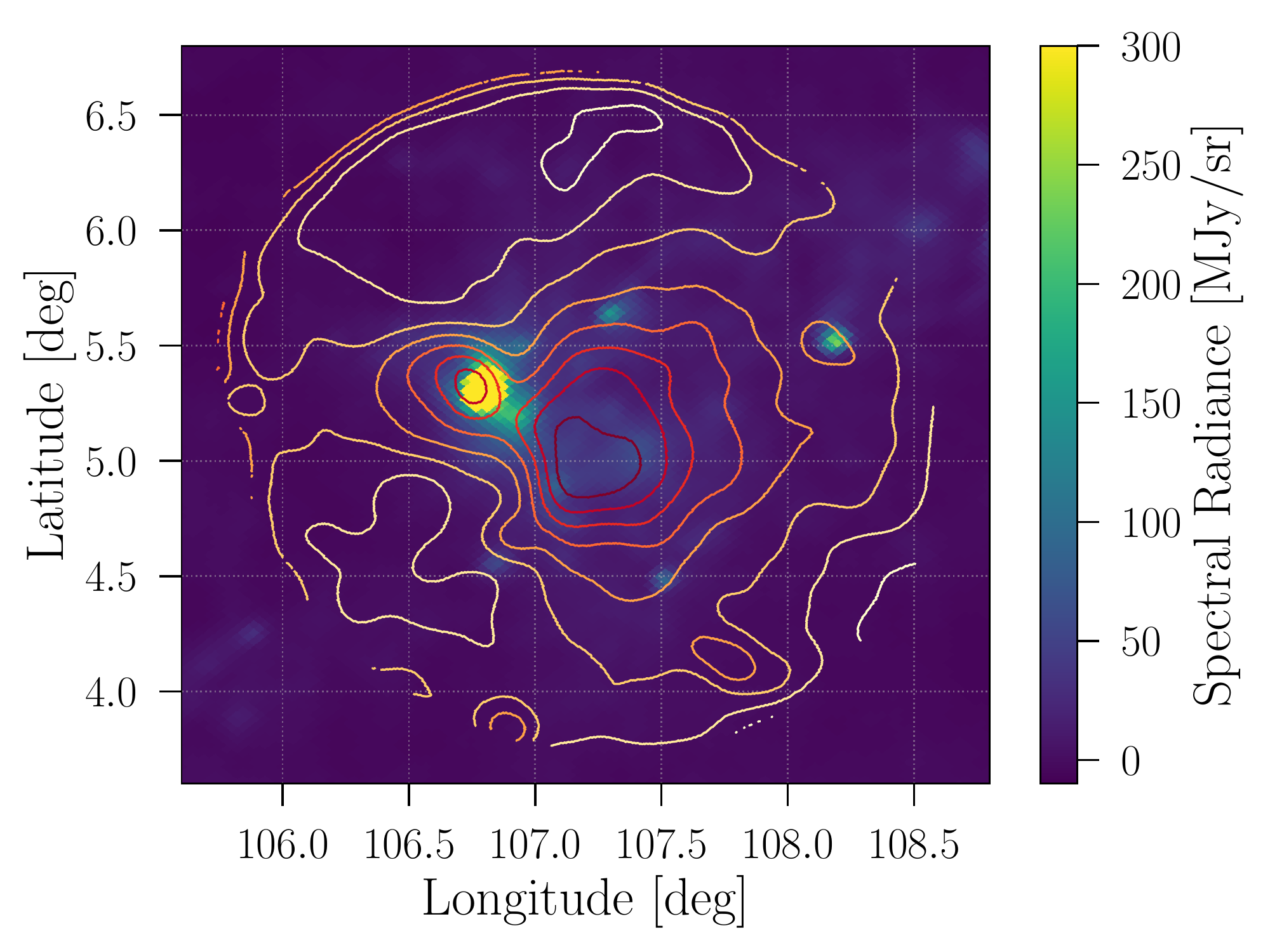}
\includegraphics[scale=0.4]{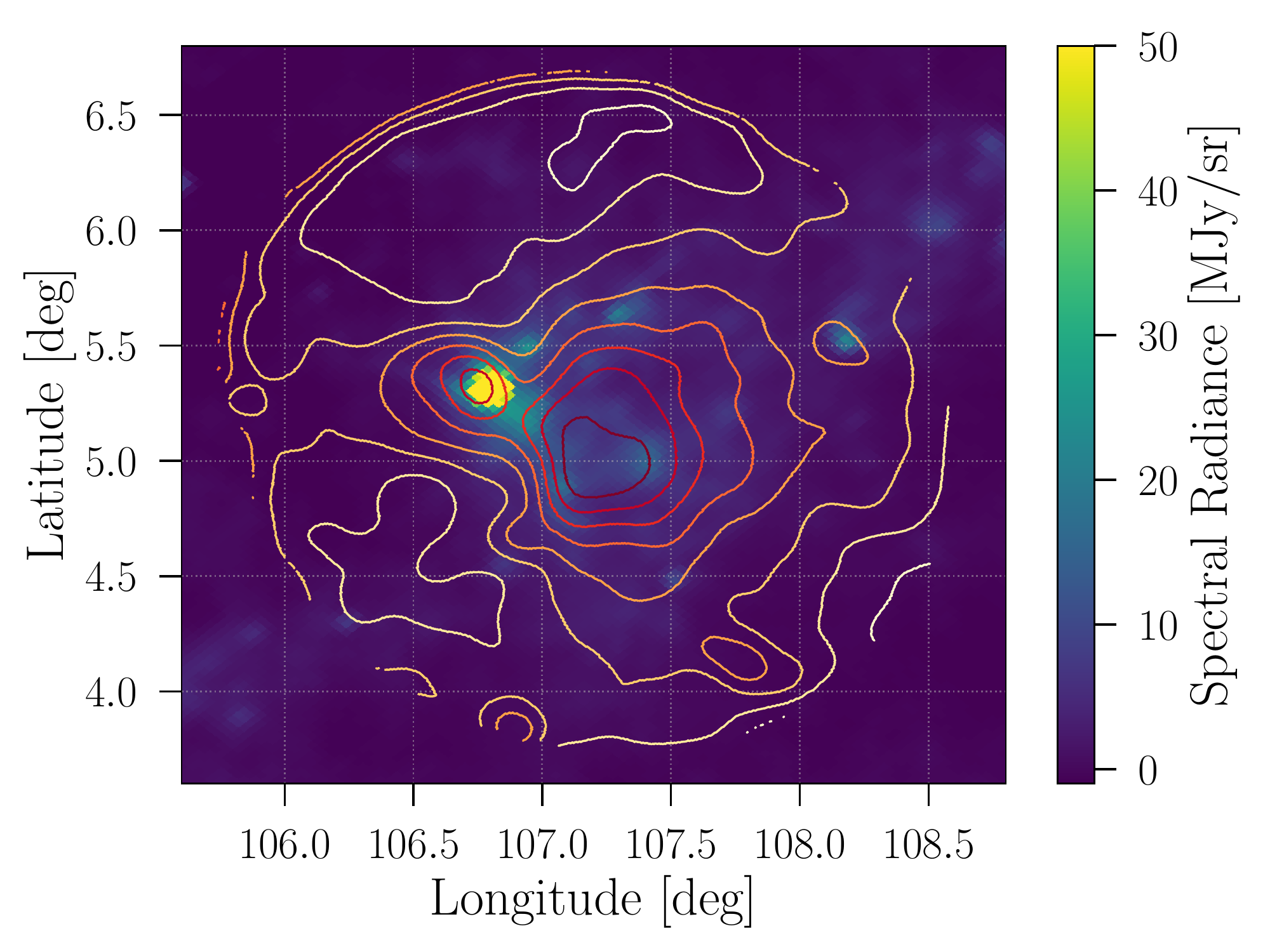}
\hspace{0.1in}
\includegraphics[scale=0.4]{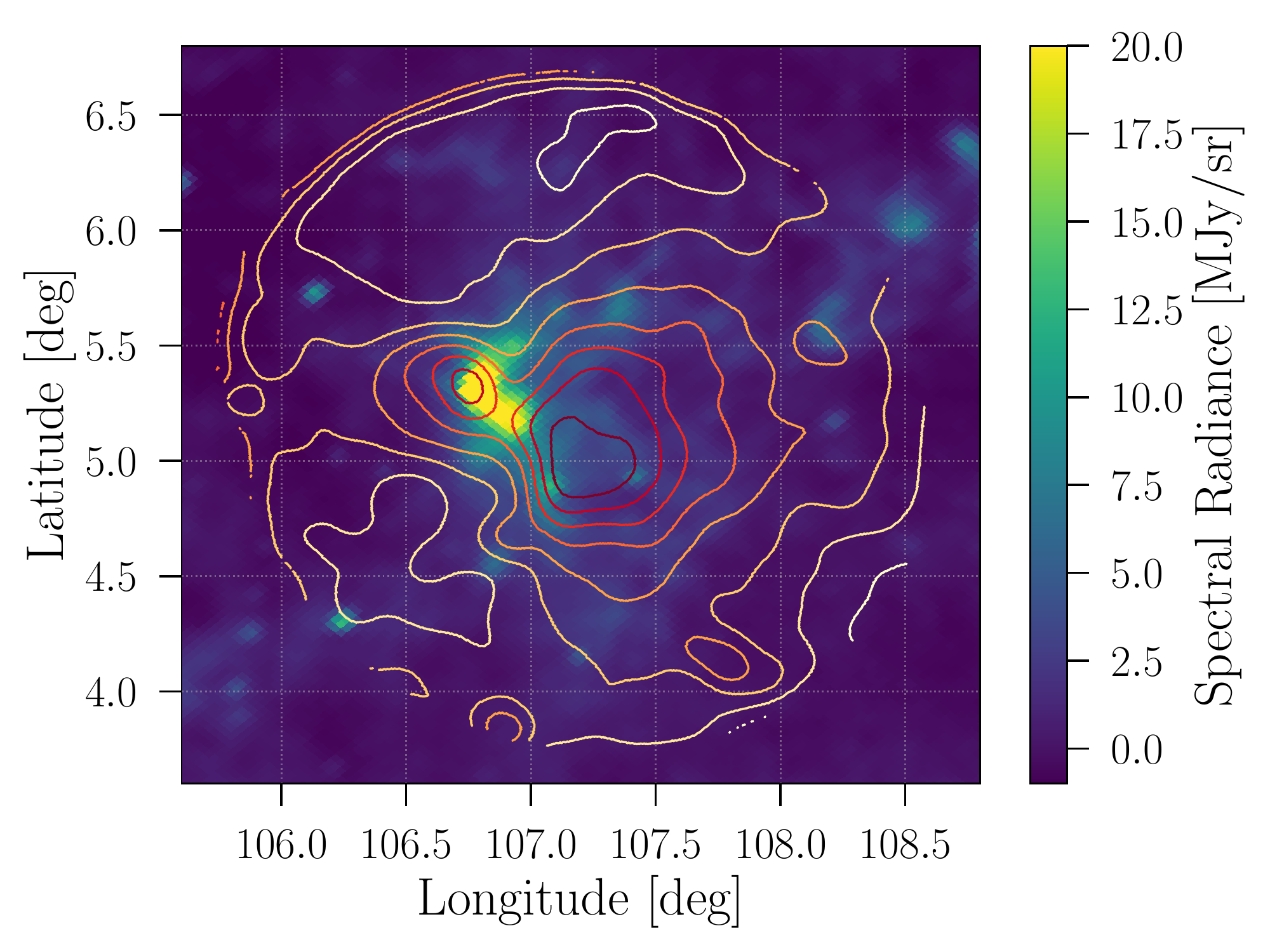}
\caption{
Morphological comparison between our GBT maps and the IRIS maps between 3~THz and 25~THz.
\textbf{Top Row:} IRIS 3~THz and 5~THz maps.
\textbf{Middle Row:}  IRIS 12~THz and 25~THz maps.
The overplotted contours come from our Bank A (4.575~GHz) GBT map.
References are given in Table~\ref{tab:datasets}.
}
\label{fig:contours5}
\end{figure*}

In this Appendix we show the posterior probability distributions from the spectral flux density model fits (see Section~\ref{sec:results}) and all of the maps used in this study.
The posteriors are shown in Figures~\ref{fig:spinningdust_fit}~and~\ref{fig:freefree_fit}, and the associated maximum-likelihood model parameter values are given in Table~\ref{tab:parameters}.
The maps are shown in Figures~\ref{fig:contours1}~to~\ref{fig:contours5}.
The left column of Figure~\ref{fig:contours2} shows the Bank A, B, and C maps from this GBT study.
These three maps are the same three maps shown in the left column of Figure~\ref{fig:bankABmaps}, but smoothed with a 10$^\prime$ FWHM beam to remove noise and point sources.
A contour plot of the smoothed Bank A map in this Figure (the top left panel) is overplotted on all of the maps in Figures~\ref{fig:contours1}~to~\ref{fig:contours5} for a morphological comparison.
This Bank A contour plot clearly shows S140 and the AME cloud centered on G107.2+5.2.
The right column of Figures~\ref{fig:contours1}~to~\ref{fig:contours3} shows the associated map in the left column smoothed to a resolution of 40$^\prime$; the photometry aperture and the zero-point annulus are overplotted for comparison.
The aperture photometry details are given in Section~\ref{sec:photometry}).
For a morphological comparison (see Section~\ref{sec:maps}), the nine highest-frequency maps (143~GHz to 25~THz) are shown in Figures~\ref{fig:contours4}~and~\ref{fig:contours5}.

%%%%%%%%%%%%%%%%%%%%%%%%%%%%%%%%%%%%%%%%%%%%%%%%%%%%%%%%%%%%%%%%%%%%%%%%%%%%%%%%

%%%%%%%%%%%%%%%%%%%%%%%%%%%%%%%%%%%%%%%%%%%%%%%%%%%%%%%%%%%%%%%%%%%%%%%%%%%%%%%%

%%%%%%%%%%%%%%%%%%%%%%%%%%%%%%%%%%%%%%%%%%%%%%%%%%%%%%%%%%%%%%%%%%%%%%%%%%%%%%%%

%%%%%%%%%%%%%%%%%%%%%%%%%%%%%%%%%%%%%%%%%%%%%%%%%%%%%%%%%%%%%%%%%%%%%%%%%%%%%%%%

%%%%%%%%%%%%%%%%%%%%%%%%%%%%%%%%%%%%%%%%%%%%%%%%%%%%%%%%%%%%%%%%%%%%%%%%%%%%%%%%

%%%%%%%%%%%%%%%%%%%%%%%%%%%%%%%%%%%%%%%%%%%%%%%%%%%%%%%%%%%%%%%%%%%%%%%%%%%%%%%%

%%%%%%%%%%%%%%%%%%%%%%%%%%%%%%%%%%%%%%%%%%%%%%%%%%%%%%%%%%%%%%%%%%%%%%%%%%%%%%%%

%%%%%%%%%%%%%%%%%%%%%%%%%%%%%%%%%%%%%%%%%%%%%%%%%%%%%%%%%%%%%%%%%%%%%%%%%%%%%%%%

\end{appendix}

%%%%%%%%%%%%%%%%%%%%%%%%%%%%%%%%%%%%%%%%%%%%%%%%%%%%%%%%%%%%%%%%%%%%%%%%%%%%%%%%

\clearpage

%%%%%%%%%%%%%%%%%%%%%%%%%%%%%%%%%%%%%%%%%%%%%%%%%%%%%%%%%%%%%%%%%%%%%%%%%%%%%%%%

\bibliographystyle{aasjournal}
\bibliography{Lit}

%%%%%%%%%%%%%%%%%%%%%%%%%%%%%%%%%%%%%%%%%%%%%%%%%%%%%%%%%%%%%%%%%%%%%%%%%%%%%%%%

\end{document}